 \documentclass[12pt]{amsart}
\usepackage[margin=1in]{geometry}
\usepackage[applemac]{inputenc}							% Change applemac --> utf8
\usepackage{amsfonts}
\usepackage{amssymb}
\usepackage{mathabx}
\usepackage{braket}
\usepackage{enumitem}
\usepackage{color}
\usepackage[colorlinks ,pdfstartview=FitB]{hyperref}

 \newtheorem{thm}{Theorem}[section]
 \newtheorem{cor}[thm]{Corollary}
 \newtheorem{lemma}[thm]{Lemma}
 \newtheorem{prop}[thm]{Proposition}

 \newtheorem{remark}[thm]{Remark}

\newcommand{\C}{\mathbb{C}}
\newcommand{\R}{\mathbb{R}}
\newcommand{\N}{\mathbb{N}}

\newcommand{\Z}{\mathbb{Z}}

\newcommand{\e}{\textrm e}
\newcommand{\dd}{\textrm{d}}

\DeclareMathOperator{\tr}{tr}
\DeclareMathOperator{\rank}{rank}

\DeclareMathOperator{\spann}{span}

\DeclareMathOperator{\id}{id}

\DeclareMathOperator{\modd}{mod}

\def\idty{{\mathchoice {\mathrm{1\mskip-4mu l}} {\mathrm{1\mskip-4mu l}} %
{\mathrm{1\mskip-4.5mu l}} {\mathrm{1\mskip-5mu l}}}}

\numberwithin{equation}{section}
\numberwithin{figure}{section}

\begin{document}
%%%%%%%%%%%%%%%%%%%%%%%%%%%%%%%%%%%%%%%%%%%%%%%%%%%%%%%%%%%%%
%%%%%%%%%%%%%%%%%%%%%%%%%%%%%%%%%%%%%%%%%%%%%%%%%%%%%%%%%%%%%

\title{Lower Bound to the Entanglement Entropy of the XXZ Spin Ring}
\author[C. Fischbacher]{Christoph Fischbacher$^1$}
\address{		$^1$ Department of Mathematics\\
					University of California, Irvine\\
					Irvine, CA, 92697, USA}
\email{fischbac@uci.edu}

\author[R.\ Schulte]{Ruth Schulte$^2$}
\address{		$^2$ Mathematisches Institut\\
					Ludwig Maximilians Universit\"at, M\"unchen\\
					M\"unchen, 80333, Germany}
\email{schulte@mathematik.uni-muenchen.de}

\maketitle

\begin{abstract} We study the free XXZ quantum spin model defined on a ring of size $L$ and show that the bipartite entanglement entropy of eigenstates belonging to the first energy band above the vacuum ground state satisfies a logarithmically corrected area law. Along the way, we show a Combes-Thomas estimate for fiber operators which can also be applied to discrete many-particle Schr\"odinger operators on more general translation-invariant graphs. 
\end{abstract}

\tableofcontents
% !TEX root = main.tex

\section{Introduction}

Considered to be one of the indicators of many-body-localization (MBL), area laws for the entanglement entropy have attracted significant interest by the physics \cite{Eisertetal, Laflorencie:2016kg} as well as the mathematics community \cite{PhysRevLett.113.150404,MR3744386,ElgartPasturShcherbina2016, Hastings, BW18, ARFS19}. In contrast, delocalization induced by long-range correlations leads to increase stronger than a mere area law. However, only a few examples, where such phenomena were observed, have been rigorously studied as of yet \cite{LeschkeSobolevSpitzer14,LeschkeSobolevSpitzer17,PfirschSobolev18,Wolf:2006ek,Mulleretal,mller2020stability}. It is noteworthy that almost all of these examples are found within the non-interacting setting. In particular, logarithmic corrections of area laws seem to be a common occurrence in physical systems that are known to be delocalized. 

An interacting system that exhibits MBL phenomena is the disordered XXZ spin chain. Recently, localization phenomena for this model have been rigorously studied in \cite{EKS,EKS2, BW17}; see also \cite{Stolz20} for a survey of the newest developments. An area law of the entanglement entropy for low energy states in such a system has been proven in \cite{BW18}. However, the situation for the infinite XXZ spin chain without disorder is fundamentally different: with the help of the Bethe ansatz, it can be shown that the lowest spectral band is purely absolutely continuous with delocalized generalized eigenfunctions \cite{NSS2007, FS14}. It is thus reasonable to also expect a different scaling behavior for the entanglement entropy. In this paper, we therefore consider finite XXZ spin chains of arbitrary size with periodic boundary conditions and constant magnetization density.  

In \cite{BW18}, Beaud and Warzel already showed for low energy states in the finite XXZ chain with droplet boundary conditions, that there is a logarithmic upper bound of the entanglement entropy, independent of non-negative background potentials. The explicit results in \cite[Prop. 1.2]{BW18} and \cite[Thm. 1.2]{ARFS19} show that the log-term is optimal in the Ising model. By developing a suitable perturbational approach, we extend this to the XXZ model in the Ising phase.

It is well-known that the XXZ Hamiltonian preserves the total magnetization and is equiva\-lent to a direct sum of discrete many-particle Schr\"odinger operators of hard-core bosons. In the Ising phase, the associated potential energetically favors clustered configurations \cite{NSS2007,NS, FS14, FS18}, commonly referred to as ``droplets". Thus, the mass of low energy states is mainly concentrated around these droplet configurations. Our perturbative result will rely on the fact that in the Ising limit, droplet configurations and low energy states coincide. This follows from a suitable Combes-Thomas bound which we will show in the first part of this paper.

While the main result of the paper contributes to the question whether the logarithmic upper bound is optimal, we also believe its perturbative approach to be of more general interest. In particular, for the disordered Ising model, it was shown in \cite{ARFS19} that it is possible to find states in the next-highest energy band whose entanglement entropy exhibits a logarithmic lower bound with arbitrary high probability. This suggests delocalization phenomena for higher energy states in the XXZ model despite disorder. 

We will proceed as follows:

In Section \ref{sec:2}, we introduce the XXZ model on the ring and review some of its basic properties. Exploiting its preservation of total magnetization, we decompose the Hamiltonian into a direct sum of operators acting on subspaces of fixed total magnetization. We will also state our main results (Theorems \ref{thm:main} and \ref{thm:EigenfktEst}).

Section \ref{sec:3} is dedicated to obtaining estimates on eigenfunctions of the XXZ Hamiltonian. To this end, we firstly exploit the ring's translational symmetry and define a suitable Fourier transform. We then introduce an equivalent formulation of the XXZ Hamiltonian using Schr\"odinger operators, which we will use to show an appropriate Combes-Thomas estimate \ref{thm:EigenfktEst}. At the end of this section, we consider the Ising model, for which we pick a suitable low-energy state which exhibits the desired logarithmic lower bound. The purpose of the remainder of the paper will therefore be to show that if one does not move too far away from the Ising limit, this logarithmic lower bound persists.

To this end, we then focus in Section \ref{sec:4} on showing that the reduced state of droplet eigenstates in the Ising phase is exponentially close to the reduced state coming from the Ising limit (with respect to a suitable distance function). The underlying geometry of the ring poses some technical difficulties which are overcome by suitable estimates, basically allowing us to treat the model with methods developed for the chain. 

After this, in Section \ref{sec:5}, we estimate the eigenvalues of the difference of these reduced states, which allows us to find suitable bounds of its Schatten-quasinorms. Using a result by Combes, Hislop and Nakamura \cite{MR1824200}, this allows us to estimate the $L^p$-norms of the associated Kre\u\i n's spectral shift function which we then use to show Theorem \ref{thm:main}.
\\\\
\noindent {\bf Acknowledgements:} C.F.~ is grateful to the Institut Mittag-Leffler in Djursholm, Sweden, where some of this work was done as part of the program Spectral Methods in Mathematical Physics in Spring 2019. 
R.S. was funded by Deutsche Forschungsgemeinschaft under Germany's Excellence Strategy - EXC-2111 - 390814868 and LMUMentoring.
It is also our pleasure to thank Peter M\"uller and G\"unter Stolz for helpful discussions as well as encouragement and support.
% !TEX root = main.tex

\section{Model and main results} \label{sec:2}
%%%%%%%%%%%%%%%%%%%%%%%%%%%%%%%%%%%%%%%%%%%%%%%%%%%%%%%%%%%%%%%%%
For any $L\in\N$, consider the XXZ model on a discrete ring of size $L$. We start by describing the ring using the graph $\mathcal{G}_L:=(\mathcal{V}_L,\mathcal{E}_L )$ with vertex set $\mathcal{V}_L:=\{0,1,\dots,L-1\}$ and edge set $\mathcal{E}_L:=\{\{j,(j+1)\modd L\}:\;j\in\mathcal V_L\}$.

The underlying $2^L$-dimensional Fock space $\mathbb{H}_L$ is given by $\mathbb{H}_L=\bigotimes_{j\in\mathcal{V}_L}\C^2$. Let  $|\uparrow\rangle:=(1\: 0)^t$ and $|\downarrow\rangle:=(0\: 1)^t$ denote the canonical basis of $\C^2$. To construct a basis for the Fock-space we define the spin lowering operator
	\begin{equation}
		S^-:=\left(\begin{array}{cc}
							0	&	0\\
							1	&	0
						\end{array}\right).
	\end{equation}
For any set $\mathcal A$ let $\mathcal P(\mathcal A)$ denote its power set. We now introduce a canonical basis $\big\{\ket{\delta_x^L}\big\}_{x\in\mathcal{P}(\mathcal{V}_L)}$ of  $\mathbb{H}_L$ by $\ket{\delta_\emptyset^L}:=\ket{\uparrow}^{\otimes L}$ and for any other $x\in\mathcal{P}(\mathcal{V}_L)$ by
	\begin{equation}
		\ket{\delta_x^L}:=\prod_{j\in x}S^-_j\ket{\delta_\emptyset^L}.
	\end{equation}
Here, and in the following, for any $A\in\C^{2\times 2}$ the notation $A_j$ refers to a spin operator acting as $A$ on the site $j\in\mathcal V_L$ and as the identity else. 

 The XXZ-Hamiltonian $H_L:\;\mathbb H_L\rightarrow \mathbb H_L$  is given by
	\begin{equation}
		H_L\equiv H_L(\Delta):=\sum_{\{j,k\}\in\mathcal{E}_L}h_{jk}(\Delta),
	\end{equation}
where the two-site operator $h_{jk}$ describes an interaction between two spins located at the two sites $\{j,k\}\in\mathcal{E}_L$. It is given by
	\begin{equation}
		h_{jk}\equiv h_{jk}(\Delta):=\left(\frac{1}{4}-S^3_jS^3_k\right)-\frac{1}{\Delta}\left(S^1_jS^1_k+S^2_jS^2_k\right),
	\end{equation}
with $S^1,\,S^2$ and $S^3$ being the standard spin--1/2 matrices
	\begin{equation}
		S^1:=\begin{pmatrix} 0 & 1/2\\ 1/2 & 0 \end{pmatrix},\quad S^2:=\begin{pmatrix} 0 & -i/2\\ i/2 & 0 \end{pmatrix}\quad\mbox{and}		\quad S^3:=\begin{pmatrix} 1/2 & 0\\ 0 & -1/2 \end{pmatrix}.
	\end{equation}
In the following, we will assume that the \emph{anisotropy parameter} $\Delta$ satisfies $\Delta>1$, which ensures that for each $\{j,k\}\in\mathcal{E}_L$, we have $h_{jk}\geq 0$ and consequently $H_L\geq 0$. The case $\Delta> 1$ is commonly referred to as the ``Ising-phase" of the XXZ model. The two-dimensional ground--state space of $H_L$ corresponding to the ground state energy $E_0=0$ is the linear span of the two vectors $\ket{\delta_\emptyset^L}$ (``all spins-up") and $\ket{\delta^L_{\mathcal V_L}}$ (``all spins-down"). 
The operator $H_L$ preserves the total magnetization (for more details see \cite{FS18}). We therefore treat each down-spin as a particle. For all $N\in\{0,\hdots,L\}$, let us define the $N$-particle subspace by
	\begin{equation}
		\mathbb H_L^N:=\spann\{\ket{\delta_x^L}:\;x\in\mathcal P(\mathcal V_L),\;|x|=N\}.
	\end{equation}
Since $H_L$ is particle number preserving, each $\mathbb H_L^N$ reduces the operator $H_L$. Hence, we express it as the direct sum
	\begin{equation}
		H_L=\bigoplus_{N=0}^LH_L^N,
	\end{equation}
where $H_L^N:=H_L\upharpoonright_{\mathbb H_L^N}$ for all $N\in\{0,1,\dots,L\}$. The operators $H_L^L$ and $H_L^0$ are identical to the zero operator on $\mathbb H_L^L=\spann\{\ket{\delta^L_{\mathcal V_L}}\}$ and $\mathbb H_L^0=\spann\{\ket{\delta^L_\emptyset}\}$ respectively.

Now, let $\Lambda\subset\mathcal V_L$ and  $\mathbb H_\Lambda:=\bigotimes_{j\in\Lambda}\C^2$. Given any normalized state $\ket{\psi}\in\mathbb H_L$, we consider its entanglement entropy with respect to the spatial decomposition $\mathbb H_L=\mathbb H_\Lambda\otimes \mathbb H_{\Lambda^c}$.
As before, we denote  by $\{\ket{\delta^\Lambda_x}\}_{x\in\mathcal P(\Lambda)}$ the canonical basis of $\mathbb H_\Lambda$. Analogously, we define the $N$-particle subspace  by
	\begin{equation}
		\mathbb H_\Lambda^N:=\textrm{span}\{\ket{\delta^\Lambda_x}:x\in\mathcal P(\Lambda),\;|x|=N\}.
	\end{equation}
for all $N\in\{0,\hdots,|\Lambda|\}$.
This choice allows the convenient identification
	\begin{equation}\label{eq:BasisDecomposition}
		\ket{\delta_{x\cup y}^L}=\ket{\delta^\Lambda_{x}}\otimes\ket{\delta^{\Lambda^c}_y}\in\mathbb H_L
	\end{equation}
for any $x\in\mathcal P(\Lambda)$ and $y\in\mathcal P(\Lambda^c)$.
 Let $\rho(\psi):=\ket{\psi}\!\bra{\psi}$ denote the density operator corresponding to $\ket{\psi}$ and moreover let $\rho_\Lambda(\psi):=\tr_{\Lambda^c}\{\rho(\psi)\}\in L(\mathbb H_{\Lambda})$ denote the respective partial trace  over $\mathbb H_{\Lambda^c}$. The entanglement entropy of $\psi$ is given by
 	\begin{equation}
		S(\psi, \Lambda):=\tr s(\rho_\Lambda(\psi)),
	\end{equation}
where $s:\,[0,1]\rightarrow \R$, $x\mapsto -x\ln x$.

Our main result concerns the entanglement entropy of low energy states whose eigenenergy belongs to the interval $I_{1}\equiv I_{1}(\Delta):=[1-\frac{1}{\Delta},2(1-\frac{1}{\Delta}))$.
%%%%%%%%%%%%%%%%%%%%%%%%%%%%%%%%%%%%%%%%%%%%%%%%%%%
\begin{thm}\label{thm:main}
Let $\epsilon\in(0,1/16)$, let $\theta\in(\epsilon,1/16)$. For $L\in\N$, let $N\equiv N(L):=\lfloor\epsilon L\rfloor$ and $\Lambda_L:=\{0,\hdots,2\lfloor \theta L\rfloor\}\subset \mathcal V_L$. Then there exists $\Delta_0\equiv\Delta_0(\epsilon)>3$ such that for all $\Delta\ge \Delta_0$, $L\in\N$ and $E\in\sigma(H_L^N)\cap I_{1}$ there exists a corresponding eigenstate $\ket{\varphi_L^N(\Delta,E)}\in\mathbb H_L^N$ such that
	\begin{equation}
		\liminf_{L\rightarrow\infty} \frac{S(\varphi_L^N(\Delta,E),\Lambda_L)}{\ln L}\ge \frac{\epsilon}{2}.
	\end{equation}
\end{thm}
%%%%%%%%%%%%%%%%%%%%%%%%%%%%%%%%%%%%%%%%%%%%%%%%%%%%
\begin{remark}
	\begin{enumerate}[label=(\roman*)]
		\item While we have made the particular choice for $\Lambda_L$ to scale proportionally to the ring size $L$ and not independently of it, our result nevertheless shows 
		that an area law could not possibly be true in the generic case. 
		Similar choices are often considered in the physics literature \cite{PhysRevB.85.094417,PhysRevB.89.115104, Vidaletal}. 
		\item While we were not able to find any reference in the literature, we expect that for almost every $\Delta>3$, the multiplicity of an eigenvalue within the droplet band is at most two. This would imply that the result of Theorem \ref{thm:main} holds for every eigenfunction in the droplet band.
	\end{enumerate}
\end{remark}
%%%%%%%%%%%%%%%%%%%%%%%%%%%%%%%%%%%%%%%%%%%%%%%%%%%%
An important ingredient for the proof of Theorem \ref{thm:main} is the following estimate showing that low-energy eigenfunctions are mainly concentrated around droplet configurations, which are given by 
	\begin{equation} 
		\mathcal{V}_{L,1}^N:=\{\{j,(j+1)\modd L,\hdots,(j+N-1)\modd L\}:\;j\in\mathcal V_L\}.
	\end{equation}
This result follows from a Combes-Thomas estimate similar to those shown in \cite{EKS, ARFS19}. The main new feature of our estimate here is that the ring's symmetry is taken into account which allows to obtain an additional factor of $L^{-1/2}$. 

%%%%%%%%%%%%%%%%%%%%%%%%%%%%%%%%%%%%%%%%%%%%%%%%%%%%
\begin{thm}\label{thm:EigenfktEst}
Let $L,N\in\N$ with $N<L$ and $\Delta>3$. For any $E\in\sigma(H_L^N)\cap I_{1}$ there exists a corresponding eigenstate $\ket{\varphi_L^N}\equiv\ket{\varphi_L^N(\Delta,E)}\in\mathbb H_L^N$ such that 
	\begin{equation}\label{eq:EigenfctEstthm}
		|\braket{\delta_x^L,\varphi_L^N}|\leq \frac{2^4}{\sqrt{L}}\cdot\e^{-\mu_1 d_L^N({x,{\mathcal{V}}_{L,1}^N})},
	\end{equation}
for all $x\in\mathcal{V}_L^N$. Here, $d_L^N$ is the $N$-particle graph distance as defined in Section~\ref{sec:Fourier} and
	\begin{equation}
		\mu_1\equiv\mu_1(\Delta):=\ln\Big(1+\frac{(\Delta-1)}{8}\Big).
	\end{equation}
\end{thm}

% !TEX root = main.tex

%%%%%%%%%%%%%%%%%%%%%%%%%%%%%%%%%%%%%%%%%%%%%%%%%%%%%%%%%%%%%%%%%%
%%%%%%%%%%%%%%%%%%%%%%%%%%%%%%%%%%%%%%%%%%%%%%%%%%%%%%%%%%%%%%%%%%
\section{Estimating eigenfunctions}\label{sec:3}
%%%%%%%%%%%%%%%%%%%%%%%%%%%%%%%%%%%%%%%%%%%%%%%%%%%%%%%%%%%%%%%%%%
%%%%%%%%%%%%%%%%%%%%%%%%%%%%%%%%%%%%%%%%%%%%%%%%%%%%%%%%%%%%%%%%%%

%%%%%%%%%%%%%%%%%%%%%%%%%%%%%%%%%%%%%%%%%%%%%%%%%%%%%%%%%%%%%%%%%%
\subsection{Fourier transform}\label{sec:Fourier}
%%%%%%%%%%%%%%%%%%%%%%%%%%%%%%%%%%%%%%%%%%%%%%%%%%%%%%%%%%%%%%%%%%
%%%%%%%%%%%%%%%%%%%%%%%%%%%%%%%%%%%%%%%%%%%%%%%%%%%%%%%%%%%%%%%%%%

For the entire section let $L,N\in\N$ with $N<L$. Since we are mainly interested in the $N$-particle subspace $\mathcal H_L^N$, we introduce the graph of $N$-particle configurations first. 

Recall that the spin ring is described by the graph $\mathcal G_L:=(\mathcal V_L,\mathcal E_L)$. The corresponding graph distance between two sites $j,k\in\mathcal V_L$  is given by
	\begin{equation}
		d_L(j,k)={L}/{2}-||j-k|-{L}/{2}|.
	\end{equation}
Following \cite{FS18}, we construct the $N$-th symmetric product $\mathcal{G}_L^N:=(\mathcal{V}_L^N,\mathcal{E}_L^N)$ of $\mathcal{G}_L$, where
	\begin{align}
		\mathcal{V}_L^N:=\{x\subseteq\mathcal{V}_L: |x|=N\}\quad\mbox{and}\quad\mathcal{E}_L^N:=\{\{x,y\}\subseteq
		\mathcal 		V_L^N: x\triangle y\in\mathcal{E}_L\}.
	\end{align}
Here, $x\triangle y$ denotes the symmetric difference between the two subsets $x,y\subseteq\mathcal{V}_L$. We also write $x\sim y$ for $\{x,y\}\in\mathcal{E}_L^N$. Finally, $d^N_L(\cdot,\cdot)$ denotes the graph distance on $\mathcal{G}_L^N$. As in \cite{FS18}, we identify $\mathbb H_L\cong\ell^2(\mathcal P(\mathcal V_L))$ and $\mathbb H_L^N\cong\ell^2(\mathcal  V_L^N)$. 

In order to exploit the ring's translational symmetry, we define a suitable Fourier-transform. To this end, for any $\gamma\in\Z$, we define the translations ${T}_L^\gamma:\,\mathcal P(\mathcal V_L)\rightarrow\mathcal P(\mathcal V_L)$ by
	\begin{equation}
		 T_L^\gamma x=\{(j+\gamma)\modd L: j\in x\}\quad\textrm{ for all }x\subseteq\mathcal V_L.
	\end{equation}
For every $\gamma\in\Z$, the unitary translation operator $\tilde T_L^\gamma:\,\ell^2(\mathcal P(\mathcal V_L))\rightarrow \ell^2(\mathcal P(\mathcal V_L))$ is given by 
	\begin{equation}
		(\tilde T_L^\gamma \psi)(x)=\psi( T_L^\gamma x)\quad\textrm{ for all }\psi\in \ell^2(\mathcal P(\mathcal V_L)),\;x\in\mathcal P(\mathcal V_L).
	\end{equation}	
Due to translational symmetry of $H_L$, we then get $[\tilde T_L^\gamma,H_L]=0$ for any $\gamma\in\Z$. 

Now, let ``$\approx$" denote the equivalence relation on $\mathcal{V}_L^N\times\mathcal{V}_L^N$ defined as 
	\begin{equation}
		x\approx y:\Leftrightarrow \exists \gamma\in\{0,1,\dots,L-1 \}\mbox{ such that }  T^\gamma_L x=y.
	\end{equation}
Moreover, let $\widehat{\mathcal V}_L^N \subset\mathcal{V}_L^N$ be a fixed set of representatives of each equivalence class induced by ``$\approx$".  For an element $\hat{x}\in\widehat{\mathcal{V}}_L^N$, we denote the corresponding equivalence class by $[\hat{x}]$.  We define $\hat d^N_L:\,\widehat{\mathcal{V}}_L^N\times\widehat{\mathcal{V}}_L^N\rightarrow{\N_0}$ by 

	\begin{equation}
		\hat d^N_L(\hat x,\hat y):=\min_{\gamma\in\Z}d_L^N(\hat x,T_L^\gamma\hat y)\textrm{ for  all }\hat x,\hat y\in 
		\widehat{\mathcal V}^N_L.
	\end{equation}
%%%%%%%%%%%%%%%%%%%%%%%%%%%%%%%%%%%%%%%
\begin{lemma}
$\hat{d}^N_L$ is a metric on $\widehat{\mathcal V}^N_L$.
\end{lemma}
\begin{proof}
Since $d^N_L$ is a metric, if $\hat{d}^N_L(\hat{x},\hat{y})=0$, this means that there exists a $\gamma\in\{0,1,\dots,L-1\}$ such that $\hat{x}=T^\gamma_L \hat{y}$. By definition of $\widehat{\mathcal V}_L^N$, this implies that $\hat{x}=\hat{y}$.
	
Now, for any $\hat{x},\hat{y}\in\widehat{\mathcal{V}}_L^N$ let us consider
	\begin{equation}
		\hat{d}_L^N(\hat{x},\hat{y})=\min_{\gamma\in \Z}d^N_L(\hat{x},T_L^\gamma \hat{y})=\min_{\gamma} d^N_L(T_L^{-\gamma}			\hat{x},\hat{y})=\min_{\gamma}d^N_L(\hat{y},T_L^{-\gamma}\hat{x})=\hat{d}^N_L(\hat{y},\hat{x}).
	\end{equation}	
Finally, for any $\hat{x},\hat{y},\hat{z}\in\widehat{\mathcal{V}}_L^N$ and any $\sigma\in\{0,1,\dots,L-1\}$ consider
	\begin{equation}\begin{split}
		\hat{d}_L^N(\hat{x},\hat{z})&=\min_{\gamma}d_L^N(\hat{x},T_L^\gamma \hat{z})\leq \min_{\gamma}(d_L^N(\hat{x},
		T_L^\sigma \hat{y})+d_L^N(T_L^\sigma \hat{y}, T_L^\gamma \hat{z}))\\
		&=d_L^N(\hat{x},T_L^\sigma \hat{y})+\min_{\gamma}d_L^N(\hat{y},T_L^{\gamma-\sigma}\hat{z})=d_L^N(\hat{x},T_L^\sigma 		\hat{y})+\hat{d}_L^N(\hat{y},\hat{z}).
	\end{split}\end{equation}
Minimizing over $\sigma\in\Z$ now yields the desired triangle inequality $\hat{d}_L^N(\hat{x},\hat{z})\leq\hat{d}_L^N(\hat{x},\hat{y})+\hat{d}_L^N(\hat{y},\hat{z})$ and thus the lemma.
\end{proof}
%%%%%%%%%%%%%%%%%%%%%%%%%%%%%%%%%%%%%%%
We note that not all equivalence classes have the same cardinality. In fact, for any  $\hat x\in\widehat{\mathcal V}_L^N$ the number of elements in $[\hat x]$ is given by
	\begin{equation}
		n_{\hat x}:=|[\hat x]|=\min\{\gamma\in\N: \;T_L^\gamma \hat x=\hat x\}.
	\end{equation}
Moreover, for any $\hat x\in\widehat{\mathcal V}_L^N$ the number $n_{\hat x}$ divides $L$.
Let us now define the unitary Fourier transform. To this end, let 
	\begin{equation}
		\mathbb S_L^N:=\bigg\{\phi\in  \ell^2(\mathcal V_L\times \widehat{\mathcal{V}}_L^N):\; \forall \hat x\in\widehat{\mathcal V}_L^N	
		\;\forall \gamma\notin \frac L{n_{\hat x}}\{0,\hdots,n_{\hat x}-1\}\textrm{ we have }\phi(\gamma,\hat x)=0\bigg\}.
	\end{equation}
The scalar product $\braket{\cdot,\cdot}_{\mathbb S_L^N}$ on this space is defined in the following way: 
	\begin{equation}
		\braket{\phi_1,\phi_2}_{\mathbb S_L^N}:=\sum_{\gamma\in\mathcal V_L}\sum_{\hat x\in\widehat{\mathcal V}_L^N}
		\frac{1}{L/n_{\hat x}}\overline{\phi_1(\gamma,\hat x)}\phi_2(\gamma,\hat x),
	\end{equation}
	for any $\phi_1,\,\phi_2\in\mathbb S_L^N$. Moreover, for any $f\in\mathbb S_L^N$, we define $\|f\|_{\mathbb S_L^N}:=\sqrt{\braket{f,f}_{\mathbb S_L^N}}$. The Fourier transform $\mathfrak{F}^N_L$ is given by
	\begin{equation}\begin{split}
		\mathfrak{F}^N_L:\hspace{1cm} \ell^2(\mathcal{V}_L^N)&\rightarrow \mathbb S_L^N\\
		(\mathfrak{F}^N_L\psi)(\gamma,\hat{x})&:=\frac{1}{\sqrt{L}}\sum_{z=0}^{L-1}e^{-\frac{2\pi i}{L}\gamma z}\psi( T_L^z
		\hat{x})
	\end{split}\end{equation}
%%%%%%%%%%%%%%%%%%%%%%%%%%%%%%%%%%%%%%%%
\begin{lemma}
The Fourier transform is well-defined. Furthermore, 
it is unitary and its adjoint is given by
	\begin{equation}\begin{split}\label{eq:FourierAdj}
		(\mathfrak{F}^N_L)^\ast:\hspace{1.5cm}  \mathbb S_L^N&\rightarrow \ell^2(\mathcal{V}_L^N)\\
		((\mathfrak{F}^N_L)^\ast \phi)(x)&:=\frac{1}{\sqrt{L}}\sum_{\gamma=0}^{L-1} e^{\frac{2\pi i}{L}\gamma z}\phi(\gamma,\hat{x})\:,
	\end{split}\end{equation}
where $\hat x\in\widehat{\mathcal V}^N_L$ and $z\in\{0,\hdots,n_{\hat x}-1\}$ are uniquely determined by $x=T_L^z\hat{x}$.
\end{lemma}
\begin{proof}
Firstly, let us prove that $\mathfrak{F}_L^N$  is well-defined by showing that it indeed maps into $\mathbb S_L^N$. For $\psi\in\ell^2(\mathcal V_L^N)$, $\hat x\in\widehat{\mathcal V}_L^N$ and $\gamma\notin (L/n_{\hat x})\Z\cap\{0,\hdots,n_{\hat x}-1\}$ consider
	\begin{align}
		(\mathfrak F_L^N\psi)(\gamma,\hat x)&=\frac{1}{\sqrt L}\sum_{\zeta=0}^{n_{\hat x}-1}\sum_{k=0}^{L/n_{\hat x}-1}
		\e^{-\frac{2\pi i}{L}(\zeta+kn_{\hat x})\gamma}\psi(T^{\zeta+kn_{\hat x}}_L\hat x)\notag\\&=\frac{1}{\sqrt L}
		\sum_{\zeta=0}^{n_{\hat x}-1}\e^{-\frac{2\pi i}{L}\zeta\gamma}\psi(T^\zeta_L\hat x)
		\Bigg[\sum_{k=0}^{L/n_{\hat x}-1}\e^{\frac{2\pi i}{L/n_{\hat x}}k\gamma}\Bigg]=0.\label{eq:wellDef1}
	\end{align}
In the first step of \eqref{eq:wellDef1} we used that for every $z\in\mathcal V_L$ there exists unique $\zeta\in\{0,\hdots,n_{\hat x}-1\}$ and $k\in\{0,\hdots,L/n_{\hat x}-1\}$ such that $z=\zeta+kn_{\hat x}$.
The last equality is due to the fact that the sum over all the $L/n_{\hat x}$-th roots of unity is equal to zero.\\
Let us now show that the adjoint of $\mathfrak F_L^N$ is indeed given by \eqref{eq:FourierAdj}. To this end let $\psi\in \ell^2(\mathcal V_L^N)$ and $\phi\in\mathbb S_L^N$. Then
	\begin{align}
		\braket{\phi,\mathfrak F_L^N\psi}_{\mathbb S_L^N}&=\frac{1}{\sqrt L}\sum_{\gamma=0}^{L-1}\sum_{\hat x\in\widehat{\mathcal V}_L^N}
		\sum_{\zeta=0}^{n_{\hat x}-1}\sum_{k=0}^{L/n_{\hat x}-1}\frac{1}{L/n_{\hat x}}\overline{\phi(\gamma,\hat x)}\e^{-\frac{2\pi i}{L}
		\gamma(\zeta+kn_{\hat x})}\psi(T^\zeta_L\hat x)\notag\\
		&=\frac{1}{\sqrt L}\sum_{\hat x\in\widehat{\mathcal V}_L^N}\sum_{\zeta=0}^{n_{\hat x}-1}\overline{\bigg[\sum_{\gamma=0}^{L-1}
		\e^{\frac{2\pi i}{L}\gamma\zeta}\phi(\gamma,\hat x)\bigg[\frac{1}{L/n_{\hat x}}\sum_{k=0}^{L/n_{\hat x}-1}
		\e^{\frac{2\pi i}{L/n_{\hat x}}\gamma k}\bigg]\bigg]}\psi(T^\zeta_L\hat x)\label{eq:Adj11}
	\end{align}
We note for any $\gamma\in (L/n_{\hat x})\Z$ that $\frac{1}{L/n_{\hat x}}\sum_{k=0}^{L/n_{\hat x}-1}\e^{\frac{2\pi i}{L/n_{\hat x}}\gamma k}=1$. Hence \eqref{eq:Adj11} is equal to
	\begin{equation}
		\sum_{x\in\mathcal V_L^N}\overline{((\mathfrak F_L^N)^\ast\phi)(x)}\psi(x)=\braket{(\mathfrak F_L^N)^\ast\phi,\psi}.
	\end{equation}
To show that indeed $(\mathfrak F_L^N)^\ast=(\mathfrak F_L^N)^{-1}$, take any $\psi\in\ell^2(\mathcal V_L^N)$ and $x\in\mathcal V_L^N$. There exist unique $\hat x\in\widehat{\mathcal V}_L^N$ and $z\in\{0,\hdots,n_{\hat x}-1\}$ such that $x=T^z_L\hat x$. We obtain
	\begin{equation}\label{eq:InversFourier1}
		((\mathfrak F_L^N)^\ast\mathfrak F_L^N\psi)(T_L^z\hat x)=\frac{1}{L}\sum_{\gamma\in\mathcal V_L
		, \gamma\in (L/n_{\hat x})\Z}\sum_{\zeta=0}^{L-1}\e^{\frac{2\pi i}{L}\gamma z}\e^{-\frac{2\pi i}{L}
		\gamma \zeta}\psi(T^\zeta_L\hat x),
	\end{equation}
where we used that $\mathfrak F_L^N\psi\in\mathbb S_L^N$. By applying the coordinate shift $\sigma=\frac{\gamma}{L/n_{\hat x}}$ we see that \eqref{eq:InversFourier1} is equal to
	\begin{equation}
		\frac{1}{L}\sum_{\sigma=0}^{n_{\hat x}-1}\sum_{\xi=0}^{n_{\hat x}-1}\sum_{k=0}^{L/n_{\hat x}-1}
		\e^{\frac{2\pi i}{n_{\hat x}}(z-(\xi+kn_{\hat x}))\sigma}\psi(T^\xi_L\hat x)
		=\frac{1}{n_{\hat x}}\sum_{\xi=0}^{n_{\hat x}-1}\bigg[\sum_{\sigma=0}^{n_{\hat x}-1}
		\e^{\frac{2\pi i}{n_{\hat x}}(z-\xi)\sigma}\bigg]\psi(T^\xi_L\hat x)=\psi(T^z_L\hat x).
	\end{equation}
It can be shown analogously that $\mathfrak F_L^N(\mathfrak F_L^N)^\ast=1$.
\end{proof}

%%%%%%%%%%%%%%%%%%%%%%%%%%%%%%%%%%%%%%%
\subsection{The Schr\"odinger operator formulation}
%%%%%%%%%%%%%%%%%%%%%%%%%%%%%%%%%%%%%%%
%%%%%%%%%%%%%%%%%%%%%%%%%%%%%%%%%%%%%%%
Let again $L,N\in\N$ with $N<L$ be fixed. In \cite{FS18} it was shown that the $N$-particle Hamiltonian is equivalent to a discrete Schr\"odinger operator acting on $\mathbb H_L^N\cong\ell^2(\mathcal V_L^N)$. More specifically, 
	\begin{equation}
		H_L^N\cong-\frac{1}{2\Delta}A_L^N+W_L^N,
	\end{equation} 
where $A_L^N$ denotes the adjacency operator on $\mathcal{G}_L^N$
	\begin{equation}
		(A_L^N\psi)(x):=\sum_{y:x\sim y}\psi(y),
	\end{equation}
while $W_L^N$ is a multiplication by the function $W:\mathcal P(\mathcal V_L)\rightarrow\N_0$ restricted to $\mathbb H_L^N$ which counts the number of connected components of a configuration $x\in\mathcal P(\mathcal V_L)$
	\begin{equation}\label{eq:DefW}
		W(x):=\frac{1}{2}|\{\{\alpha,\beta\}\in\mathcal{E}_L: \alpha\in x, \beta\notin x\}|.
	\end{equation}
	
Let us now consider the Fourier transform of the Hamiltonian $\hat{H}_L^N:=\mathfrak{F}^N_LH_L^N(\mathfrak{F}^N_L)^\ast$, with $\hat A_{L}^N$ and $\hat W_{L}^N$ being defined analogously.
%%%%%%%%%%%%%%%%%%%%%%%%%%%%%%%%%%%%%%%
\begin{lemma}\label{lem:Fibres1}
For any $\phi\in \mathbb S_L^N$, $\hat x\in\widehat{\mathcal V}_L^N$ and $\gamma\in\mathcal V_L$ we have
	\begin{equation}
			(\hat H_L^N\phi)(\gamma,\hat x)=-\frac{1}{2\Delta}\sum_{\hat y\in\widehat{\mathcal{V}}_L^N}
			a_{L,\gamma}^N(\hat x,\hat y)\phi(\gamma, \hat y)+W(\hat{x})\phi(\gamma,\hat x),
	\end{equation}
where the matrix elements of $a_{L,\gamma}^N$ are given by
	\begin{equation}\label{eq:agamma}
		a_{L,\gamma}^N(\hat x,\hat y)=\sum_{z\in\{0,\hdots,n_{\hat y}-1\}\atop T^z_L\hat y\sim \hat x}\mbox{\emph{e}}^{\frac{2\pi i}{L}\gamma z}.
	\end{equation}
\end{lemma}
\begin{proof}
Firstly, for the potential $W_L^N$, observe that for any $\phi\in \mathbb S_N^L$ we get
	\begin{equation}
		(\mathfrak{F}^N_LW_L^N(\mathfrak{F}^N_L)^\ast\phi)(\gamma,\hat{x})=W(\hat{x})\phi(\gamma,\hat{x}).
	\end{equation}
Let us now consider the adjacency operator $A^N_L$.
	\begin{align}
		(\mathfrak{F}^N_LA_L^N(\mathfrak{F}^N_L)^\ast\phi)(\gamma,\hat x)&=\frac{1}{\sqrt L}\sum_{z=0}^{L-1}
		\e^{-i\frac{2\pi}{L}\gamma z}(A_L^N(\mathfrak F_L^N)^\ast\phi)( T^z_L\hat x)\notag\\
		&=\frac{1}{\sqrt L}\sum_{z=0}^{L-1}\sum_{y:\,y\sim T^z_L\hat x}\e^{-i\frac{2\pi}{L}\gamma z}(
		(\mathfrak F_L^N)^\ast\phi)(y)\label{eq:FourierAdj11}
	\end{align}
For any $y\in\mathcal V^N_L$ there exist unique $\hat y\in\widehat{\mathcal V}_L^N$ and $\sigma\in\{0,\hdots,n_{\hat y}-1\}$ such that $y=\tilde T^{\sigma}_L\hat y$.	 Hence,
	\begin{equation}\label{eq:FourierAdj21}
		(\mathfrak{F}^N_LA_L^N(\mathfrak{F}^N_L)^\ast\phi)(\gamma,\hat x)=\frac{1}{L}\sum_{z=0}^{L-1}\sum_{\hat y}\sum_{\sigma\in\{0,\hdots,n_{\hat y}-1\}\atop
		  T^{\sigma-z}_L\hat y\sim\hat x}\sum_{\xi=0}^{L-1}\e^{-i\frac{2\pi}{L}z(\gamma -\xi)}\e^{i\frac{2\pi}{L}\xi(\sigma-z)}
		 \phi(\xi,\hat y).
	\end{equation}
Since $\phi\in\mathbb S_L^N$, we have $\phi(\xi,\hat y)=0$ for any $\hat y\in\widehat{\mathcal V}_L^N$ and $\xi\notin (L/n_{\hat y})\{0,\hdots,n_{\hat y}-1\}$. We therefore consider only $\xi\in (L/n_{\hat y})\{0,\hdots,n_{\hat y}-1\}$. The second factor in \eqref{eq:FourierAdj21} is subsequently given by
	\begin{equation}
		\e^{i\frac{2\pi}{L}\xi(\sigma-z)}=\e^{i\frac{2\pi}{n_{\hat y}}\frac{\xi}{L/n_{\hat y}}(\sigma-z)\modd n_{\hat y}}.
	\end{equation}
By changing the summation index in \eqref{eq:FourierAdj21} from $\sigma$ to $\zeta:=(\sigma-z)\textrm{mod }n_{\hat y}$ we  conclude that \eqref{eq:FourierAdj21} is equal to
	\begin{equation}
		\frac{1}{L}\sum_{\hat y}\sum_{\zeta\in\{0,\hdots,n_{\hat y}-1\}\atop
		  T^{\zeta}_L\hat y\sim\hat x}\sum_{\xi=0}^{L-1}\bigg[\sum_{z=0}^{L-1}\e^{-i\frac{2\pi}{L}z(\gamma -\xi)}\bigg]
		 \e^{i\frac{2\pi}{L}\xi\zeta}
		 \phi(\xi,\hat y)=\sum_{\hat y}\sum_{\zeta\in\{0,\hdots,n_{\hat y}-1\}\atop
		  T^{\zeta}_L\hat y\sim\hat x} \e^{i\frac{2\pi}{L}\gamma\zeta}
		 \phi(\gamma,\hat y).
	\end{equation}
This concludes the proof.
\end{proof}
%%%%%%%%%%%%%%%%%%%%%%%%%%%%%%%%%%%%%%
\begin{remark}
The operator $\hat A_L^N$ is selfadjoint on $\mathbb S_L^N$, since it is unitarily equivalent to the selfadjoint operator $A_L^N$. This implies in particular that for all $\gamma\in\mathcal V_L$ and $\hat x,\,\hat y\in\widehat{\mathcal V}_L^N$ we obtain
	\begin{equation}\label{eq:SelfAdjA}
		\frac{1}{L/n_{\hat x}}a_{L,\gamma}^N(\hat x,\hat y)=\frac{1}{L/n_{\hat y}}a_{L,\gamma}^N(\hat y,\hat x).
	\end{equation}
\end{remark}
%%%%%%%%%%%%%%%%%%%%%%%%%%%%%%%%%%%%%%
Now, we decompose $\mathbb S_L^N$ into fiber spaces corresponding to the fiber index $\gamma\in\mathcal V_L$. We obtain
	\begin{equation}
		\mathbb S_L^N=\bigoplus_{\gamma=0}^{L-1}\mathbb S_{L,\gamma}^N\:,
	\end{equation}
where
	\begin{equation}
		\mathbb S_{L,\gamma}^N:=\{\phi\in\mathbb S_L^N:\forall \hat x\in\widehat{\mathcal V}_L^N,
		\,\forall \sigma\in\mathcal  V_L,\, \sigma\neq\gamma,\textrm{ we have }\phi(\sigma,\hat x)=0\}.
	\end{equation}
In Lemma \ref{lem:Fibres1} it is shown that for each $\gamma\in\mathcal V_L$ the subspace $\mathbb S_{L,\gamma}^N$ reduces $\hat H_L^N$. Consequently, we  decompose  
	\begin{equation}
		\hat{H}_L^N=\bigoplus_{\gamma=0}^{L-1}\hat{H}_{L,\gamma}^N,
	\end{equation}
where $\hat{H}_{L,\gamma}^N:=\hat{H}_L^N\upharpoonright_{\mathbb S_{L,\gamma}^N}$. Analogously, we set $\hat{A}_{L,\gamma}^N:=\hat{A}_L^N\upharpoonright_{\mathbb S_{L,\gamma}^N}$ and $\hat{W}_{L,\gamma}^N:=\hat{W}_L^N\upharpoonright_{\mathbb S_{L,\gamma}^N}$ and thus obtain
	\begin{equation} \label{eq:fibers}
		\hat{H}_L^N=\bigoplus_{\gamma=0}^{L-1}\hat{H}_{L,\gamma}^N=\bigoplus_{\gamma=0}^{L-1}\Big(-\frac{1}{2\Delta}\hat{A}_{L,\gamma}^N
		+\hat W_{L,\gamma}^N\Big)
	\end{equation}
and 
	\begin{equation}\label{eq:sumofspectrum}
		\sigma({H}_L^N)=\sigma(\hat{H}_L^N)=\bigcup_{\gamma=0}^{L-1}\sigma(\hat{H}^N_{L,\gamma}).
	\end{equation}

%%%%%%%%%%%%%%%%%%%%%%%%%%%%%%%%%%%%%%%%
\subsection{Combes--Thomas estimate on fiber operators and proof of Theorem~\ref{thm:EigenfktEst}}\label{sec:CombesThomas}
%%%%%%%%%%%%%%%%%%%%%%%%%%%%%%%%%%%%%%%%
%%%%%%%%%%%%%%%%%%%%%%%%%%%%%%%%%%%%%%%%
Let again $L,N\in\N$ with $N<L$ be fixed. For the reader's convenience we will omit the indices $N$ and $L$ in the following proofs. However, every quantity may depend on $N$ and $L$ unless stated otherwise.
\begin{lemma}\label{lem:WAW} 
For all $\gamma\in\mathcal V_L$ the operator $\hat A_{L,\gamma}^N$ satisfies 
	\begin{equation} \label{eq:fiberrelbound}
		-2\hat W_{L,\gamma}^N\leq \hat A_{L,\gamma}^N\leq 2\hat W_{L,\gamma}^N.
	\end{equation}
\end{lemma}
\begin{proof} It is sufficient to prove only the upper bound. The lower bound follows analogously by considering $-\hat A_{\gamma}$.\\
Let $\hat{x}\in\ell^2(\widehat{\mathcal{V}})$. Equation \eqref{eq:agamma} implies 
	\begin{equation}
		\sum_{\hat{y}\in\widehat{\mathcal V}}|a_{\gamma}(\hat{x},\hat{y})|
		\le\sum_{\hat{y}\in\widehat{\mathcal{V}}}\sum_{z\in\{0,\hdots,n_{\hat y}\}:\atop T^z\hat{y}\sim\hat{x}}
		\big|\e^{\frac{2\pi i}{L}\gamma z}\big|=\sum_{y\in\mathcal V:\atop y\sim \hat{x}}1.
	\end{equation}
According to \eqref{eq:DefW} we get  
	\begin{equation}\label{eq:Sumsmaller2W}
		\sum_{\hat{y}\in\widehat{\mathcal V}}|a_{\gamma}(\hat{x},\hat{y})|\le2W(\hat{x}).
	\end{equation} 
Now, consider an arbitrary $\phi\in\mathbb S_{\gamma}$. Then 
	\begin{align}
		\braket{\phi,\hat A_{\gamma}\phi}_{\mathbb S}&=\sum_{\hat x,\,\hat y\in\widehat{\mathcal V}}
		\overline{\phi(\gamma,\hat x)}\frac{1}{L/n_{\hat x}}a_{\gamma}(\hat x,\hat y)\phi(\gamma,\hat y)\notag\\
		&\le\Big[\sum_{\hat x,\,\hat y\in\widehat{\mathcal V}}|\phi(\gamma,\hat x)|^2\frac{1}{L/n_{\hat x}}|a_\gamma(\hat x,\hat y)|\Big]^{1/2}
		\Big[\sum_{\hat x,\,\hat y\in\widehat{\mathcal V}}|\phi(\gamma,\hat y)|^2\frac{1}{L/n_{\hat x}}|a_\gamma(\hat x,\hat y)|\Big]^{1/2}.\label{eq:Asmaller2W}
	\end{align}
By the identity \eqref{eq:SelfAdjA}, we obtain
	\begin{equation}
		\braket{\phi,\hat A_{\gamma}\phi}_{\mathbb S}\le\sum_{\hat x,\,\hat y\in\widehat{\mathcal V}}|\phi(\gamma,\hat x)|^2\frac{1}{L/n_{\hat x}}|a_\gamma(\hat x,\hat y)|.
	\end{equation}
Hence by applying \eqref{eq:Sumsmaller2W} we arrive at
	\begin{equation}
		\braket{\phi,\hat A_{\gamma}\phi}_{\mathbb S}\le 2\sum_{\hat x}|\phi(\gamma,\hat x)|^2
		\frac{1}{L/n_{\hat x}}W(\hat x)=2\braket{\phi,\hat W_{\gamma}\phi}_{\mathbb S}.
	\end{equation}
\end{proof}
%%%%%%%%%%%%%%%%%%%%%%%%%%%%%%%%%%%%%%%%
We are now able to prove a Combes-Thomas estimate on a fiber. The following is an adaptation of the proof of a similar result on the unbounded XXZ-chain \cite{ARFS19,EKS,EKS2}.
%%%%%%%%%%%%%%%%%%%%%%%%%%%%%%%%%%%%%%%%
\begin{thm}
For any $\gamma\in\{0,1,\dots,L-1\}$ and any multiplication operator $\hat Y_{L,\gamma}^N:\;\mathbb S_{L,\gamma}^N\rightarrow\mathbb S_{L,\gamma}^N$, consider the Hamiltonian $\hat O_{L,\gamma}^N=-\frac{1}{2\Delta}\hat A_{L,\gamma}^N+\hat W_{L,\gamma}^N+\hat Y_{L,\gamma}^N$. Moreover, let $z\notin\sigma(\hat O_{L,\gamma}^N)$ be such that 
	\begin{equation} \label{initbound}
		\big\|(\hat W_{L,\gamma}^N)^{1/2}(\hat O_{L,\gamma}^N-z)^{-1}(\hat W_{L,\gamma}^N)^{1/2}\big\|
		\leq \frac{1}{\kappa_{L}^N(z)}<\infty,
	\end{equation}
for some $\kappa_{L}^N(z)>0$.
Then for all  $\mathcal{A,\,B}\subset \widehat{\mathcal{V}}_L^N$, we have 
	\begin{equation}
		\big\|1_\mathcal{A}\big(\hat O_{L,\gamma}^N-z\big)^{-1}1_\mathcal{B}\big\| %\leq  \big\|1_\mathcal{A}
		%(\hat W_{L,\gamma}^N)^{1/2}\big(\hat O_{L,\gamma}^N-z\big)^{-1}(\hat W_{L,\gamma}^N)^{1/2}
		%1_\mathcal{B}\big\|
		\leq \frac{2}{\kappa_L^N(z)}\,\e^{-\eta_L^N(z) \hat{d}(\mathcal{A,B})},
	\end{equation}
where $\hat d(\mathcal A,\mathcal B):=\inf\{\hat d(\hat x,\hat y):\;\hat x\in\mathcal A,\,\hat y\in\mathcal B\}$ for all $\mathcal{A,\,B}\in\mathcal P(\widehat{\mathcal V}_L^N)$ and
	\begin{equation} \label{etaz}
		\eta_L^N(z)= \ln\bigg(1+\frac{\kappa_L^N(z)\Delta}2\bigg).
	\end{equation}
\label{prop:ct1}
\end{thm}
\begin{proof}
Firstly, observe that \eqref{eq:fiberrelbound} implies that for any $\gamma\in\{0,\hdots,L-1\}$
	\begin{equation}\label{eq:relbound2}
		-2 \leq (\hat W_{\gamma})^{-1/2}\hat A_{\gamma}(\hat W_{\gamma})^{-1/2}\leq 2.
	\end{equation}
Now, for any $\mathcal{A}\subseteq \widehat{\mathcal{V}}$, let $\rho_{\mathcal{A},\gamma}:\, \mathbb S_{\gamma}\rightarrow\mathbb S_{\gamma}$ be the operator of multiplication by $\hat{d}(\mathcal{A},\cdot)$, i.e., $(\rho_{\mathcal{A},\gamma}\phi)(\gamma, \hat{x}):=\hat{d}(\mathcal{A},\hat{x})\phi(\gamma,\hat{x})$ for any $\phi\in\mathbb S_{\gamma}$. For any $\eta>0$ let us define 
	\begin{equation}
		\hat O_{\eta,\gamma}:=\e^{-\eta\rho_{\mathcal{A},\gamma}}\hat O_{\gamma} \e^{\eta\rho_{\mathcal{A},\gamma}}
	\end{equation}	
and  $\hat B_{\eta,\gamma}:=\hat O_{\eta,\gamma}-\hat O_\gamma.$ Observe that 
	\begin{equation}
		\hat B_{\eta,\gamma}=-\frac{1}{2\Delta}\big(\e^{-\eta\rho_{\mathcal{A},\gamma}}
		\hat A_\gamma \e^{\eta\rho_{\mathcal{A},\gamma}}-\hat A_\gamma\big).
	\end{equation}
Now, for any $\phi\in \mathbb S_\gamma$, consider
	\begin{align}
		\big\|&\hat W_\gamma^{-1/2}\hat B_{\eta,\gamma} \hat W_\gamma^{-1/2}\phi\big\|^2_{\mathbb S}\notag \\
		&=\frac{1}{4\Delta^2} \sum_{\hat{x}} \frac{1}{L/n_{\hat x}}\bigg| \sum_{\hat{y}}
		\hat W^{-1/2}(\hat{x})\hat{W}^{-1/2}(\hat{y})
		\big(\e^{\eta(\rho_{\mathcal{A},\gamma}(\hat{y})-\rho_{\mathcal{A},\gamma}(\hat{x}))}-1\big) 
		a_\gamma(\hat{x},\hat{y}) \phi(\gamma,\hat{y}) \bigg|^2 \label{eq:EstQ1}
	\end{align}
We note that for all $\gamma\in\{0,\hdots,L-1\}$ and all $\hat x,\hat y\in\widehat{\mathcal{V}}$ we have $|a_\gamma(\hat x,\hat y)|\le a_0(\hat x,\hat y)$ which follows from \eqref{eq:agamma}. Furthermore we have $|\e^{\eta(\hat d(\hat x,\mathcal A)-\hat d(\hat y,\mathcal A))}-1|\le (\e^\eta-1)$ for all $\hat x,\hat y\in\widehat{\mathcal V}$ with $\hat d(\hat x,\hat y)=1$. Hence \eqref{eq:EstQ1} is bounded by
	\begin{align}
		& \frac{1}{4\Delta^2}\big(\e^{\eta }-1\big)^2\sum_{\hat{x}} \frac{1}{L/n_{\hat x}}\bigg[ \sum_{\hat{y}}
		\hat W^{-1/2}(\hat{x})\hat{W}^{-1/2}(\hat{y})a_0(\hat{x},\hat{y}) |\phi(\gamma,\hat{y})| \bigg]^2 \notag\\
		 \le\;& \frac{1}{4\Delta^2}\big(\e^{\eta }-1\big)^2 \big\|\hat W_0^{-1/2}\hat A_0 \hat W_0^{-1/2}
		\tilde\phi\big\|^2,
	\end{align}	
where $\tilde\phi\in\mathbb S_0$ is defined by $\tilde \phi(\tilde\gamma,\hat x):=\delta_{\tilde\gamma,0}|\phi(\gamma,\hat x)|$ for all $\hat x\in\widehat{\mathcal V}$ and $\tilde\gamma\in\{0,\hdots,L-1\}$. The function $\tilde \phi$ is indeed an element of $\mathbb S_0$, since for all $\hat x\in\widehat{\mathcal V}$ we have $0\in L/n_{\hat x}\{0,\hdots,n_{\hat x}-1\}$. Clearly, $\|\tilde \phi\|=\|\phi\|$. By using \eqref{eq:relbound2} we further estimate the left hand side of \eqref{eq:EstQ1} and eventually get
	\begin{equation} \label{eq:dildiffbound}
		\big\|\hat W_\gamma^{-1/2}\hat B_{\eta,\gamma}\hat W_\gamma^{-1/2}\big\|\leq \frac{1}{\Delta}(e^{\eta}-1).
	\end{equation}	
For $\eta \equiv \eta(z)$ as in \eqref{etaz} it now follows that
	\begin{equation}\begin{split} \label{halfbound}
		\|\hat W_\gamma^{-1/2} \hat B_{\eta,\gamma} (\hat O_\gamma-z)^{-1} \hat W_\gamma^{1/2} \| & 
		=  \| \hat W_\gamma^{-1/2} \hat B_{\eta,\gamma}\hat  W_\gamma^{-1/2}\hat  W_\gamma^{1/2} 
		(\hat O_\gamma-z)^{-1} \hat W_\gamma^{1/2} \| \\
		& \le  \frac{ (\e^{\eta} -1)}{\Delta\kappa(z)} = \frac{1}{2}. 
	\end{split}\end{equation}
Using the resolvent identity we get
	\begin{equation}
		\hat W_\gamma^{1/2} (\hat O_{\eta,\gamma} -z)^{-1} \hat W_\gamma^{1/2} (I + \hat W_\gamma^{-1/2} \hat B_{\eta,\gamma} 
		(\hat O_{\gamma}-z)^{-1} \hat W_\gamma^{1/2}) = \hat W_\gamma^{1/2} (\hat O_\gamma-z)^{-1}\hat  W_\gamma^{1/2}.
	\end{equation}
By further applying the elementary inequality $\|(I+C)^{-1}\| \le (1-\|C\|)^{-1}$ for any $C\in L(\mathbb S_\gamma)$, $\|C\|<1$, we obtain from \eqref{initbound} and \eqref{halfbound} that
	\begin{align}
		\|&\hat W_\gamma^{1/2} (\hat O_{\eta,\gamma}-z)^{-1} \hat W_\gamma^{1/2}\| \notag\\ 
		& \le   \|\hat W_\gamma^{1/2} (\hat O_\gamma-z)^{-1} \hat W_\gamma^{1/2}\| \|(I+ \hat W_\gamma^{-1/2} 
		\hat B_{\eta,\gamma} (\hat O_{\gamma}-z)^{-1} \hat W_\gamma^{1/2})^{-1}\| \le \frac{2}{\kappa(z)}. 
	\end{align}
We conclude
	\begin{align}
		\big\| & 1_{\mathcal{A}}\hat W_\gamma^{1/2}(\hat O_\gamma-z)^{-1}\hat W_\gamma^{1/2}1_{\mathcal{B}}\big\|  =  \big\|1_\mathcal{A}
		\e^{\eta\rho_\mathcal{A}}\hat W_\gamma^{1/2}(O_{\eta,\gamma}-z)^{-1}\hat W_\gamma^{1/2}
		\e^{-\eta\rho_\mathcal{A}}1_\mathcal{B}\big\|  \notag\\
		& \leq \big\|\hat W_\gamma^{1/2}(\hat O_{\eta,\gamma}-z)^{-1}\hat W_\gamma^{1/2}\big\|\big\|
		\e^{-\eta\rho_{\mathcal{A}}}1_{\mathcal{B}}\big\|\leq\frac{2}{\kappa(z)}\e^{-\eta \hat{d}(\mathcal{A,B})}, 
	\end{align}
which is the desired result.
\end{proof}

%%%%%%%%%%%%%%%%%%%%%%%%%%%%%%%%%%%%%%%%%%
We use the Combes-Thomas estimate to deduce pointwise upper bounds to eigenfunctions of the fiber operators. These estimates apply uniformly to all eigenstates corresponding to eigenvalues in a certain energy range.
These energy ranges are associated with configurations of $K$ or less clusters $\widehat{\mathcal{V}}_{L,K}^N:=\{\hat{x}\in\widehat{\mathcal{V}}_L^N: W(\hat{x})\leq K\}$ and are given by
	\begin{equation}
		I_{K,\delta}:=\Big[1-\frac{1}{\Delta}, (K+1-\delta)\Big(1-\frac{1}{\Delta}\Big)\Big],
	\end{equation}
where $\delta\in(0,1)$ and $K\in\N$.

 For $K\in\N$ with $K\le\|\hat W_{L}^N\|$ and $\gamma\in\mathcal V_L$, let $\hat P_{L,K,\gamma}^N: \mathbb S_{L,\gamma}^N\rightarrow \mathbb S_{L,\gamma}^N$ be the orthogonal projection given by
	\begin{equation}
		\hat P_{L,K,\gamma}^N:=1_{\le K}(\hat W_{L,\gamma}^N).
	\end{equation}
Let further the whole projection $\hat P_{L,K}^N:\,\mathbb S_L^N\rightarrow \mathbb S_L^N$ be defined by
	\begin{equation}
		\hat P_{L,K}^N:=\bigoplus_{\gamma\in\mathcal V_L}\hat P_{L,K,\gamma}^N.
	\end{equation}
	
%%%%%%%%%%%%%%%%%%%%%%%%%%%%%%%%%%%%%%%%%%%
\begin{thm}\label{thm:EigenFkt1}
Let $K\in\N$ with $K\le\|\hat W_{L}^N\|$, $\delta\in(0,1)$ and $\gamma\in\mathcal V_L$. For any $E\in\sigma(\hat H_{L,\gamma}^N)\cap I_{K,\delta}$ let ${\phi_{L,\gamma}^N}\equiv{\phi_{L,\gamma}^N(\Delta,E)}\in\mathbb S_{L,\gamma}^N$ be a corresponding eigenstate. Then, for any $\mathcal A\subseteq\widehat{\mathcal V}_L^N$ we obtain
	\begin{equation}
		\|1_{{\mathcal{A}}}\phi_{L,\gamma}^N\|\leq \frac{2(K+1)^2}{\delta}\cdot\e^{-\mu_K\hat{d}^N_L
		({\mathcal{{A}}},\widehat{\mathcal{V}}_{L,K}^N)}\|\hat P_{L,K,\gamma}^N\phi_{L,\gamma}^N\|,
	\end{equation}
where 
	\begin{equation}\label{eq:muk}
		\mu_K\equiv\mu_K(\delta,\Delta):=\ln\Big(1+\frac{\delta(\Delta-1)}{2(K+1)}\Big).
	\end{equation}
\end{thm}
\begin{proof} Let us define the following multiplication operator $\hat Y_{K,\gamma}:=(K+1)(1-{1}/{\Delta})\hat P_{K,\gamma}$. Then
	\begin{align}
	&\hat W_\gamma^{-1/2}\big(\hat H_\gamma+\hat Y_{K,\gamma}-E\big)
	\hat W_\gamma^{-1/2}\notag\\
	=&-\frac{1}{2\Delta}\hat W_\gamma^{-1/2}\hat A_\gamma \hat W_\gamma^{-1/2}+1+(K+1)\Big(1-\frac{1}{\Delta}\Big)
	\hat P_{K,\gamma}\hat W_\gamma^{-1}-E \hat W_\gamma^{-1}\label{eq:CrazyOp}
	\end{align}
By using the result of Lemma \ref{lem:WAW} as well as $E\in I_{K,\delta}$ we estimate
	\begin{equation}
		-\frac{1}{2\Delta}\hat W_\gamma^{-1/2}\hat A_\gamma \hat W_\gamma^{-1/2}+1\ge \Big(1-\frac{1}{\Delta}\Big).
	\end{equation}
Moreover,
	\begin{equation}
		(K+1)\Big(1-\frac{1}{\Delta}\Big)
	\hat P_{K,\gamma}\hat W_\gamma^{-1}-E \hat P_{K,\gamma}\hat W_\gamma^{-1}\ge \delta \Big(1-\frac 1\Delta\Big) \hat P_{K,\gamma}
	\end{equation}
and
	\begin{equation}
		-E (1-\hat P_{K,\gamma})\hat W_\gamma^{-1}\ge -\frac{E}{K+1}(1-\hat P_{K,\gamma})\ge \Big(1-\frac{1}{\Delta}\Big)
		\Big(-1+\frac\delta{K+1}\Big)(1-\hat P_{K,\gamma})
	\end{equation}
Hence \eqref{eq:CrazyOp} is estimated from below by
	\begin{equation}
		\hat W_\gamma^{-1/2}\big(\hat H_\gamma+\hat Y_{K,\gamma}-E\big)\hat W_\gamma^{-1/2}\geq \frac{\delta}{K+1}\Big(1-\frac{1}{\Delta}\Big).
	\end{equation}
This implies that $E \notin\sigma(\hat H_\gamma+\hat Y_{K,\gamma})$ and in particular, we get
	\begin{equation}
		\|\hat W_{\gamma}^{1/2}\big(\hat H_\gamma+\hat Y_{K,\gamma}-E\big)^{-1}\hat W_\gamma^{1/2}\|
		\leq \frac{(K+1)\Delta}{\delta(\Delta-1)}.
	\end{equation}
By Theorem \ref{prop:ct1}, this implies that 
	\begin{equation}
		\|1_{\mathcal{{A}}}(\hat H_\gamma+\hat Y_{K,\gamma}-E)^{-1}
		1_{\mathcal{{B}}}\|\leq 
		\frac{2\Delta(K+1)}{\delta(\Delta-1)}\cdot\Big(1+\frac{\delta(\Delta-1)}{2(K+1)}\Big)^{-\hat{d}({{\mathcal{A}}},{{\mathcal{B}}})}
	\end{equation}
for any $\mathcal{{{A}}}, {\mathcal{{B}}}\in\widehat{\mathcal{V}}_{L}^N$. Now, consider
	\begin{equation}\begin{split}\label{eq:CBfor Eigenfkt}
		&\;\|1_{{\mathcal{A}}}\phi_\gamma\|_{\mathbb S}\\
		=&\;\big\|1_{{\mathcal{A}}}\big(\hat H_\gamma+\hat Y_{K,\gamma}-E\big)^{-1}\big(\hat H_\gamma
		+\hat Y_{K,\gamma}-E\big)\phi_\gamma\Big\|_{\mathbb S}.
	\end{split}\end{equation}
We note that since $\phi_\gamma$ is an eigenfunction of $\hat H_\gamma$ we have $(\hat H_\gamma-E)\phi_\gamma=0$. Hence \eqref{eq:CBfor Eigenfkt} is equal to 
	\begin{equation}\begin{split}
		&\;(K+1)\Big(1-\frac{1}{\Delta}\Big)\big\|1_{{\mathcal{A}}}\big(\hat H_\gamma+\hat Y_{K,\gamma}-E\big)^{-1}
		\hat P_{K,\gamma}\phi_\gamma\big\|_{\mathbb S}\\
		\leq&\; \frac{2(K+1)^2}{\delta}\cdot\Big(1+\frac{\delta(\Delta-1)}{2(K+1)}\Big)^{-\hat{d}({{\mathcal{A}}},
		{\widehat{\mathcal{V}}_{K}})}\|\hat P_{K,\gamma}\phi_\gamma\|_{\mathbb S},
	\end{split}\end{equation}
which is the desired result.
\end{proof}
%%%%%%%%%%%%%%%%%%%%%%%%%%%%%%%%%%%%%%%%%%%%%
Applying this result on fiber operators to the full $N$-particle Hamiltonian yields Theorem~\ref{thm:EigenfktEst}. In fact, our result can be applied to obtain estimates for eigenfunctions with eigenenergy in the $K$-cluster band $I_{K,\delta}$ for any $K$ and not just $K=1$.
%%%%%%%%%%%%%%%%%%%%%%%%%%%%%%%%%%%%%%%%%%%%%
\begin{cor}\label{cor:EigenfktEst}
Let $K\in\N$. For every $E\in I_{K,\delta}\cap \sigma(H_L^N)$ there exists an eigenstate $\ket{\psi_L^N}\equiv\ket{\psi_L^N(\Delta,E)}\in\mathbb H_L^N$ such that for all $x\in\mathcal{V}_L^N$ we obtain
	\begin{equation}\label{eq:EigenfctEst}
		|\braket{\delta_x^L,\psi_L^N}|\leq \frac{2(K+1)^2}{\delta\sqrt{L}}\cdot\e^{-\mu_K d_L^N({x,{\mathcal{V}}_{L,K}^N})},
	\end{equation}
where $\mu_K(\delta,\Delta)$ was defined in \eqref{eq:muk}.
\end{cor}
\begin{proof}
According to \eqref{eq:sumofspectrum}, for every $E\in I_{K,\delta}\cap \sigma(H)$ there exists a fiber index $\gamma\in\{0,\hdots,L\}$ such that $E\in \sigma(\hat H_{\gamma})$. Let $\phi(\Delta,E)\in\mathbb S_\gamma$ be a normalized eigenvector of $\hat H_{\gamma}$ to $E$. Let $\ket{\psi(\Delta,E)}\cong (\mathfrak F)^\ast\phi (\Delta,E)$ be the corresponding eigenstate of $H$ to $E$.

Since $\phi\in\mathbb S_\gamma$ and by \eqref{eq:FourierAdj} we have 
	\begin{equation}\label{eq:natureofphi}
		\psi (T^z\hat x)=\frac{1}{\sqrt{L}}\e^{\frac{2\pi i}{L}z\gamma} \phi(\gamma,\hat x)\quad
		\textrm{ for all }z\in\Z\textrm{ and }\hat x\in\widehat{\mathcal V}.
	\end{equation}
The result now follows from Theorem~\ref{thm:EigenFkt1}, since
	\begin{equation}
		|\braket{\delta_x,\psi}|=\frac{1}{\sqrt{L}}|\phi(\gamma,\hat{x})|=\frac{1}{\sqrt{L}} \|1_{\{\hat{x}\}}
		\phi\|\leq \frac{2(K+1)^2}{\delta\sqrt{L}}\cdot\e^{-\mu_K\hat{d}(\hat{x},{\widehat{\mathcal{V}}_{K}})}
	\end{equation}
	and 
	\begin{equation}
	\hat{d}(\hat{x},\widehat{\mathcal{V}}_{K})=\min_\gamma d(T^\gamma\hat{x},\widehat{\mathcal{V}}_{K})=\min_\gamma d(x,T^\gamma\widehat{\mathcal{V}}_{K})=d(x,\mathcal{V}_{K}),
	\end{equation}
	where we used that $\bigcup_{\gamma}T^\gamma\widehat{\mathcal{V}}_{K}=\mathcal{V}_{K}$.
\end{proof}

\begin{proof}[Proof of Theorem~\ref{thm:EigenfktEst}]
	Recall that $I_1=[1-\frac1\Delta,2(1-\frac1\Delta)){}$. According to Lemma~\ref{lem:Fasereinduetigkeit} we have $\sigma(H_L^N)\cap{}(1,2(1-1/\Delta)){}=\emptyset$, since $\Delta>3$. Hence $I_1\cap \sigma(H_L^N)= I_{1,1/2}\cap \sigma(H_L^N)$. The claim now follows immediately from Corollary~\ref{cor:EigenfktEst} with $\delta=1/2$, $K=1$ and $\mu_1(\Delta)=\tilde\mu_1(\Delta,1/2)$.
\end{proof}
%%%%%%%%%%%%%%%%%%%%%%%%%%%%%%%%%%%%%%%%%%%%%%%%%
\begin{remark}\label{rem:allarethesame}
\begin{enumerate}[label=(\roman*)]
	\item In Lemma~\ref{lem:Fasereinduetigkeit}, it was shown that for $\Delta>2$ and all $\gamma\in\mathcal V_L$, 
	each fiber operator $\hat H_{L,\gamma}^N$ has exactly one eigenvalue $E_\gamma\in[(1-1/\Delta),2(1-1/\Delta))$. 
	Let $\{\ket{\varphi_{L,\gamma}^N(\Delta)}\}_{\gamma\in\mathcal V_L}\subset\mathbb H_L^N$ be the orthonormal set of corresponding eigenstates, 
	which is unique up to phase factors.
	\item From Lemma~\ref{lem:absolutergroundstate} it follows that $E_0<E_\gamma$ for any $\gamma\neq 0$. This implies in particular that 
	$\ket{\varphi_{L,0}^N(\Delta)}$ is the unique ground state of $H_L^N$.
\end{enumerate}
\end{remark}

%%%%%%%%%%%%%%%%%%%%%%%%%%%%%%%%%%%%%%%%%%%%%%%%%
%%%%%%%%%%%%%%%%%%%%%%%%%%%%%%%%%%%%%%%%%%%%%%%%%
\subsection{The Ising-limit}\label{cap:Isinglimit}
%%%%%%%%%%%%%%%%%%%%%%%%%%%%%%%%%%%%%%%%%%%%%%%%%

Let again $N,L\in\N$ with $N<L$. The main idea of Theorem~\ref{thm:main} is to view it as a perturbative result of the Ising limit ``$\Delta=\infty$". From Corollary~\ref{cor:EigenfktEst} it readily follows that for all $\gamma\in\mathcal V_L$ the density $\rho(\varphi_L^N(\gamma,\Delta))$ converges weakly to
	\begin{equation}\label{eq:rhoinfinity}
		\rho^N_{L,\gamma}:=\rho\bigg(\sum_{\zeta\in\mathcal V_L}\frac{1}{\sqrt L}\e^{\frac{2\pi i}{L}\zeta\gamma}\ket{\delta^L_{T^\zeta_L\hat x_0}}\bigg)=
		\sum_{\xi,\zeta\in\mathcal V_L}\frac{1}{L}\e^{\frac{2\pi i}{L}(\zeta-\xi)\gamma}\ket{\delta^L_{T^\zeta_L\hat x_0}}\!\bra{\delta^L_{T^\xi_L\hat x_0}},
	\end{equation}
where $\hat x_0\in\widehat{\mathcal V}_L^N\cap\mathcal V_{L,1}^N$ is the unique representative of all droplets in $\mathcal V_L^N$. As we will see in the following, the entanglement entropy of $\rho_{L,\gamma}^N$ satisfies the desired logarithmic correction to the area law. 

In order to calculate the entanglement entropy of a  given pure  state $\ket{\psi}\in\mathbb H_L$ recall that it is necessary to determine its partial trace first.

%%%%%%%%%%%%%%%%%%%%%%%%%%%%%%%%%%%%%%%
\begin{lemma}\label{lem:EstimateReducedDensity}
Let $\Lambda\subset\mathcal V_L$. For any state $\ket{\psi}\in\mathbb H_L^N$ and for all $n\in\big\{0,\hdots,\min\{|\Lambda|,N\}\big\}$ there exists $\rho^n_\Lambda(\psi)\in L(\mathbb H^n_\Lambda)$ such that
	\begin{equation}
		\tr_{\Lambda^c}\big\{\rho(\psi)\big\}=\bigoplus_{n=0}^{\min\{|\Lambda|,N\}}\rho^n_\Lambda(\psi).
	\end{equation}
Furthermore for any $n\in\big\{0,\hdots,\min\{|\Lambda|,N\}\big\}$ and any $y,y^\prime \in \mathbb H_L^n$ we have
	\begin{equation}\label{eq:EstimateReducedDensity}
		\braket{\delta_y^\Lambda,\rho^n_\Lambda(\psi)\delta_{y^\prime}^\Lambda}=\sum_{\genfrac..{0pt}{2}{z\in\mathcal{P}
		(\Lambda^c),}{|z|=N-n}}\braket{\delta^L_{y\cup z},\rho(\psi)\delta^L_{y^\prime\cup z}}.
	\end{equation}
\end{lemma}
\begin{proof}
This statement is shown by applying the definition of the partial trace, since $\mathcal P(\mathcal V_L)=\big\{y\cup z:\;y\in\mathcal P(\Lambda)\textrm{ and }z\in\mathcal P(\Lambda^c)\big\}$ and therefore by \eqref{eq:BasisDecomposition}
	\begin{equation}\label{eq:partialTrace}
		\tr_{\Lambda^c}\big\{\rho(\psi)\big\}=\sum_{y,y^\prime\in\mathcal P(\Lambda)}\bigg[\sum_{z\in\mathcal P(\Lambda^c)}
		\braket{\delta^L_{y\cup z},\rho(\psi)\delta^L_{y\cup z}}\bigg]\ket{\delta^\Lambda_y}\!\bra{\delta^\Lambda_{y^\prime}}.
	\end{equation}
\end{proof}
%%%%%%%%%%%%%%%%%%%%%%%%%%%%%%%%%%%%%%%%

\begin{remark}
	In the following let $N<L/2$. Furthermore, for $\lambda_-,\lambda_+\in\mathcal V_L$ with $\lambda_+-\lambda_-\in(N,L/2)$, let $\Lambda\equiv\Lambda(\lambda_-,\lambda_+)=\{\lambda_-,\hdots,\lambda_+\}\subset \mathcal V_L$. 

\begin{enumerate}[label=(\roman*)]
	\item Let us consider the reduced density matrix of $\rho_{L,\gamma}^N$. By Lemma~\ref{lem:EstimateReducedDensity}, for each $n\in\{0,\hdots, N\}$ there exists 
	$\rho_{L,\Lambda,\gamma}^n\in L(\mathbb H^n_{\Lambda})$ such that
		\begin{equation}
			\rho_{L,\Lambda,\gamma}:=\tr_{\Lambda^c}\big\{\rho_{L,\gamma}^N\big\}=\bigoplus_{	n=0}^{N}
			\rho_{L,\Lambda,\gamma}^n.
		\end{equation}
	For $n\in\{1,\hdots,N-1\}$ these operators are given by	
		\begin{equation}\label{eq:rhoLambdaDeltainfty}
			\rho_{L,\Lambda,\gamma}^n:=\frac{1}{L}(\ket{\delta_{y_+^n}}\!\bra{\delta_{y_+^n}}+
			\ket{\delta_{y_-^n}}\!\bra{\delta_{y_-^n}}),
		\end{equation}
	where $y_\pm^n:=\lambda_\pm\mp\{0,\hdots,n-1\}$.
	\item The entanglement entropy of $\rho_{L,\Lambda,\gamma}^n$ for any $n\in\{1,\hdots,N-1\}$ is given by
		\begin{equation}\label{eq:Srho0}
			\tr\big\{s(\rho_{L,\Lambda,\gamma}^n)\big\}=\frac{2\ln L}{L},
		\end{equation}
	where $s:\,[0,1]\rightarrow\R$, $t\mapsto-t\ln t$. Hence
		\begin{equation}
			\tr_{\Lambda}\big\{s(\rho_{L,\Lambda,\gamma})\big\}\ge\sum_{n=1}^{N-1}\tr\big\{s(\rho_{L,\Lambda,\gamma}^n)\big\}=2\frac{N-1}{L}\ln L.
		\end{equation}
	\end{enumerate}
\end{remark}

% !TEX root = main.tex

\section{Estimating the reduced state} \label{sec:4}
\subsection{Some technical preliminaries}
For the entirety of this section, let $L,N\in \N$ with $N<L$ be fixed. Let $ I(N):=\{1,\hdots,N\}$.
We first concern ourselves with the peculiar geometry of the graph $\mathcal V_L^N$ and its graph norm. 

For any two points $x,y\in\mathcal V_L^N$ let 
	\begin{equation}
		{\bf u}:=(u^{(0)},\hdots,u^{(k)})\in\bigtimes_{l=0}^k\mathcal V_L^N=(\mathcal V_L^N)^{k+1}
	\end{equation} 
be called a path from $x$ to $y$ of length $k\in\N_0$, if and only if $u^{(0)}=x$, $u^{(k)}=y$ and $d_L^N(u^{(l-1)},u^{(l)})=1$ for all $l\in\{1,\hdots,k\}$.
If $k=d_L^N(x,y)$, then we call ${\bf u}$ a shortest path from $x$ to $y$. 

%%%%%%%%%%%%%%%%%%%%%%%%%%%%%%%%%%%%%%%%%%%%%%%%
\begin{lemma}\label{lem:decompositionOfPaths}
Let $x,y\in\mathcal V_L^N$ and let ${\bf u}$ be a shortest path from $x$ to $y$ of length $k=d_L^N(x,y)$. Let $k_0\in\{0,\hdots,k\}$,
	\begin{equation}
		{\bf v}:=(u^{(0)},\hdots,u^{(k_0)})\quad\textrm{ and }\quad {\bf w}:=(u^{(k_0)},\hdots,u^{(k)}).
	\end{equation}
Then ${\bf v}$ is a shortest path from $x$ to $u^{(k_0)}$ and ${\bf w}$ is a shortest path from $u^{(k_0)}$ to $y$. Moreover,
	\begin{equation}\label{eq:PathTriangleEquality}
		d_L^N(x,y)=d_L^N(x,u^{(k_0)})+d_L^N(u^{(k_0)},y).
	\end{equation}
\end{lemma}	
\begin{proof}
The path ${\bf v}$ is a path from $x$ to $u^{(k_0)}$ of length $k_0$ and therefore $d_L^N(x,u^{(k_0)})\le k_0$. Analogously, ${\bf w}$ is a path from $u^{(k_0)}$ to $y$ of length $k-k_0$ with $d_L^N(u^{(k_0)},y)\le k-k_0$. Hence,
	\begin{equation}\label{eq:triangulareuality}
		k=d_L^N(x,y)\le d_L^N(x,u^{(k_0)})+d_L^N(u^{(k_0)},y)\le k_0+(k-k_0)=k.
	\end{equation}
This implies equality in \eqref{eq:triangulareuality} and consequently $d_L^N(x,u^{(k_0)})=k_0$ and $d_L^N(u^{(k_0)},y)=k-k_0$.
\end{proof}

%%%%%%%%%%%%%%%%%%%%%%%%%%%%%%%%%%%%%%%%%%%%%%%	
In what follows, it will be useful to consider each element $z\in\mathcal V_L^N$ as a set of $N$ distinguishable, hard-core particles. We use the following convention to label each individual particle: For each $z\in\mathcal V_L^N$ there exists a unique $(z_1,\hdots,z_N)\in(\mathcal V_L)^N$ with $z_1<z_2<\hdots<z_N$ such that $z=\{z_j:\;j\in I(N)\}$.
This enables us to track each individual particle along the path ${\bf u}$ from $x$ to $y$. To this end, we now construct a sequence $(\tilde{u}^{(l)})_{l\le k}\subseteq(\mathcal V_L)^N$ with the property that $u^{(l)}=\{\tilde u^{(l)}_j:\; j\in I(N)\}$ for all $0\le l\le k$. Firstly, we set $\tilde u^{(0)}:=(z_1,\hdots,z_N)$. For all $1\le l\le k$, we then define
	\begin{equation}\label{eq:Ztilde}
		\Bigg\{\begin{array}{lll}
		\tilde u^{(l)}_j:=\tilde u^{(l-1)}_j&&\textrm{for all }j\in I(N)\textrm{ with }\tilde u^{(l-1)}_j\in u^{(l)},\\
		\tilde u^{(l)}_j\in u^{(l)}\backslash u^{(l-1)}&&\textrm{else.}
		\end{array}
	\end{equation}
Note that  $\tilde u^{(l)}$ is well-defined for all $1\le l\le k$. The configuration $u^{(l)}$ is obtained by moving exactly one particle in $u^{(l-1)}$ to an unoccupied neighbouring site in $\mathcal{V}_L$. Hence there is always exactly one $j_0\in I(N)$ such that $\tilde{u}^{(l+1)}_{j_0}\neq \tilde{u}^{(l)}_{j_0}$. 

For the next lemma we require the following definition. For any $j\in I(N)$ we denote by
	\begin{equation}\label{eq:Lj}
		L_j^{\bf u}:=\sum_{l=1}^{k}d_L(\tilde u_j^{(l-1)},\tilde u_j^{(l)})
	\end{equation}
the distance that the $j$-th particle has traveled along the path ${\bf u}$.

%%%%%%%%%%%%%%%%%%%%%%%%%%%%%%%%%%%%%%%%
\begin{lemma}\label{lem:cycthing}
For any $x,y\in\mathcal{V}_L^N$ the graph distance is given by
	\begin{equation}\label{eq:CycRules}
		d^N_L(x,y)=\min_{\sigma\in\mathfrak{S}^{\textrm{cyc}}_N}\sum_{j=1}
		^Nd_L(x_j,y_{\sigma(j)})=\min_{\sigma\in\mathfrak{S}_N}\sum_{j=1}^Nd_L(x_j,y_{\sigma(j)}).
	\end{equation}  
Here, $\mathfrak{S}_N^{(\textrm{cyc})}$ denotes the set of (cyclic) permutations of the set $I(N)$.
\end{lemma}
\begin{proof}
Let ${\bf u}$ be an arbitrary shortest path from $x$ to $y$ of length $k:=d_L^N(x,y)$. For any $l\in\{0,\hdots, k\}$ let $\tau_l\equiv\tau_l({\bf u})\in\mathfrak{S}_N$ be the uniquely defined permutation such that
	\begin{equation} \label{eq:cyclicordering}
		0\leq \tilde{u}^{(l)}_{\tau_l(1)}<\tilde{u}^{(l)}_{\tau_l(2)}<\dots<\tilde{u}^{(l)}_{\tau_l(N)}\leq L-1.
	\end{equation}
We now claim that $\tau_l\in\mathfrak{S}_N^{cyc}$ for all $0\le l\le k$, which we prove by induction. For the base case $l=0$, the statement is true since $\tau_0=\id\in\mathfrak S^{cyc}_N$. Now assume that for $l< k$, there exists a $\tau_l\in \mathfrak{S}_N^{cyc}$ such that \eqref{eq:cyclicordering} is satisfied. To show that the statement is then also true for $l+1$, we distinguish three cases:
\begin{itemize}
	\item First case: $\tilde{u}^{(l)}_{\tau_{l}(1)}=0$ and $\tilde u^{(l+1)}_{\tau_l(1)}=L-1$. This implies $\tilde u^{(l+1)}
	_{\tau_l(N)}<L-1$. According to the induction hypothesis we have
	$0= \tilde{u}^{(l)}_{\tau_l(1)}<\tilde{u}^{(l)}_{\tau_l(2)}<\dots<\tilde{u}^{(l)}_{\tau_l(N)}<L-1$, therefore we conclude
		\begin{equation}
			0< \tilde{u}^{(l+1)}_{\tau_l(2)} < \tilde{u}^{(l+1)}_{\tau_l(3)} < \dots < \tilde u^{(l+1)}_{\tau_l(N)}
			<\tilde{u}^{(l+1)}_{\tau_l(1)}= L-1.
		\end{equation}
	The permutation $\tau_{l+1}=\tau_l\circ\sigma$ satisfies \eqref{eq:cyclicordering} at the position $l+1$, where $\sigma\in\mathfrak{S}_N^{cyc}$ 
	is the uniquely defined cyclic permutation with $\sigma(1)=2$. Clearly, this implies that $\tau_{l+1}\in\mathfrak{S}_N^{cyc}$, 
	since the composition of two cyclic permutations is cyclic.   
	\item Second case: $\tilde u^{(l)}_{\tau(N)}=L-1$ and $\tilde u^{(l+1)}_{\tau_l(N)}=0$. By a completely analogous 
	argument as for the first case, we get $\tau_{l+1}=\tau_l\circ\sigma^{-1}\in\mathfrak S^{cyc}_N$.
	\item The third case covers any other situation. Let $j_0\le N$ be the unique index for which 
	$\big\{\tilde u^{(l)}_{\tau_l(j_0)}, \tilde u^{(l+1)}_{\tau(j_0)}\big\}\in \mathcal E_L$. Since the previous two cases have 
	been excluded, observe that the only two possibilities are $ \tilde u^{(l+1)}_{\tau(j_0)}= \tilde u^{(l)}_{\tau_l(j_0)}\pm 1
	\neq \tilde u^{(l)}_{\tau_l(j_0\pm1)}$. 
	In either case, it is important to note that this implies 
		\begin{equation} \label{eq:order}
			\tilde u^{(l)}_{\tau_l(j_0-1)}=\tilde u^{(l+1)}_{\tau_l(j_0-1)}<\tilde u^{(l+1)}_{\tau_l(j_0)}<\tilde u^{(j+1)}_{\tau_l(j_0+1)}
			=\tilde u^{(l)}_{\tau_l(j_0+1)}.
		\end{equation}
	Hence, $\tau_{l+1}=\tau_l\in\mathfrak S_N^{cyc}$.
\end{itemize}
Since each step on the path moves exactly one particle to a neighboring position we have
	\begin{equation}\label{eq:sumLjeqk}
		k=\sum_{l=1}^k\sum_{j=1}^N d_L(\tilde u_j^{(l)},\tilde u_j^{(l-1)})=\sum_{j=1}^NL_j^{\bf u}.
	\end{equation}
Moreover, for any $j\in I(N)$ we have
	\begin{equation} \label{eq:1}
		d_L(x_j,y_{\tau_k^{-1}(j)})=d_L(\tilde u_j^{(0)},\tilde u_j^{(k)})\leq \sum_{l=1}^k d_L(\tilde u_j^{(l-1)},\tilde u_j^{(l)})=L_j^{\bf u}.
	\end{equation} 	
Since $\tau_k^{-1}\in\mathfrak S^{cyc}_N$, this implies -- together with \eqref{eq:sumLjeqk}:
	\begin{equation}\label{eq:202}
		\min_{\tau\in\mathfrak S^{cyc}_N}\sum_{j=1}^N d_L(x_j,y_{\tau(j)})\le\sum_{j=1}^NL_j^{\bf u}=k.
	\end{equation}
On the other hand, it was shown in \cite[Appendix A]{FS18} that $d^N_L(x,y)=\min_{\sigma\in\mathfrak{S}_N}\sum_{j=1}^Nd_L(x_j,y_{\sigma(j)})$. 
Since $\mathfrak{S}_N^{cyc}\subseteq\mathfrak{S}_N$, we immediately obtain
	\begin{equation} \label{eq:0}
		k=d^N_L(x,y)=\min_{\sigma\in\mathfrak{S}_N}\sum_{j=1}^Nd_L(x_j,y_{\sigma(j)})\leq 
		\min_{\tau\in\mathfrak{S}^{cyc}_N}\sum_{j=1}^Nd_L(x_j,y_{\tau(j)})\:,
	\end{equation}
which concludes the proof.
\end{proof}

%%%%%%%%%%%%%%%%%%%%%%%%%%%%%%%%%%%%%%%%%%
\begin{cor}\label{cor:SamlerL/2}
 Let $x,y\in\mathcal V_L^N$, $k:=d_L^N(x,y)$, and let ${\bf u}$ be a shortest path between $x$ and $y$. Then 
	\begin{equation}\label{eq:LjkleinerLHalbe}
		L^{\bf u}_j=d_L(\tilde u^{(0)}_j,\tilde u^{(k)}_j) \le L/2
	\end{equation}
and  $\tilde u^{(k)}_j \in\{ (x_j \pm L_j^{\bf u})\modd L\}$ for all $j\in I(N)$. 
\end{cor}
\begin{proof}
Equation \eqref{eq:CycRules} immediately implies \eqref{eq:1} and \eqref{eq:202}. Due to the definition of $d_L$ we have
	\begin{equation}\label{eq:Lzequality}
		L^{\bf u}_j=d_L(\tilde u_j^{(0)},\tilde u_j^{(k)})=d_L(x_j, y_{\tau_k^{-1}(j)})\le L/2.
	\end{equation}
This already yields $\tilde u^{(k)}_j \in\{ (x_j \pm L_j^{\bf u})\modd L\}$ for all $j\in I(N)$.
\end{proof}
%%%%%%%%%%%%%%%%%%%%%%%%%%%%%%%%%%%%%%%%%%

Corollary~\ref{cor:SamlerL/2} implies that along any shortest path ${\bf u}$ from $x$ to $y$ of length $k$, each individual particle moves, if at all, either clockwise or counter-clockwise. We therefore define
	\begin{align}
		I_\pm^{\bf u}&:=\{j\le N:\; L_j^{\bf u}\neq 0\textrm{ and }\exists l\in\{0,\hdots,k\}\textrm{ with }\tilde u^{(l)}_j=(\tilde u^{(0)}_j\pm 1)\modd L\}
		\quad\textrm{ and}\\
		I_0^{\bf u}&:=\{j\le N:\; L_j^{\bf u}=0\}\:.
	\end{align}
Note that $I(N)=I_+^{\bf u}\cup I_-^{\bf u}\cup I^{\bf u}_0$ for any shortest path ${\bf u}$, where the union is disjoint.
The definition of $L_j^{\bf u}$ as well as \eqref{eq:LjkleinerLHalbe} imply that 
	\begin{equation}\label{eq:zhochls}
		\{u^{(l)}_j:\;0\le l\le k\}=\{(x_j\pm \xi)\modd L:\;0\le \xi\le L^{\bf u}_j\}\quad\textrm{ for all }j\in I_\pm^{\bf u}\cup I_0^{\bf u}.
	\end{equation}

Let $\kappa_j^{(l)}\equiv\kappa_j^{(l)}({\bf u})\in\{0,\hdots, L_j^{\bf u}\}$ for any  $j\in I(N)$ and any $ l\in\{0,\hdots, k\}$ such that
	\begin{equation}\label{eq:kappal}
		\tilde u_j^{(l)}=(\tilde u^{(0)}_j\pm\kappa_j^{(l)})\modd L 	\quad \textrm{ for all }j\in I_\pm^{\bf u}\cup I_0^{\bf u}.
	\end{equation}
%The quantity $\kappa_j^{(l)}$ describes the number of steps the $j$-th particle moves until the $l$-th step of the path ${\bf u}$. 
It follows from the previous observations that the quantity $\kappa_j^{(l)}$ is well-defined.

The next lemma further establishes that we can indeed consider ${\bf u}$ as a path of hard core particles.
%%%%%%%%%%%%%%%%%%%%%%%%%%%%%%%%%%%%%%%%%%

\begin{lemma}[hard-core particle property] \label{lem:hard-coreProperty}
Let $e_0:=\{0,L-1\}$, $c\in\mathcal{V}_{L,1}^N$ with $e_0\nsubseteq c$ and $x\in\mathcal{V}_L^N$. Moreover, let ${\bf u}$ be a shortest path from $c$ to $x$ of length $k:=d_L^N(c,x)$.\\
Then,
\begin{enumerate}[label=(\roman*)]
	\item If $i,j\in I_-^{\bf u}$ with $i<j$ then $\kappa^{(l)}_{i}\ge \kappa^{(l)}_{j}$  for all $l\in\{0,\hdots, k\}$.
	\item If $i,j\in I_+^{\bf u}$ with $i<j$ then $\kappa^{(l)}_{i}\le \kappa^{(l)}_{j}$ for all $l\in\{0,\hdots, k\}$.
	\item We have the following inequalities:
		\begin{equation}\label{eq:hardcore}
			\sup I_-^{\bf u}\le \inf (I_0^{\bf u}\cup I_+^{\bf u})\quad \textrm{ and }\quad\sup (I^{\bf u}_-\cup I_0^{\bf u})\le \inf I_+^{\bf u},
		\end{equation}
	where we use the convention that $\inf \emptyset =\infty$ and $\sup\emptyset =-\infty$.

	\item If $1\in I_-^{\bf u}\cup I_0^{\bf u}$ and $N\in I_+^{\bf u}\cup I_{0}^{\bf u}$ then 
		\begin{equation}
			L_1^{\bf u}+L_N^{\bf u}\le L-N.
		\end{equation}
\end{enumerate}
\end{lemma}
\begin{proof}

Since $c\in\mathcal{V}_{L,1}^N$ with $e_0\nsubseteq c$ we have $c_j=c_1+j-1$ for all $j\in I(N)$. Let $J:=\{0,\hdots,k\}$.
\begin{enumerate}[label=(\roman*)]
	\item It suffices to show that if $j,j+1\in  I_-^{\bf u}$, then $\kappa^{(l)}_{j}\ge \kappa^{(l)}_{j+1}$ for all $l\in J$.
	Let us define $f:\;J\rightarrow\Z$, $l\mapsto  \kappa^{(l)}_{j}- \kappa^{(l)}_{j+1}$. 
	Suppose there exists a $k_0\in J$ such that $f(k_0)<0$. Due to the definition of the path we have $f(0)=0$ and $|f(l)-f(l-1)|\in\{0,1\}$ 
	for all $l\in J\setminus\{0\}$. Now, suppose there exists a $k_0\in J$ such that $f(k_0)<0$. This would imply that there exist a $k_1\in J$, $k_1\le k_0$, with $f(k_1)=-1$. Hence
		\begin{equation}
			u^{(k_1)}_{j}=(c_{j}-\kappa_{j}^{(k_1)})\modd L=(c_{j}-f(k_1)-\kappa_{j+1}^{(k_1)})\modd L=u^{(k_1)}_{j+1},
		\end{equation}
	since $c_{j}-f(k_1)=c_{j+1}$, which is a contradiction.
	\item Analogous to (i).
	\item We only prove the first inequality in \eqref{eq:hardcore}. The right-hand side follows analogously. If $I_-^{\bf u}=\emptyset$, then the result is trivial. So, from now on, we assume that $I_-^{\bf u}\neq\emptyset$.
	 Suppose $\inf ( I_0^{\bf u}\cup I_+^{\bf u})<\sup I_-^{\bf u}$. This implies that there exists a 
	$j\in I_-^{\bf u}$ such that $j-1\in I_0^{\bf u}\cup I_+^{\bf u}$.  
	Consider the function $g:\;J\rightarrow\Z$, $l\mapsto \kappa_j^{(l)}+\kappa_{j-1}^{(l)}$. We again have
	$g(0)=0$, $g(k)\ge L^{\bf u}_j\ge 1$ and 
	$|g(l)-g(l-1)|\in\{0,1\}$ for all $l\in J\setminus\{0\}$. Hence, there exists $k_0\in J$ with $g(k_0)=1$. This implies
		\begin{equation}
			\tilde u^{(k_0)}_{j-1}=(c_{j-1}+\kappa_{j-1}^{k_0})\modd L=(c_{j-1}+g(k_0)-\kappa_{j}^{k_0})\modd L=\tilde u^{(k_0)}_{j},
		\end{equation}
	which is a contradiction.
	\item Let us consider the function $h:\;J\rightarrow\N_0$, $l\mapsto\kappa^{(l)}_{1}+\kappa^{(l)}_N$. Suppose $L_1^{\bf u}+L_N^{\bf u}>L-N$. Since 
	$h(k)=L_1^{\bf u}+L_N^{\bf u}$ and $h(l)-h(l-1)\in\{0,1\}$ for all $l\in J\setminus\{0\}$, we conclude that there exists a $k_0\in J$ with $h(k_0)=L-N+1$.
	Hence, by using $c_N=c_1+N-1=c_1-h(k_0)+L$, we obtain
		\begin{equation}
			u^{(k_0)}_1=(c_1-\kappa^{(k_0)}_1)\modd L=(c_1-h(k_0)+\kappa^{(k_0)}_N)\modd L=u^{(k_0)}_N,
		\end{equation}
	which is a contradiction.
	\end{enumerate}
\end{proof}
%%%%%%%%%%%%%%%%%%%%%%%%%%%%%%%%%%%%%%%%%%

To calculate the distance of two configurations in the graph norm on the ring, we rely on the results obtained for the line. The only difference between the graph of the ring $\mathcal G_L$ and the graph of the line is, that the edge $e_0=\{0,L-1\}$ is not an element of the edge set of the line. We therefore are interested in paths, for which no particle crosses this particular edge.

Given a path ${\bf u}$  of length $k$ and an edge $e\in\mathcal E_L$, we say that ${\bf u}$ does not cross $e$, if  $u^{(l-1)}\triangle u^{(l)}\neq e$ for all $l\in\{1,\hdots,k\}$.

%%%%%%%%%%%%%%%%%%%%%%%%%%%%%%%%%%%%%%%%%%
\begin{lemma}[cutting lemma]\label{lem:SchnipSchnap}
Let $N<L/2$. Let $c\in\mathcal{V}_{L,1}^N$ and $x\in\mathcal{V}_L^N$. Then, there exists a shortest path ${\bf u}$ from $c$ to $x$ and an edge $e\in\mathcal{E}_L$ with $\min_{m\in c}d_L(m,e)\ge \lceil L/2\rceil-N$ such that ${\bf u}$ does not cross $e$. 
\end{lemma}
\begin{proof}
Due to the translational symmetry of the system we may assume w.l.o.g. that $c=\{0,\hdots,N-1\}$. Let ${\bf u}$ be a shortest path of length $k:=d_L^N(c,x)$ connecting $c$ and $x$. To prove the lemma we have to show that there exists an $e\in\mathcal E_L$ such that for all $l\in\{1,\hdots,k\}$ we have $u^{(l-1)}\triangle u^{(l)}\in\mathcal E_L\setminus\{e\}$. However, it suffices to show that for all $j\in I(N)$ we have $e\nsubseteq\big\{u^{(l)}_j:\;0\le l\le k\big\}$. We distinguish between three cases.
\begin{itemize}
	\item First case: $I_-^{\bf u}=\emptyset$, which means that no particle moves clockwise. Let $e:=\{N-1+\lfloor L/2\rfloor,N+ \lfloor L/2\rfloor\}$.
	For all $l\in\{0,\hdots,k\}$ and for all $j\in I(N)$ we have 
	\begin{equation}\label{eq:Iminusempty}
		\{u^{(l)}_j:\;0\le l\le k\}=j-1+\{0,\hdots,L_j^{\bf u}\}\subseteq\{0,\hdots,N-1+\lfloor L/2\rfloor\},
	\end{equation} 
	where we used Corollary~\ref{cor:SamlerL/2} for the second inclusion.  
	Since $e$ is not a subset of the right-hand side of 
	\eqref{eq:Iminusempty}, this proves the claim for this case.
	\item Second case: $I_+^{\bf u}=\emptyset$ or all particle move either clockwise or not at all. Analogously to the first case, we see that 
	the edge $e:=\{\lceil L/2\rceil-1,\lceil L/2\rceil\}$ satisfies the claim.
	\item Third case: both $I_-^{\bf u}$ and $I_+^{\bf u}$ are non-empty. Let us first note that due to 
	Lemma~\ref{lem:hard-coreProperty}~(iii)
	$\max I_-^{\bf u}\le\min (I_0^{\bf u}\cup I_+^{\bf u})$ and $\max (I_0^{\bf u}\cup I_-^{\bf u})\le\min I_+^{\bf u}$. This implies that $N\in  I_+^{\bf u}$ and 
	$1\in  I_-^{\bf u}$. 	According to Lemma~\ref{lem:hard-coreProperty}~(iv) we have $L_1^{\bf u}+L_N^{\bf u}\le L-N$ and hence
		\begin{equation}
			\tilde u_1^{(k)}=L-L_1^{\bf u}> N+L_N^{\bf u}-1=\tilde u_N^{(k)}.
		\end{equation}
	Let $e\in\mathcal E_L$ with $e\subset\{\tilde u_N^{(k)},\hdots,\tilde u_1^{(k)}\}$. For any $j\in I_+^{\bf u}\cup I_0^{\bf u}$ 
	and any $l\in\{0,\hdots,k\}$ we have 
	\begin{equation}
		\tilde u^{(l)}_j\in j-1+\{0,\hdots,\kappa_{j}^{(l)}\}\subseteq\{0,\hdots, \tilde u_N^{(k)}\}=:\mathcal A_+,
	\end{equation} 
   where we used Lemma~\ref{lem:hard-coreProperty}~(ii) for the second inclusion. 
	 Analogously, for any $j\in I_-^{\bf u}\cup I_0^{\bf u}$ and any $l\in\{0,\hdots, k\}$ we have 
	 \begin{equation}\label{eq:nichteineTeilmenge2}
	 	u^{(l)}_j\in\{0,\hdots,N\}\cup\{L-L_1^{\bf u},\hdots,L-1\}=:\mathcal A_-.
	\end{equation} 
Since $e\nsubseteq\mathcal A_{\pm}$, this shows that ${\bf u}$ does not cross $e$.
	
	Moreover, since $\tilde u^{(k)}_N\le N-1+\lfloor L/2\rfloor$ and $\tilde u^{(k)}_1\ge \lceil L/2\rceil$ there exists at least 
	one edge $e\in\mathcal E_L$ with $e\subseteq\{\lceil L/2\rceil,\hdots,N-1+\lfloor L/2\rfloor\}\subseteq 
	\{\tilde u_1^{(k)},\hdots, \tilde u_N^{(k)}\}$ such that $e$ has a distance of at least $\lceil L/2\rceil-N$ to any $m\in c$.
	\end{itemize}
\end{proof}
%%%%%%%%%%%%%%%%%%%%%%%%%%%%%%%%%%%%%%%%%%%%%%%%%%%%%%%

Thus, when trying to find a shortest path between an arbitrary configuration to a given droplet, the previous lemma tells us that we can always cut the ring open alongside an edge $e$ and rather consider the graph of the thus obtained line. This is useful, since it enables us to draw from the results in \cite{ARFS19}. Let us summarize these results in the following proposition:

\begin{prop}\label{rem:theline}
\begin{enumerate}[label=(\roman*)]
	\item	Let $\mathcal V:=\Z$ and $\mathcal E:=\big\{\{j,j+1\}:\;j\in\Z\big\}$. Moreover, for any $N\in\N$, let the graph of $N$-particle configurations be given by 
	$\mathcal G^N:=(\mathcal V^N,\mathcal E^N)$ where $\mathcal V^N:=\{x\subseteq\mathcal V:\;|x|=N\}$ and 
	$\mathcal E^N:=\big\{\{x,y\}\subseteq \mathcal V^N:\; x\triangle y\in\mathcal E\big\}$. 
	Then, for any $x=\{x_1<x_2<\hdots<x_N\}, y=\{y_1<y_2<\hdots<y_N\}\in\mathcal V^N$, the graph distance $d^N$ is explicitly given by:
		\begin{equation}
			d^N(x,y)=\sum_{j=1}^N|x_j-y_j|.
		\end{equation}
	\item For any $m\in\mathcal V$, and $N\in\N$ let us define the droplet centered around $m$ by
		\begin{equation}
			c_m^N:=m+\Big\{-\Big\lfloor\frac{N-1}{2}\Big\rfloor,\hdots,\Big\lceil\frac{N-1}{2}\Big\rceil\Big\},
		\end{equation}
	and let $\mathcal V_1^N:=\{c_m^N:\;m\in\mathcal V\}$ denote the set of all droplets. We are interested in the droplets closest to a given 
	$ x\subseteq\mathcal V$, $|x|>0$. Let us define 
		\begin{equation}
			\mathcal W(x):=\{m\in\mathcal V:\;d^N(x,c_m^{|x|})=d^N(x,\mathcal V_1^{|x|})\}.
		\end{equation}

	According to \cite[Lemma~A.1]{ARFS19}, we have 
		\begin{equation}
			\mathcal W(x)=\left\{	\begin{array}{ll}
												\{x_\kappa\}										&	\textrm{if }N\textrm{ is odd},\\
												\{x_\kappa,\hdots, x_{\kappa+1}-1\}	&	\textrm{if }N\textrm{ is even},
											\end{array}\right.
		\end{equation}
	for all $x\in\mathcal V^N$, where $\kappa:=\lfloor (N+1)/2\rfloor$.
\end{enumerate}
\end{prop}
%%%%%%%%%%%%%%%%%%%%%%%%%%%%%%%%%%%%%%%%%%%%%%%%%%%%%%%
Let us now introduce a notation for a droplet in $\mathcal V_L^N$ centered around a site $m\in\mathcal V_L$
	\begin{equation}\label{def:droplet}
		c_{L,m}^N:=\{j\modd L:\;j\in c_m^N\}.
	\end{equation}
Analogously to Remark~\ref{rem:theline}~(ii), for any $x\subseteq V_L$, $|x|>0$, we define the set of centers of droplets thet are closest closest to this configuration
	\begin{equation}
		\mathcal W_L(x):=\big\{m\in\mathcal V_L:\;d_L^{|x|}(x,c_{L,m}^{|x|})=d_L^{|x|}(x,\mathcal V_{L,1}^{|x|})\big\}.
	\end{equation}

%%%%%%%%%%%%%%%%%%%%%%%%%%%%%%%%%%%%%%%%%%%%%%%%%%%
\begin{lemma} \label{prop:chaindrop} Let $N<L/2$ and $x\in\mathcal V_L^N$. Then,
	\begin{equation}
		\mathcal W_L(x)\cap x\neq \emptyset.
	\end{equation}
Furthermore, let $m\in\mathcal W_L(x)\cap x$. If there exists a shortest path ${\bf u}$ from $c_{L,m}^N$ to $x$ that does not cross $e_0:=\{0,L-1\}$ and  $e_0\nsubseteq c_{L,m}^N$, then $m=x_\kappa$, where $\kappa:=\lfloor(N+1)/2\rfloor$.
\end{lemma}
\begin{proof} 
Let $\nu\in \mathcal W_L(x)$ and define $c:=c_{L,\nu}^N$. According to Lemma~\ref{lem:SchnipSchnap}, there exists a shortest path ${\bf u}$ from $c$ to $x$ and an edge $e\in\mathcal E_L$ with $\min_{j\in c}d_L(j,e)> \lceil L/2\rceil-L/2\ge 0 $ such that ${\bf u}$ does not cross $e$. This implies $e\nsubseteq c$. Let us assume w.l.o.g. that $e=e_0$. 
In any other case we can choose a suitable translation by $\gamma\in\Z$ such that $T^\gamma_Le=e_0$ and consider $T^\gamma_L x$, $T^\gamma_Lc$ and the path ${\bf u}_\gamma:=(T^\gamma_Lu^{(0)},\hdots,T^\gamma_Lu^{(k)})$ instead.

Since ${\bf u}$ does not cross $e_0$, it can also be viewed as a path on the infinite line $\mathcal G^N$. Therefore, according to Proposition~\ref{rem:theline} we have
$d_L^N(x,c)=d^N(x,c)$. Let us consider the droplet $c^\prime:=c_{x_\kappa}^N=c_{L,x_\kappa}^N$. Then, Proposition~\ref{rem:theline}~(i) states that $\{x_\kappa\}=\mathcal W(x)\cap x$. Hence,
	\begin{equation}\label{eq:WilliamBush}
		d_L^N(x,\mathcal V_{L,1}^N)=d^N(x,c)\ge d^N(x,c^\prime)= \sum_{j=1}^N|c^\prime_j-x_j|\ge \sum_{j=1}^Nd_L(c^\prime_j,x_j)\ge d_L^N(c^\prime,x),
	\end{equation}
where we applied Lemma~\ref{lem:cycthing} to achieve the final estimate. Since $c^\prime\in\mathcal V_{L,1}^N$ we have equality in \eqref{eq:WilliamBush} and therefore $x_\kappa\in\mathcal W_L(x)\cap x$. Moreover, if $\nu\in x$ it follows immediately that $\nu=x_\kappa$, since  $d^N(x,c)=d^N(x,c^\prime)$ and the set $\mathcal W(x)\cap x$ contains only one element.
\end{proof}
%%%%%%%%%%%%%%%%%%%%%%%%%%%%%%%%%%%%%%%%%%%%%%%%%%%%

By making further assumptions on the configuration $x$, we are able to determine $\mathcal W_L(x)\cap x$ precisely. For now, let us only consider configurations that are contained in a sufficiently small sector of the ring.

Here, a sector of size $\theta\in(0,1/2)$ around a site $m\in\mathcal V_L^N$ is given by
	\begin{equation}\label{Def:S}
		\mathcal S_{L,m}(\theta):=\{k\in\mathcal V_L:\; d_L(k,m)<\theta L\}.
	\end{equation}
%%%%%%%%%%%%%%%%%%%%%%%%%%%%%%%%%%%%%%%%%%%%%%%%%%%%%

\begin{lemma} \label{lemma:middle}
Let $M:=\lfloor(L-1)/2\rfloor$, $\beta\in(0,1/4)$, $N<\beta L$ and  $x\in\mathcal{V}^N\big(\mathcal{S}_{L,M}(1/4-\beta/2)\big)$. Then, 
	\begin{equation}\label{eq:kappaismiddle}
		\{x_{\kappa}\}=\mathcal W_L(x)\cap x,
	\end{equation}
where $\kappa:=\lfloor (N+1)/2\rfloor$. 

Furthermore, no shortest path ${\bf u}$ from $c_{L,x_\kappa}^N$ to $x$ crosses $e_0:=\{0,L-1\}$. Moreover, we have $\{1,\hdots, \kappa\}\subseteq I_0^{\bf u}\cup I_-^{\bf u}$ and $\{\kappa,\hdots,N\}\subseteq I_0^{\bf u}\cup I_+^{\bf u}$.
\end{lemma}
\begin{proof} 
Let $\nu\in \mathcal{S}_{L,M}(1/4-\beta/2)$. We claim that no shortest path ${\bf u}$ from $c_{L,\nu}^N$ to $x$ crosses the edge $e_0$.
Let $c:=c_{L,\nu}^N$ and $k:=d_L^N(x,c_{L,\nu}^N)$. As in the proof of Lemma~\ref{lem:SchnipSchnap}, it is sufficient to show that for any $j\in I(N)$ we have $e_0\nsubseteq\{\tilde u^{(l)}_j:\;0\le l\le k\}$. 

Suppose there exists a $j\in I(N)$ such that $e_0\subseteq\mathcal Z_j:=\{\tilde u^{(l)}_j:\;0\le l\le k\}$, which readily implies $j\notin I_0^{\bf u}$. But according to \eqref{eq:zhochls}, we know that for any $j\in I_\pm^{\bf u}$, we have
	\begin{equation}\label{eq:Zj}
		|\mathcal Z_j|=|\{c_j,\hdots,(c_j\pm L_j^{\bf u})\modd L\}|=L^{\bf u}_j+1\le\lfloor L/2\rfloor+1,
	\end{equation}
where we used Corollary~\ref{cor:SamlerL/2} to estimate $L^{\bf u}_j$. W.l.o.g., let us assume $j\in I_-^{\bf u}$.
Since we assumed $e_0\subseteq \mathcal Z_j$, we deduce from \eqref{eq:zhochls} that
	\begin{equation}
		\{c_j,(c_j- L^{\bf u}_j)\modd L\}\cup(\mathcal S_{L,M}(1/4))^c\subseteq\mathcal Z_j.
	\end{equation}
Notice that both $c_j\in \mathcal S_{L,M}(1/4)$ and $c_j-L_j^{\bf u}\in x\subseteq \mathcal S_{L,M}(1/4)$.
Hence,
	\begin{equation}
		|\mathcal Z_j|\ge 2+|(\mathcal S_{L,M}(1/4))^c|\ge 2+\lfloor L/2\rfloor,
	\end{equation}
which is a contradiction. 

Since no shortest path ${\bf u}$ from $c_{L,\nu}^N$ to $x$ crosses $e_0$, observe that ${\bf u}$ can also be viewed as a path on the graph induced by $N$ particles on the infinite line $\mathcal G^N$. Hence, 	\begin{equation}\label{eq:dLgleichdohneL}
		d_L^N(x,c_{L,x_j}^N)=d^N(x,c_{L,x_j}^N)
	\end{equation}
 and $\mathcal W_L(x)\cap x=\mathcal W(x)\cap x=\{x_\kappa\}$ according to Proposition~\ref{rem:theline}~(ii).

Let us now consider a shortest path ${\bf v}$ from $c^\prime:=c_{L,x_\kappa}^N$ to $x$. Lemma~\ref{lem:hard-coreProperty} implies that $\tilde v^{(k)}_j=x_j$ and $L^{\bf v}_j=d_L(x_j,c^\prime_j)$ for all $j\in I(N)$. Hence $L_\kappa^{\bf v}=0$ and therefore $\kappa\in I^{\bf z}_0$. The rest of the statement follows from Lemma~\ref{lem:hard-coreProperty}~(iii).
\end{proof}
%%%%%%%%%%%%%%%%%%%%%%%%%%%%%%%%%%%%%%%%%%%%%%%%%%%%%

%%%%%%%%%%%%%%%%%%%%%%%%%%%%%%%%%%%%%%%%%%

\begin{lemma}\label{lem:summingeverythingup}
Let $N< L/2$ and $\mu\ge \ln 2$. Then, 
	\begin{equation}
		\sum_{x\in \mathcal V^N_L}\e^{-\mu d_L^N(x,\mathcal V_{L,1}^N)}\le L(1+ 2^9\e^{-\mu}).
	\end{equation}
\end{lemma}
\begin{proof}
For any $m\in\mathcal V_L$ let
	\begin{equation}
		\mathcal B_{L,m}^N:=\{x\in\mathcal V_L^N:\;m\in\mathcal W_L(x)\cap x\}.
	\end{equation}
By Lemma~\ref{prop:chaindrop} we have
	\begin{equation}\label{eq:HoratioHornblower}
		\bigcup_{m\in\mathcal V_L}\mathcal B_{L,m}^N=\mathcal V_L^N.
	\end{equation}

Let $x\in \mathcal B_{L,m}^N$. According to Lemma~\ref{lem:SchnipSchnap} there exists an edge $e$ with $\max_{j\in c}d_L(j,e)>0$ and a shortest path ${\bf u}$ from $c:=c_{L,m}^N$ to $x$ that does not cross $e$. Pick $\gamma\in\Z$ such that $T^\gamma_Le=e_0:=\{0,L-1\}$. Let $x^\prime:= T^\gamma_Lx$, $c^\prime:=T^\gamma_Lc$ and ${\bf v}:=\{T^\gamma_Lu^{(0)},\hdots,T^\gamma_Lu^{(k)}\}$, where $k:=d_L^N(x,c)$.
 Let us define $\chi_-\equiv\chi_-(x)\in\N_0^{\kappa-1}$, $\chi_+\equiv \chi_+(x)\in\N_0^{N-\kappa}$ by 
	\begin{align}
		\chi_{-,j}&:=d_L(x^\prime_{\kappa-j},c^\prime_{\kappa-j})&&\textrm{for }j\le \kappa-1,\\
		\chi_{+,j}&:=d_L(x^\prime_{\kappa+j},c^\prime_{\kappa+j})&&\textrm{for }j\le N-\kappa.
	\end{align}
We want to show that $\chi_+\in\mathcal X^{N-\kappa}$ and $\chi_-\in\mathcal X^{\kappa-1}$. Since ${\bf v}$ does not cross $e_0$ we have $v^{(k)}_j=x^\prime_j$ for all $j\in I(N)$. By Lemma~\ref{prop:chaindrop} we have $x^\prime_\kappa=c^\prime_\kappa$. Hence $\kappa\in I^{\bf v}_0$, since $L_\kappa^{\bf v}=d_L(x_\kappa^\prime,c_\kappa^\prime)=0$. As a consequence of Lemma~\ref{lem:hard-coreProperty}~(iii) we conclude $\{1,\hdots,\kappa\}\subseteq I^{\bf v}_0\cup I^{\bf v}_-$ and $\{\kappa,\hdots,N\}\subseteq I^{\bf v}_0\cup I^{\bf v}_+$

By Lemma~\ref{lem:hard-coreProperty}~(i), we get
	\begin{equation}
		\chi_{-,j}=L^{\bf v}_{\kappa-j}\le L^{\bf v}_{\kappa-j-1}=\chi_{-,j+1}\quad \textrm{for all }1\le j<\kappa-1,
	\end{equation}
and therefore $\chi_-\in\mathcal X^{\kappa-1}$. Analogously $\chi_+\in\mathcal X^{N-\kappa}$. 
Furthermore
	\begin{equation}
		d_L^N(x,c)=d_L^N(x^\prime,c^\prime)=\sum_{j=1}^{N}L^{\bf v}_j=|\chi_-|_1+|\chi_+|_1.
	\end{equation}
Note that each pair $(\chi_-,\chi_+)\in\mathcal X^{\kappa-1}\times\mathcal X^{N-\kappa}$ corresponds to one $x\in \mathcal B_{L,m}^N$ only, since
	\begin{equation}
		x=\{(m+j+\chi_{+,j})\modd L:\;j\le N-\kappa\}\cup\{m\}\cup\{(m-j-\chi_{-,j})\modd L:\;j\le \kappa-1\}.
	\end{equation}
		\begin{equation}
			\sum_{x\in\mathcal B_{L,m}^N}\e^{-\mu d_L^N(x,c)}\le \sum_{\chi_-\in\mathcal X^{\kappa-1}} 
			\sum_{\chi_+\in\mathcal X^{N-\kappa}}\e^{-\mu(|\chi_-|_1+|\chi_+|_1)}\le (1+30\e^{-\mu})^2\le  1+2^9\e^{-\mu}
		\end{equation}
	where we applied Lemma~\ref{lem:geomsum} and $\mu\ge\ln 2$. Together with \eqref{eq:HoratioHornblower}, this concludes the proof.
\end{proof}

%%%%%%%%%%%%%%%%%%%%%%%%%%%%%%%%%%%%%%%%%%
%%%%%%%%%%%%%%%%%%%%%%%%%%%%%%%%%%%%%%%%%%
\subsection{Moving particles to the boundary}
%%%%%%%%%%%%%%%%%%%%%%%%%%%%%%%%%%%%%%%%%%
Let us first introduce some notation. In the following, let $\epsilon\in(0,1/16)$ and $\theta\in(\epsilon,1/16)$ be fixed. Let $L\in\N$ and $N\equiv N(\epsilon, L):=\lfloor \epsilon L \rfloor$. 
Let 
	\begin{equation}
		\Lambda_L\equiv \Lambda_L(\theta):=\mathcal S_{L,M}(\theta)=\{\lambda_-,\hdots,\lambda_+\}\subseteq \mathcal V_L,
	\end{equation}
where $M:=\lfloor(L-1)/2\rfloor$, $\lambda_-:=\min\Lambda_L$ and $\lambda_+:=\max\Lambda_L$.

We are ultimately interested in taking the partial trace over the Fock-space associated with $\Lambda_L^c$. This entails summing up all contributions of configurations in $\mathcal P(\Lambda_L^c)$, as we have seen in Lemma~\ref{lem:EstimateReducedDensity}.  Thus, let us introduce some additional notation to classify configurations in $\mathcal P(\Gamma^c)$, for any connected subset $\Gamma=\{\gamma_-,\hdots,\gamma_+\}\subseteq\mathcal S_{L,M}(1/4)$ with $\gamma_-<\gamma_+$. 

	For $x\subseteq\mathcal V_L$ let $x^{in}\equiv x^{in}(\Gamma):=x\cap \Gamma$ and $x^{out}\equiv x^{out}(\Gamma):=x\setminus \Gamma$.
	If $|x^{in}|,|x^{out}|>0$, let $\ell,r\in\N_0$ be arbitrary such that $\ell+r=|x^{out}|$. The idea is to further split $x^{out}$ into a configuration of $\ell$ particles which are thought of as being close to $\gamma_-$ and a configuration of $r$ particles close to $\gamma_+$. Note that for every such $\ell$ and $r$, there exists a unique permutation $\sigma_{\ell,r}\equiv\sigma_{\ell,r}(x,\Gamma)\in\mathfrak S^{cyc}_N$ with the property that 
	\begin{equation}
	x^{in}=\{x_{\sigma_{\ell,r}(j)}:\;\ell<j\le N-r\}.
	\end{equation}
		
We then define
	\begin{equation}
		x^{\ell,out}_-\equiv x_-^{\ell,out}(x,\Gamma):=\{x_{\sigma_{\ell,r}(j)}:j\le\ell\}\quad\textrm{ and }\quad 
		x^{r,out}_+\equiv x_+^{r,out}(x,\Gamma):=\{x_{\sigma_{\ell,r}(j)}:j> N-r\}.
	\end{equation}
We have $x^{out}=x^{\ell,out}_-\cup x^{r,out}_+$. 
	
For any $j\in I(N)$ let
	\begin{equation}
		a_{\pm,j}(\Gamma):=\gamma_\pm\pm j.
	\end{equation}	
Then there exist unique $\chi_-^{\ell}\equiv \chi^{\ell}_-(x,\Gamma) \in\mathcal X^{\ell}$ and $\chi_+^{r}\equiv \chi^{r}_+(x,\Gamma) \in\mathcal X^{r}$  with $\chi_{-,\ell}^{\ell},\chi_{+,r}^{r}<L$ such that
	\begin{align}\label{eq:chiminus}
		x_{\sigma_{\ell,r}(\zeta)}&=(a_{-,\ell-\zeta+1}-\chi_{-,\ell-\zeta+1}^{\ell})\modd L \quad\textrm{for all }1\le \zeta\le\ell,\\
		x_{\sigma_{\ell,r}(N+1-\xi)}&=(a_{+,r-\xi+1}+\chi_{+,r-\xi+1}^{r})\modd L \quad\textrm{for all }1\le\xi\le r.\label{eq:chiplus}
	\end{align}

Finally let us denote the special configuration in $\mathcal V_{L}^{\ell+r}(\Gamma^c)$ that consists of two clusters of size $\ell$ and $r$ at the boundaries of $\Gamma^c$ by
	\begin{align}\label{eq:defb}
		b_{\ell,r}\equiv b_{\ell,r}(\Gamma):=\{a_{-,j}:\;j\le\ell\}\cup\{a_{+,j}:\;j\le r\}.
	\end{align}

%%%%%%%%%%%%%%%%%%%%%%%%%%%%%%%%%%%%%%%%%%%%%%	
\begin{lemma}\label{lem:averytechnicallemma}
Let $\Gamma=\{\gamma_-,\hdots,\gamma_+\}\subseteq\mathcal S_{L,M}(\theta+2\epsilon)$. Moreover, let $n\in\N$ with $n<N$ and $x\in\mathcal V_L^N$ with $|x^{in}(\Gamma)|=n$. Let $r,\ell\in\N_0$ such that $\ell+r=N-n$ and $c\in\mathcal V_{L,1}^N$ with  $\{c_j:\;\ell<j\le N-r\}\subseteq\Gamma$. Assume that ${\bf u}$ is a shortest path from $c$ to $x$ of length $k:=d_L^N(c,x)$ such that 
	\begin{equation}\label{eq:cyclicrlpermutation}
		\tilde u^{(k)}_j=x_{\sigma_{\ell,r}(j)}\quad\textrm{for all }j\in I(N)
	\end{equation}
 and in addition that
	\begin{align}
		\{1,\hdots,\ell\}\subseteq I^{\bf u}_0\cup I^{\bf u}_-&&\textrm{and}&&c_{\zeta}\ge a_{-,\ell-\zeta+1}\quad\textrm{for }\zeta\le\ell \label{eq:leftboundary}\\
		\{N-r+1,\hdots,N\}\subseteq I^{\bf u}_0\cup I^{\bf u}_+&&\textrm{and}&&c_{N+1-\xi}\le a_{+,r-\xi+1}\quad\textrm{for }\xi\le r \label{eq:rightboundary}.
	\end{align}
Then there exists a shortest path ${\bf v}$ from $c$ to $x$ with
	\begin{equation}
		v^{(k_0)}=x^{in}\cup b_{\ell,r}(\Gamma)\subseteq\mathcal S_{L,M}(1/4-\epsilon/2)
	\end{equation}
	for some $k_0\in\{0,\hdots, k\}$.
Furthermore, 
	\begin{equation}
		k-k_0=\sum_{\zeta=1}^{ \ell}\chi_{-,\zeta}^\ell(x,\Gamma)+\sum_{\xi=1}^r\chi_{+,\xi}^r(x,\Gamma).
	\end{equation}
\end{lemma}
\begin{proof}
For all $j\in I(N)$ let
	\begin{equation}
		\mathcal Z_j:=\{\tilde u^{(l)}_j:\;0\le l\le k\}.
	\end{equation}
	
We claim that for all $j\in \{\ell+1,\hdots,N-r\}\cap(I^{\bf u}_0\cup I^{\bf u}_\pm)$ one has
	\begin{equation}\label{eq:everythingstaysinside}
		\mathcal Z_j=\{c_j,\hdots,(c_j\pm L^{\bf u}_j)\modd L\}\subseteq \Gamma.
	\end{equation}	
Suppose this is not true. This would imply thet there exists a $j\in I(N)$ with $\Gamma^c\cup\{\tilde u_j^{(0)},\tilde u_j^{(k)}\}\subseteq \mathcal Z_j$. By Assumption \eqref{eq:cyclicrlpermutation} we have  $\tilde u^{(k)}_j=x_{\sigma_{\ell,r}(j)}\in\Gamma$ and $\tilde u^{(0)}_j=c_j\in\Gamma$ for all $\ell<j\le N-r$. Hence,
	\begin{equation}
		|\mathcal Z_j|=|\Gamma^c|+2> L/2+1\ge L_j^{\bf u}+1=|\mathcal Z_j|,
	\end{equation}
where we used $|\Gamma^c|\ge|(\mathcal S_{L,M}(1/4))^c|> L/2-1$. This is a contradiction.

We claim that for for  all $\zeta\le \ell$ and for all $\xi\le r$, it holds that
	\begin{align}\label{eq:ainsetuminus}
		a_{-,\ell-\zeta+1}\equiv a_{-,\ell-\zeta+1}(\Gamma)&\in\mathcal Z_\zeta=\{c_\zeta,\hdots,(c_\zeta-L_\zeta^{\bf u})\modd L\},\\
		\label{eq:ainsetuplus}
		a_{-,r-\xi+1}\equiv a_{-,r-\xi+1}(\Gamma)&\in\mathcal Z_{N+1-\xi}=\{c_{N+1-\xi},\hdots,(c_{N+1-\xi}+L_{N+1-\xi}^{\bf u})\modd L\}.
	\end{align}
We present a proof for \eqref{eq:ainsetuminus}, since \eqref{eq:ainsetuplus} follows analogously.
For $\zeta=\ell$ we have $\tilde u_\ell^{(k)}\in x^{out}(\Gamma)\subseteq\Gamma^c$ according to Assumption \eqref{eq:cyclicrlpermutation}. Together with Assumption \eqref{eq:leftboundary} this implies $\gamma_--1=a_{-,1}\in\mathcal Z_\ell$. The claim now follows from an inductive argument as well as from Lemma~\ref{lem:hard-coreProperty}~(i). 
	
Let us define
	\begin{equation}
		K_{-,\ell}:=\left\{	\begin{array}{ll}
								d_L(c_{\ell},a_{-,1})							&	\textrm{for }\ell>0,\\
								0														&	\textrm{for }\ell=0,
							\end{array}\right.
		\quad\quad\textrm{ and }\quad\quad		
		K_{+,r}:=\left\{	\begin{array}{ll}
								d_L(c_{N+1-r},a_{+,1})							&	\textrm{for }r>0,\\
								0															&	\textrm{for }r=0.
							\end{array}\right.			
	\end{equation}
Then, \eqref{eq:ainsetuminus} and \eqref{eq:ainsetuplus} imply that for all $\zeta\le \ell$ and for all $\xi\le r$
	\begin{equation}
		K_{-,\ell}=d_L(c_{\zeta},a_{-,\ell-\zeta+1})\le L^{\bf u}_\zeta\quad\textrm{and}\quad K_{+,r}=d_L(c_{N+1-\xi},a_{-,r-\xi+1})\le L^{\bf u}_{N+1-\xi}.
	\end{equation}

Let us now give an iterative construction of a path ${\bf v}=(v^{(0)},\hdots,v^{(k)})$ starting from $c$. 
To this end, set $\tilde v^{(0)}:=(c_1,\hdots,c_N)$. For $\zeta\in\{1,\hdots,\ell\}$ and $l\in (\zeta-1)K_{-,\ell}+ \{1,\hdots, K_{-,\ell} \}$, let
	\begin{equation}
		\tilde v^{(l)}:=(\tilde v^{(l-1)}_1,\hdots,\tilde v^{(l-1)}_{\zeta}-1,\hdots,\tilde v^{(l-1)}_N).
	\end{equation}
Let $k_1:=\ell K_{-,\ell}$. For $\xi\in\{1,\hdots,r\}$ and $l\in  k_1+(\xi-1)K_{+,r}+\{1,\hdots, K_{+,r}\}$ let
 	\begin{equation}
		\tilde v^{(l)}:=(\tilde v^{(l-1)}_1,\hdots,\tilde v^{(l-1)}_{N+1-\xi}+1,\hdots,\tilde v^{(l-1)}_N).
	\end{equation}
The path ${\bf v}$ has the property that it moves all particles of the configuration $\{c_j:\;j\le \ell \textrm{ or }j\ge N-r+1\}$ into the configuration $b_{\ell,r}(\Gamma)$ outside the boundary of $\Gamma$. 

In the next step, we move the particles that are still remaining inside of $\Gamma$ into the configuration $x^{in}=x^{in}(\Gamma)$. Let $k_2:= k_1+rK_{+,r}$. Let $\ell^\prime:=|\{j\in I_-^{\bf u}: j>\ell\}|$, $r^\prime:=|\{j\in I_+^{\bf u}: j\le N-r\}|$. For $\zeta\in \{1,\hdots, \ell^\prime\}$ and $l\in k_2+\sum_{j=1}^{\zeta-1} L_{\ell+j}^{\bf u}+\{1,\hdots,L_{\zeta}^{\bf u}\}$ we set
	\begin{equation}
		\tilde v^{(l)}:=(\tilde v^{(l-1)}_1,\hdots,\tilde v^{(l-1)}_{\ell+\zeta}-1,\hdots,\tilde v^{(l-1)}_N).
	\end{equation}
Let $k_3:=k_2+\sum_{j=1}^{\ell^\prime}L^{\bf u}_{\ell+j}$. For $\xi\in\{1,\hdots, r^\prime\}$ and $l\in k_3+\sum_{j=1}^{\xi-1} L_{N-r-j}^{\bf y}+\{1,\hdots,L^{N-r-\xi}\}$ we set
	\begin{equation}
		\tilde v^{(l)}:=(\tilde v^{(l-1)}_1,\hdots,\tilde v^{(l-1)}_{N-r-\xi}+1,\hdots,\tilde v^{(l-1)}_N).
	\end{equation}
The fact that this construction is well-defined follows from the statement in \eqref{eq:everythingstaysinside}, since no particle with index $j\in\{\ell+1,\hdots,N-r\}$ leaves $\Gamma$ and therefore does not intersect the configuration $b_{\ell,r}$.
In the last step, we move the configuration $b_{\ell,r}$ into $x^{out}(\Gamma)$. Let $k_0:=k_3+\sum_{j=1}^{r^\prime}L^{\bf u}_{\ell+j}$. By construction we have $ v ^{(k_0)}=b_{\ell,r}\cup x^{in}(\Gamma)$. Notice that by definition of $\chi^\ell_-\equiv \chi^\ell_-(x,\Gamma)$ and $\chi^r_+\equiv \chi^r_+(x,\Gamma)$, we obtain for all $\zeta\le\ell$ and  $\xi\le r$ that
	\begin{align}
		\chi_{-,\ell-\zeta+1}^\ell&=d_L(\tilde u^{(k)}_{\zeta},a_{-,\ell-\zeta+1})=L_\zeta^{\bf u}-K_{-,\ell},\label{eq:chinactminus}
	\end{align}
	\begin{align}\label{eq:chinactplus}
		\chi_{+,r-\xi+1}^r&=d_L(\tilde u^{(k)}_{N+1-\xi},a_{+,r-\xi+1})=L_{N+1-\xi}^{\bf u}-K_{+,r}.
	\end{align}
For all $\zeta\in\{1,\hdots,\ell\}$ and $l\in k_0+\sum_{j=1}^{\zeta-1}\chi_{-,\ell-j+1}^\ell+\{1,\hdots,\chi_{-,\ell-\zeta+1}^\ell\}$ we set
	\begin{equation}
		\tilde v^{(l)}:=(\tilde v^{(l-1)}_1,\hdots,(\tilde v^{(l-1)}_{\zeta}-1)\modd L,\hdots,\tilde v^{(l-1)}_N).
	\end{equation}
Let $k_4:=k_0+\sum_{\zeta=1}^{ \ell}\chi_{-,\zeta}^\ell$. For $\xi\in\{1,\hdots,r\}$ and $l\in k_4+\sum_{j=1}^{\xi-1}\chi_{+,r-j+1}^r+\{1,\hdots,\chi_{+,r-\xi+1}^r\}$ we set
 	\begin{equation}
		\tilde v^{(l)}:=(\tilde v^{(l-1)}_1,\hdots,(\tilde v^{(l-1)}_{N+1-\xi}+1)\modd L,\hdots,\tilde v^{(l-1)}_N).
	\end{equation}
By construction, ${\bf v}$ is a shortest path from $c$ to $x$, since  it has a length of $k=\sum_{j=1}^NL^{\bf u}_{j}$ and $\tilde v^{(k)}=x$. 
Moreover, by construction as well as \eqref{eq:chinactminus} and \eqref{eq:chinactplus}, we get
	\begin{equation}
		k-k_0=\sum_{\zeta=1}^{ \ell}\chi_{-,\zeta}^\ell+\sum_{\xi=1}^{ r}\chi_{+,\xi}^r
	\end{equation}
	
Finally, we note that $v^{(k_0)}\subseteq \mathcal S_{L,M}(1/4-\epsilon/2)$. This follows from the fact that for all $m\in b_{\ell,r}$, one has $d_L(m,M)< (\theta+2\epsilon) L+(\ell+r)\le (1/4-\epsilon/2)L$, where we used $\ell+r\le N/2<\epsilon/2$ and $\epsilon<\theta<1/16$.
\end{proof}

%%%%%%%%%%%%%%%%%%%%%%%%%%%%%%%%%%%%%%%%%%%%%%
\begin{lemma} \label{lemma:newpath}
Let $\Lambda_L^\prime:=\mathcal S_{L,M}(\theta+2\epsilon)$. Fix $n\in\N$ such that $N/2<n< N$ and $x\in\mathcal V_L^N$ with 
$x^{in}(\Lambda^\prime_L)\in\mathcal V^n(\Lambda^\prime_L)$. Let $c\in \mathcal V_{L,1}^N$, with $c\subseteq\Lambda_L^\prime$ and set $k:=d_L^N(x,c)$. Then there exists $r,\ell\in\N_0$ with $r+\ell=N-n$ and a shortest path ${\bf v}$ from $c$ to $x$ such that $\tilde v^{(k)}_j=x_{\sigma_{\ell,r}(j)}$ for all $j\in I(N)$. Moreover, there exists a $k_0\in\{0,\hdots, k\}$  such that 
	\begin{equation}
		v^{(k_0)}=x^{in}(\Lambda^{\prime}_L)\cup b_{\ell,r}(\Lambda^{\prime}_L)\subseteq \mathcal S_{L,M}(1/4-\epsilon/2).
	\end{equation}
\end{lemma}
\begin{proof}
Let $\lambda_-^\prime:=\min\Lambda^\prime_L$ and $\lambda^\prime_+:=\max\Lambda_L^\prime$. Then $\Lambda_L^\prime=\{\lambda_-^\prime,\hdots,\lambda_+^\prime\}$.

Let ${\bf  u}$  be a shortest path from $c$ to $x$. According to Lemma~\ref{lem:SchnipSchnap} there exists an edge $e\in\mathcal E_L$ with $\max_{j\in I( N)}d_L(c_j,e)\ge (1/2-\epsilon)L$, which is not crossed by the path ${\bf u}$. Notice that this implies $e\subseteq (\Lambda_L^{\prime})^{c}$.
 We define the index sets 
	\begin{equation}\label{eq:Jinoutdefi}
		J^{{\bf u},in}_\pm:=\{j\in I_{\pm}^{\bf u}:\;\tilde u^{(k)}_j\in x^{in}(\Lambda_L^\prime)\} \quad\textrm{ and }\quad J^{{\bf u},out}_\pm:=\{j\in I_{\pm}^{\bf u}: 
		\;\tilde u^{(k)}_j\in x^{out}(\Lambda_L^\prime)\}.
	\end{equation}
We claim that 
	\begin{equation}\label{eq:Jpm}
		\max J_+^{{\bf u},in}\le\min J_+^{{\bf u},out}\quad\textrm{ and }\quad\max J^{{\bf u},out}_-\le\min J^{{\bf u},in}_-.
	\end{equation}
These statements are a consequence of Lemma~\ref{lem:hard-coreProperty}. Here, we only prove the first inequality, the other one follows analogously. Suppose that $\max J^{{\bf u},in}_x>\min J_+^{{\bf u},out}$. This implies that there exists a $j\in J^{{\bf u},out}_+$ such that $j+1\in J^{{\bf u},in}_+$. According to Lemma~\ref{lem:hard-coreProperty}~(i) we have $L^{\bf u}_j\le L^{\bf u}_{j+1}$ and therefore also
	\begin{equation}
		\lambda_+^{\prime}\ge \tilde u^{(k)}_{j+1}=c_{j+1}+L^{\bf u}_{j+1}>c_{j}+L^{\bf u}_{j}=\tilde u^{(k)}_j\ge\lambda_-^{\prime},
	\end{equation}
which is a contradiction to $j\in J^{{\bf u},out}_+$. 

Let $\ell:=|J^{{\bf u},out}_-|$ and $r:=|J^{{\bf u},out}_+|$. Together with \eqref{eq:Jpm} this implies $\{j:j\le\ell\}=J^{{\bf u},out}_-\subseteq I^{\bf u}_-$ and $\{j:j>N-r\}=J^{{\bf u},out}_+\subseteq I^{\bf u}_+$. Furthermore, Lemma~\ref{lem:hard-coreProperty} yields $\tilde u_j^{(k)}=x_{\sigma_{\ell,r}(j)}$ for all $j\in I( N)$. Moreover, for all $\zeta\in\{1,\hdots,\ell\}$ holds $c_{\zeta}\ge\lambda_-> a_{-,\ell-\zeta+1}(\Lambda_L^\prime)$ and for all $\xi\in\{1,\hdots, r\}$ holds $c_{N+1-\xi}\le\lambda_+< a_{+,r-\xi+1}(\Lambda_L^\prime)$, since $c\subseteq\Lambda_L^\prime$.

 Lemma~\ref{lem:averytechnicallemma} now yields the proposition.
\end{proof}
%%%%%%%%%%%%%%%%%%%%%%%%%%%%%%%%%%%%%%%%%%%%%%%%%%%

Let us now define the set of negligible configurations that are in a large enough distance to the set of cluster configurations. This set is given by
	\begin{equation}
		\mathcal C_L^N:=\{x\in\mathcal V_L^N:\; d_L^N(x,\mathcal V_{L,1}^N)\ge L^{3/2}\}.
	\end{equation}

%%%%%%%%%%%%%%%%%%%%%%%%%%%%%%%%%%%%%%%%%%%%%%%%%%%
\begin{lemma} \label{lemma:almostringdrop}
There exists a $L_0\equiv L_0(\epsilon)>0$ such that for all $L\ge L_0$, $n\in\N$ with $N/2<n< N$ and $x\in\mathcal V_L^N\setminus \mathcal C^N_L$ with $x\cap\Lambda_L\in\mathcal V^n(\Lambda_L)$ one has
	\begin{equation}\label{eq:WLsubsetSthetaepsilon}
		\mathcal W_L(x)\subseteq \mathcal S_{L,M}(\theta+\epsilon)\:.
	\end{equation}
Moreover, for all $m\in\mathcal W_L(x)$:
	\begin{equation}
		c_{L,m}^N\subseteq \mathcal S_{L,M}(\theta+2\epsilon).
	\end{equation}
\end{lemma}
\begin{proof} 
Let us first show \eqref{eq:WLsubsetSthetaepsilon}. Let $\nu\in (\mathcal S_{L,M}(\theta+\epsilon))^c$. Then for all $\xi\in c_{L,\nu}^N$ one has
	\begin{equation}
		d_L(\xi,\Lambda_L)\ge d_L(\nu,\Lambda_L)-\lceil (N+1)/2\rceil\ge \epsilon L-2\epsilon L/3
	\end{equation}
for all $L\ge L_1$ with $L_1\equiv L_1(\epsilon):=9/\epsilon$. Since $n$ particles of $x$ are located inside of $\Lambda_L$ we have
	\begin{equation}
		d_L^N(x,c_{L,\nu}^N)\ge n \epsilon L/3\ge \epsilon^2 L^2/6,
	\end{equation}
where we applied Lemma~\ref{lem:cycthing}.
Let $L_0\equiv L_0(\epsilon)>L_1$, such that $\epsilon^2 L_0^{1/2}/6>1$. For any $L\ge L_0$ this implies $d_L^N(x,c_{L,\nu}^N)\ge L^{3/2}$. Since $x\notin\mathcal C_L^N$ by assumption, we conclude that $\nu\notin \mathcal W_L(x)$. 

For all  $L\ge L_0$ and $m\in \mathcal W_L(x)\subseteq\mathcal S_{L,M}(\theta+\epsilon)$, observe that for all $\xi\in c_{L,m}^N$, we have
	\begin{equation}
		d_L(\xi,M)\le d_L(\xi,m)+d_L(m,M)<\lceil (N+1)/2\rceil +(\theta+\epsilon)L\le (\theta+2\epsilon)L\:,
	\end{equation}
	from which we conclude $c_{L,m}^N\subseteq\mathcal S_{L,M}(\theta+2\epsilon)$. This shows the lemma.
\end{proof}

%%%%%%%%%%%%%%%%%%%%%%%%%%%%%%%%%%%%%%%%%%%%%%%%%%%%%
\begin{lemma}\label{lem:DasSchweigenDerLemma}
Let $L\ge L_0$, with $L_0$ as in Lemma~\ref{lemma:almostringdrop}. Let $n\in\N$ with $N/2<n< N$, $y\in\mathcal V^n(\Lambda_L)$, $x\in\mathcal V_L^N\setminus \mathcal C^N_L$ with $x^{in}(\Lambda_L)=y$. Then there exist $\ell,r\in\N_0$ with $r+\ell=N-n$ such that $y_{\kappa-\ell}\in\mathcal W_L(x)$ for $\kappa:=\lfloor(N+1)/2\rfloor$ and 
	\begin{equation}
		c_{L,y_{\kappa-\ell}}^N\subseteq \mathcal S_{L,M}(1/4-\epsilon/2).
	\end{equation}
Furthermore, there exists a shortest path ${\bf v}$ from $c_{L,y_{\kappa-\ell}}^N$ to $x$ and a $k_0\in\{1,\hdots, k\}$ with $k:=d_L^N(x,\mathcal V_{L,1}^N)$ such that
	\begin{equation}
		v^{(k_0)}=y\cup b_{\ell,r}(\Lambda_L)\subseteq\mathcal S_{L,M}(1/4-\epsilon/2),
	\end{equation}
and
	\begin{equation}
		k-k_0=\sum_{\zeta=1}^{ \ell}\chi_{-,\zeta}^\ell(x,\Lambda_L)+\sum_{\xi=1}^r\chi_{+,\xi}^r(x,\Lambda_L).
	\end{equation}
\end{lemma}
\begin{proof}
By Lemma~\ref{prop:chaindrop} there exists $m\in\mathcal W_L(x)\cap x$. From Lemma~\ref{lemma:almostringdrop}, we know 
	\begin{equation}
		c:=c_{L,m}^N\subseteq\Lambda_L^\prime:=\mathcal S_{L,M}(\theta+2\epsilon)\subseteq \mathcal S_{L,M}(1/4-\epsilon/2).	
	\end{equation}
According to Lemma~\ref{lemma:newpath} there exists a shortest path ${\bf u}$ from $c$ to $x$ and $\ell^\prime,r^\prime\in\N_0$ with $\ell^\prime+r^\prime=N-|x^{in}(\Lambda_L^\prime)|$ and $k_1\in\{0,\hdots, k\}$ such that 
	\begin{equation}
		z:=u^{(k_1)}=b_{\ell^\prime,r^\prime}(\Lambda^\prime_L)\cup x^{in}(\Lambda_L^\prime)\subseteq \mathcal S_{L,M}(1/4-\epsilon/2),
	\end{equation}
and $\tilde u^{(k)}_j=x_{\sigma_{\ell^\prime,r^\prime}(j)}$ for all $j\in I(N)$.

Let us define
	\begin{align}
		\ell&:=\ell^\prime+|\{\nu\in x^{in}(\Lambda_L^\prime):\;\nu<\lambda_-\}|=|\{j:z_j<\lambda_-\}|\quad\textrm{ and }\\
		 r&:=r^\prime+|\{\nu\in x^{in}(\Lambda_L^\prime):\;\nu>\lambda_+\}|=|\{j:z_j>\lambda_+\}|.
	\end{align}
These quantities satisfy $\ell+r=N-|\{j:\;z_j\in\Lambda_L\}|=N-n$. Note that by this definition $\sigma_{\ell^\prime,r^\prime}(x,\Lambda_L^\prime)=\sigma_{\ell,r}(x,\Lambda_L)$. Hence, $\tilde u^{(k)}_j=x_{\sigma_{\ell,r}(j)}$ for all $j\in I(N)$ and $y_j=z_{j+\ell}$  for all $j\in\{1,\hdots, n\}$.

Next, we show that $m=z_\kappa$. First, we claim that
	\begin{equation}\label{eq:minWLy}
		m\in\mathcal W_L(z).
	\end{equation}
Take any $\nu\in\mathcal W_L(z)$ and any shortest path ${\bf v}$ from $c^\prime:=c_{L,\nu}^N$ to $z$. Lemma~\ref{lem:decompositionOfPaths} indicated that 
	\begin{equation}\label{eq:k1k2}
		k_2:=d_L^N(z,c^\prime)=d_L^N(z,\mathcal V_{L,1}^N)\le d_L^N(z,c)=k_1.
	\end{equation}
The path $\{v^{(0)},\hdots,v^{(k_2)},u^{(k_1+1)},\hdots,u^{(k)}\}$ is therefore a path from $c^\prime$ to $x$ of length $k+(k_2-k_1)$ and therefore -- using \eqref{eq:k1k2} -- we get
	\begin{equation}
		k=d_L^N(x,\mathcal V_{L,1}^N)\le k+(k_2-k_1)\le k.
	\end{equation}
Hence  $k_1=k_2$. Equality in \eqref{eq:k1k2} implies $m\in\mathcal W_L(z)$.
According to Lemma~\ref{lemma:almostringdrop} we have $\mathcal W_L(z)\subseteq \mathcal S_{L,M}(\theta+\epsilon)\subseteq \Lambda^\prime_L$. Hence 
	\begin{equation}
		m\in x\cap\mathcal W_L(z)=x\cap \mathcal W_L(z)\cap \Lambda^\prime_L=z\cap\mathcal W_L(z)=\{z_\kappa\},
	\end{equation}
where we used  $z\subseteq \mathcal S_{L,M}^N(1/4-\epsilon/2)$ together with Lemma~\ref{lemma:middle}. Therefore $m=z_\kappa=y_{\kappa-\ell}$.

Lemma~\ref{lem:decompositionOfPaths} states that ${\bf w}:=(u^{(0)},\hdots,u^{(k_1)})$ is a shortest path from $c$ to $z$. 
According to Lemma~\ref{lemma:middle} we have $\kappa\in I^{{\bf w}}_0$, since $m=z_\kappa$. This also implies $\kappa\in I^{{\bf u}}_0$, since by construction of the path ${\bf u}$, for all $l\in\{k_1,\hdots,k\}$ one has $\tilde{u}^{(l)}_\kappa=\tilde{u}^{(k_1)}_\kappa$. 
In this case, it follows from Lemma~\ref{lem:hard-coreProperty}~(iii) that $\{j:j\le\ell\}\subseteq\{0,\hdots,\kappa\}\subseteq I_0^{{\bf u}}\cup I^{{\bf u}}_-$ 
 and for all $\zeta\le\ell$, we have
	\begin{equation}\label{eq:aminusbiggeryell}
		a_{-,\ell-\zeta+1}\le z_{\ell+1}-(\ell-\zeta+1)\le c_{\ell+1}-(\ell-\zeta+1)=c_\zeta.
	\end{equation}
Analogously, we have $\{j:\;j>N-r\}\subseteq\{\kappa,\hdots,N\}\subseteq I_0^{{\bf u}}\cup I^{{\bf u}}_+$ 
and for all $\xi\le r$ one therefore gets
	\begin{equation}\label{eq:yxiplus}
		a_{+,r-\xi+1}\ge c_{N+1-\xi}.
	\end{equation}
According to Lemma~\ref{lem:averytechnicallemma} there therefore exists a path ${\bf v}$ with all the properties stated in the proposition.
\end{proof}

%%%%%%%%%%%%%%%%%%%%%%%%%%%%%%%%%%%%%%%%%%%%%%%%%%%
%%%%%%%%%%%%%%%%%%%%%%%%%%%%%%%%%%%%%%%%%%%%%%%%%%%
\subsection{Estimates based on geometric series}
%%%%%%%%%%%%%%%%%%%%%%%%%%%%%%%%%%%%%%%%%%%%%%%
In this section, we compute and estimate various geometric sum, which will be necessary for estimating the partial trace later.
%%%%%%%%%%%%%%%%%%%%%%%%%%%%%%%%%%%%%%%%%%%%%%%
\begin{lemma}\label{lem:thebeesknees}
There exists $L_0\equiv L_0(\epsilon)>0$ such that for all $L\ge L_0$, $n\in\N$ with $N/2< n<N$, $y\in\mathcal V^{n}(\Lambda_L)$ and $\mu\ge \ln 2$ one has
	\begin{equation}
		\sum_{\genfrac..{0pt}{2}{z\in\mathcal V^{N-n}(\Lambda_L^c),}{ z\cup y\notin \mathcal V_{L,1}^N}}\e^{-\mu d^N_L(y\cup z,\mathcal V_{L,1}^N)}\le 
		333\e^{-\mu}\e^{-\mu h^n_L(y)},
	\end{equation}
with $h_L^n:\;\mathcal V^n(\Lambda_L)\rightarrow(0,\infty)$,
	\begin{equation}\label{eq:hl}
		h_L^n(y):=\left\{	\begin{array}{ll}
									\min\big\{d_L^{n+1}(y\cup\{a_{\pm,1}(\Lambda_L)\},\mathcal V_{L,1}^{n+1}), L^{5/4}\big\}-1		&	\textrm{for }y\notin\{y^n_+,y_-^n\},\\
									0																																		&	\textrm{for }y\in\{y^n_+,y_-^n\},
					\end{array}\right.
	\end{equation}
where $y_\pm^n:=\lambda_\pm\mp\{0,\hdots,n-1\}$.
\end{lemma}
\begin{proof}
Let
	\begin{align}
		\mathcal A^{\prime}(y)&:=\big\{z\in\mathcal V^{N-n}(\Lambda_L^c):\;y\cup z\notin \mathcal C_L^N\}\subseteq \mathcal V^{N-n}(\Lambda_L^c),\\
		\mathcal A(y)&:=\big\{z\in\mathcal A^{\prime}(y):\;y\cup z\notin \mathcal V_{L,1}^N\}\subseteq \mathcal A^{(\prime)}(y).
	\end{align}
There exists $L_1$ such that $L^{1/2}-\ln L/\ln 2\ge L^{1/4}$ for all $L\ge L_1$. Hence for all $L\ge L_1$ we get 
	\begin{equation}\label{eq:alphaatitswork}
		\sum_{z\in (\mathcal A(y))^c}\e^{-\mu d_L^N(y\cup z,\mathcal V_{L,1}^N)}
		\le |(\mathcal A(y))^c|\e^{-\mu L^{3/2}}\le\e^{-\mu L^{5/4}}\le\e^{-\mu}\e^{-\mu h^n_L(y)},
	\end{equation}
where we used $|(\mathcal A(y))^c|\le |\Lambda_L^c|^{N-n}\le L^L$ as well as $\mu\ge\ln 2$.
We partition $\mathcal A^{(\prime)}(y)$ into  smaller subsets. For any $\ell,r\in\N_0$ with $\ell+r=N-n$ let $c^\ell:=c_{L,y_{\kappa-\ell}}^N$ with $\kappa:=\lfloor(N+1)/2\rfloor$. Let us further define 
	\begin{align}
		\mathcal A_{\ell,r}^{(\prime)}(y):=\big\{z\in\mathcal A^{(\prime)}(y):\;&
		d_L^N(y\cup z,\mathcal V_{L,1}^N)=d_{L}^N(c^\ell,y\cup b_{\ell,r})
		+d_L^N(y\cup b_{\ell,r},y\cup z),\notag\\\label{eq:defAlrprime}
		&d_L^N(y\cup b_{\ell,r},y\cup z)=\sum_{\zeta=1}^{ \ell}\chi_{-,\zeta}^\ell(x,\Lambda_L)+\sum_{\xi=1}^r\chi_{+,\xi}^r(x,\Lambda_L)\big\},
	\end{align}
where $\chi_-^\ell(x,\Lambda_L)\in\mathcal X^\ell$ and $\chi_+^r(x,\Lambda_L)\in\mathcal X^r$ where defined in \eqref{eq:chiminus} and \eqref{eq:chiplus}.

Lemma~\ref{lem:decompositionOfPaths} and Lemma~\ref{lem:DasSchweigenDerLemma} imply immediately, that there exists a $L_2\equiv L_2(\epsilon)>L_1$ such that  for all $L\ge L_2$, we get the equality
	\begin{equation}\label{eq:Apartition}
		\mathcal A(y)=\bigcup_{\genfrac..{0pt}{2}{\ell,r\in\N_0,}{ \ell+r=N-n}}\mathcal A_{\ell,r}(y)=\mathcal A_{N-n,0}(y)
		\cup\mathcal A_{0,N-n}(y)\cup\bigcup_{\genfrac..{0pt}{2}{\ell,r\in\N,}{ \ell+r=N-n}}\mathcal A^\prime_{\ell,r}(y).
	\end{equation}
Let us first consider the case $\ell,r\in\mathbb \N$.
Definition \eqref{eq:defAlrprime} implies, together with Lemma~\ref{lem:geomsum} for $\mu\ge \ln 2$, that
	\begin{align}
		\sum_{z\in\mathcal A^\prime_{\ell,r}(y)}\e^{-\mu d_L^N(y\cup z,\mathcal V_{L,1}^N)}&\le \e^{-\mu d_{L}^N(c^\ell,y\cup b_{\ell,r})}
		\Big(\sum_{\chi^\ell\in\mathcal X^\ell}\e^{-\mu|\chi^\ell|_1}\Big)\Big(\sum_{\chi^r\in\mathcal X^r}\e^{-\mu|\chi^r|_1}\Big)\notag\\
		&\le \e^{-\mu d_{L}^N(c^\ell,y\cup b_{\ell,r})} (1+30\e^{-\mu})^2.\label{eq:handystep}
	\end{align}
Now, we estimate the first factor on the right hand side of \eqref{eq:handystep} uniformly in $\ell,r$. Both $c^\ell$ and $y\cup b_{\ell,r}$ are subsets of $\mathcal S_{L,M}(1/4-\epsilon/2)$. By Lemma~\ref{lemma:middle} we have $y_{\kappa-\ell}\in\mathcal W_L(y\cup b_{\ell,r})$ with
	\begin{equation}\label{eq:ellrgroesser0}
		d_L^N(y\cup b_{\ell,r},c^\ell)=d^N(y\cup b_{\ell,r},c^\ell)\ge\sum_{j=1}^n|y_j-c_{j+\ell}^\ell|+|a_{-,1}-c_{\ell}^\ell|+|a_{+,1}-c_{N-r+1}^\ell|,
	\end{equation}
where we applied \eqref{eq:dLgleichdohneL} and Remark~\ref{rem:theline}.
For all $\ell\in\{1,\hdots, N-n-1\}$ we have
	\begin{equation}
		|a_{-,1}-c_{\ell}^\ell|+|a_{+,1}-c_{N-r+1}^\ell|\ge|a_{+,1}-a_{-,1}|-|c_{N-r+1}^\ell-c_{\ell}^\ell|\ge 2\theta L-(n+1)\ge \epsilon L,
	\end{equation}
where we used that $a_{+,1}-a_{-,1}\ge d_L(a_{+,1},M)-d_L(a_{-,1},M)\ge2\theta L$, as well as $n+1\le N <\epsilon L$ and $\theta>\epsilon$.
Hence for all $y\in\mathcal V^n(\Lambda_L)$ we have either $|a_{-,1}-c_{\ell}^\ell|\ge\epsilon L/4+1$ or $|a_{+,1}-c_{r+\ell+1}^\ell|\ge \epsilon L/4+1$ for all $L\geq L_0=L_0(\epsilon)=\max\{L_2,4/\epsilon\}$. This implies, together with \eqref{eq:ellrgroesser0} that
	\begin{equation}\label{eq:rqllnotzero}
		d_L^N(y\cup b_{\ell,r},c^\ell)-1\ge h^n_L(y)+\epsilon L/4.
	\end{equation}
Hence, by combining  \eqref{eq:handystep} and \eqref{eq:rqllnotzero}  we find
	\begin{equation}\label{eq:allmostthere}
		\sum_{\genfrac..{0pt}{2}{\ell,r\in\N,}{ \ell+r=N-n}}\sum_{z\in\mathcal A^\prime_{\ell,r}(y)}\e^{-\mu d_L^N(y\cup z,\mathcal V_{L,1}^N)}
		\le (N-n)\e^{-\mu\epsilon L/4}(1+30\e^{-\mu})^2\e^{-\mu}e^{-\mu h^n_L(y)}\le 272\e^{-\mu}\e^{-\mu h_L^n(y)},
	\end{equation}
where we used that $(N-n)\e^{-\mu\epsilon L/4}\le(\epsilon L/2)2^{-\epsilon L/4}\le \frac{2}{\e\ln2}$ for all $\mu\ge\ln 2$.

Let us now consider the case $\ell=0$ or $r=0$. There are only two configurations $y\in \mathcal V^n(\Lambda_L)$ such that there exists a $z\in \mathcal V^{N-n}(\Lambda_L^c)$ with $y\cup z\in\mathcal V_{L,1}^N$, namely $y_+^n$ and $y_-^n$. The configurations $z_-^n:=b_{N-n,0}$ and $z_+^n:=b_{0,N-n}$ satisfy $y_\pm^n\cup z_\pm^n\in\mathcal V_{L,1}^n$. There are no other configurations in $\mathcal V^{N-n}(\Lambda_L^c)$ with this property.
We further restrict ourselves to the case $\ell=N-n$ and $r=0$. The other case can be treated analogously. Now, our approach depends on whether $y=y^n_-$ or not. We have
	\begin{equation}
		\mathcal A_{N-n,0}(y)\subseteq  \left\{		\begin{array}{ll}
																											\mathcal A^\prime_{N-n,0}(y)								&	\textrm{ for }y\neq y^n_-,\\
																											\mathcal A^\prime_{N-n,0}(y)\setminus\{z^n_-\}	&	\textrm{ for }y=y^n_-.
																										\end{array}\right.
	\end{equation}
Analogous to \eqref{eq:handystep} we obtain
	\begin{equation}
		\sum_{z\in\mathcal A_{N-n,0}(y)}\e^{-\mu d_L^N(y\cup z,\mathcal V_{L,1}^N)}\le \e^{-\mu d_L^N(y\cup b_{N-n,0},c^{N-n})}\left\{		
								\begin{array}{ll}
									\sum_{\chi\in\mathcal X^{N-n}}\e^{-\mu|\chi|_1}							&	\textrm{ for }y\neq y^n_-,\\
									\sum_{\chi\in\mathcal X^{N-n}\setminus\{0\}}\e^{-\mu|\chi|_1}		&	\textrm{ for }y=y^n_-,											
								\end{array}\right.\\
	\end{equation}
and for all $y\in\mathcal V_L^n(\Lambda_L)$ 
	\begin{equation}
		d_L^N(y\cup b_{N-n,0},c^{N-n})\ge d_L^{n+1}(y\cup\{a_{-,1}\},\mathcal V_{L,1}^{n+1})
			\ge\left\{	\begin{array}{ll} 
								h_L^n(y)+1		&	\textrm{for }y\neq y^n_-,\\
								h_L^n(y)	&	\textrm{for }y= y^n_-.
							\end{array}\right.
	\end{equation}
Lemma~\ref{lem:geomsum} then implies
	\begin{equation}\label{eq:allthespecialcases}
		\sum_{z\in\mathcal A_{N-n,0}(y)}\e^{-\mu d_L^N(y\cup z,\mathcal V_{L,1}^N)}\le\left\{		
								\begin{array}{ll}
									(1+30\e^{-\mu})	\e^{-\mu}\e^{-\mu h^n_L(y)}				&	\textrm{ for }y\neq y^n_-,\\
									30\e^{-\mu}\e^{-\mu h^n_L(y)}									&	\textrm{ for }y=y^n_-.				
								\end{array}\right.\\
	\end{equation}
Hence, by \eqref{eq:Apartition}, \eqref{eq:allmostthere} and \eqref{eq:allthespecialcases}, as well as the definition of $h_L$
	\begin{equation}
		\sum_{z\in\mathcal A(y)}\e^{-\mu d_L^N(y\cup z,\mathcal V_{L,1}^N)}\le 332\e^{-\mu}\e^{-\mu h^n_L(y)}
	\end{equation}
where we used $\mu\ge \ln 2$. Together with \eqref{eq:alphaatitswork}, this concludes the proposition.
\end{proof}
%%%%%%%%%%%%%%%%%%%%%%%%%%%%%%%%%%%%%%%%%%%%%%%%
\begin{lemma}\label{lem:Sumtrace}
There exists $L_0\equiv L_0(\epsilon)>0$ such that for all $L\ge L_0$, $n\in\N$ with $N/2< n<N$ and $\mu\ge \ln 2$ holds
	\begin{equation}
		\sum_{y\in\mathcal V^n(\Lambda_L)\setminus\{y_\pm^n\}}\e^{-\mu h^n_L(y)}\le  2^{11}\e^{-\mu}.
	\end{equation}
\end{lemma}
\begin{proof}
For any $y\in\mathcal V^n(\Lambda_L)$ holds $y^\prime:=y\cup\{a_{+,1}\}\subseteq \mathcal S_{L,M}(1/4-\epsilon/2)$. Hence, according to Lemma~\ref{lemma:middle}, we have $y_\kappa\in\mathcal W_L(y^\prime)$ with $\kappa:=\lfloor(n+2)/2\rfloor$. 
For any $j\in\{0,\hdots,|\Lambda_L|-n\}$ let 
	\begin{equation}
		\mathcal B_j^n:=\{y\in\mathcal V^n(\Lambda_L):\;y_\kappa={\lambda_+-(n-\kappa+j)}\}.
	\end{equation}
Hence
	\begin{equation}\label{eq:mathcalBln}
		\mathcal V^n(\Lambda_L)=\bigcup_{j=0}^{|\Lambda_L|-n}\mathcal B^n_j.
	\end{equation}
Let us now consider a $y\in\mathcal B_j^n$ for a $j\in\{0,\hdots,|\Lambda_L|-n\}$. Let $c:=c_{L,y_\kappa}^{n+1}$. We define $\chi^\prime_-\equiv \chi^\prime_-(y)\in\mathcal X^{\kappa-1}$ and $\chi^\prime_+\equiv \chi^\prime_+(y)\in\mathcal X^{n-\kappa}$ such that
	\begin{align}
	\chi^\prime_{-,j}(y)&:=|(y_\kappa-j)-c_{\kappa-j}| \quad\textrm{for }1\le j\le \kappa-1,\\
	\chi^\prime_{+,j}(y)&:=|(y_\kappa+j)-c_{\kappa+j}| \quad\textrm{for }1\le j\le n-\kappa.
	\end{align}
Then,
	\begin{equation}
		d_L^{n+1}(y\cup\{a_{+,1}\},\mathcal V_{L,1}^{n+1})=\sum_{j=1}^{\kappa-1}\chi^\prime_{-,j}+\sum_{j=1}^{n-\kappa}\chi^\prime_{+,j}+j.
	\end{equation}
Hence, according to Lemma~\ref{lem:geomsum}, we obtain for all $j\in\{0,\hdots,|\Lambda_L|-n\}$ and all $\mu\ge\ln 2$,
	\begin{equation}
		\sum_{y\in\mathcal B_j^n}\e^{-\mu d_L^{n+1}(y\cup\{a_{+,1}\},\mathcal V_{L,1}^{n+1})}\le \e^{-\mu j}(1+30\e^{-\mu})^2.
	\end{equation}
Therefore, by \eqref{eq:mathcalBln} we have
	\begin{equation}\label{eq:1022}
		\sum_{y\in\mathcal V^n(\Lambda_L)\setminus\{y_+^n\}}\e^{-\mu d_L^{n+1}(y\cup\{a_{+,1}\},\mathcal V_{L,1}^{n+1})}\le 
		\Big(1+\frac{\e^{-\mu}}{1-\e^{-\mu}}\Big)(1+30\e^{-\mu})^2-1
		\le 1022\e^{-\mu}
	\end{equation}
where we used, that $\mu\ge\ln 2$. By an analogous method we obtain the same bound for the sum over $\exp(d_L^{n+1}(y\cup\{a_{-,1}\},\mathcal V_{L,1}^{n+1}))$.

Let $L_0>0$ such that $L_0^{1/4}-\ln L_0>2$. By using $n\le N<\mu L$, we get for all $L\ge L_0$ that
	\begin{equation}
		\sum_{y\in\mathcal V^n(\Lambda_L)\setminus\{y_\pm^n\}}\e^{-\mu(L^{1+\alpha/2}-1)}\le L^n\e^{-\mu (L^{1+\alpha/2}-1)}\le
		 \e^{-\mu (L(L^{\alpha/2}-\ln L)-1)}\le \e^{-\mu}
	\end{equation}
By the definition of $h^n_L$ in \eqref{eq:hl} we obtain
	\begin{equation}\label{eq:theendisnear}
		\sum_{y\in\mathcal V^n(\Lambda_L)\setminus\{y_\pm^n\}}\e^{-\mu h^n_L(y)}\le \sum_{\eta\in\{\pm\}}\sum_{y\in\mathcal V^n(\Lambda_L)\setminus\{y_\eta^n\}}\e^{\mu}
		\e^{-\mu d_L^{n+1}(y\cup\{a_{\eta,1}\},\mathcal V_{L,1}^{n+1})}
		+\e^{-\mu}\le 2^{11}\e^{-\mu},
	\end{equation}
 where we used \eqref{eq:1022}. 
\end{proof}
% !TEX root = main.tex

\section{Perturbation of the Ising limit} \label{sec:5}
%%%%%%%%%%%%%%%%%%%%%%%%%%%%%%%%%%%%%%%%%%%%%%%

For the whole section let $\epsilon\in(0,1/16)$ and $\theta\in(\epsilon,1/16)$. For $L\in\N$, let $N\equiv N(L):=\lfloor\epsilon L\rfloor$. As before,  for the reader's convenience, we will omit the indices $N$ and $L$ in the following proofs. 
Let $\Lambda_L:=\mathcal S_{L,M}(\theta)$, where  $M:=\lfloor(L-1)/2\rfloor$ and the sector $\mathcal S_{L,M}(\theta)$ was defined in \eqref{Def:S}.

%%%%%%%%%%%%%%%%%%%%%%%%%%%%%%%%%%%%%%%%%%%%%%%
%%%%%%%%%%%%%%%%%%%%%%%%%%%%%%%%%%%%%%%%%%%%%%%
\subsection{The mass of the droplet configurations}
%%%%%%%%%%%%%%%%%%%%%%%%%%%%%%%%%%%%%%%%%%%%%%%
Firstly, given any low-energy eigenstate let us examine the contribution of the droplet configurations.
%%%%%%%%%%%%%%%%%%%%%%%%%%%%%%%%%%%%%%%%%%%%%%%
\begin{lemma}\label{lem:bigestEntry}
Let  $\Delta>3$ such that $\mu_1(\Delta)\ge \ln 2$, where $\mu_1$ was defined in Corollary~\ref{cor:EigenfktEst}. Moreover, let $\gamma\in\mathcal{V}_L$. Then
	\begin{equation}\label{eq:UperLowerBoundClusterState}
		\frac{1}{L}\Big(1-{2^{17}}\e^{-2\mu_1}\Big)\le|\braket{\delta_x^L,{\varphi_{L,\gamma}^N}}|^2\le \frac{1}{L}
	\end{equation}
for all $x\in\mathcal V_{L,1}^N$, where $\ket{\varphi_{L,\gamma}^N(\Delta)}$ was defined in Remark~\ref{rem:allarethesame}.
\end{lemma}
\begin{proof}
Analogously to \eqref{eq:natureofphi}, the definition of $\ket{\varphi_{\gamma}}$ implies that 
	\begin{equation}
		|\braket{\delta_{x},\varphi_{\gamma}}|=|\braket{\delta_{x_0},\varphi_{\gamma}}|
	\end{equation}
for all droplets $x\in[x_0]$, where $x_0\in\widehat{\mathcal V}\cap{\mathcal V}_1$ is the unique representative in $\widehat{\mathcal V}$ of a droplet.
Hence, by the results of Corollary~\ref{cor:EigenfktEst}, we have
	\begin{equation}\label{eq:EVareNormalized}
		1=L|\braket{\delta_{x_0},\varphi_\gamma}|^2+\sum_{x\in\mathcal V\setminus\mathcal V_{1}}|
		\braket{\delta_{x},\varphi_\gamma}|^2\le L|\braket{\delta_{x_0},\varphi_\gamma}|^2+\sum_{x\in\mathcal V
		\setminus\mathcal V_{1}}\frac{2^6}{ L\delta^2}
		\e^{-2\mu_1 d(x,\mathcal V_{1})}.
	\end{equation}
The first equality already yields the upper bound in \eqref{eq:UperLowerBoundClusterState}. For the lower bound, we still need to estimate the second term of the right hand side in \eqref{eq:EVareNormalized}. 

Lemma~\ref{lem:summingeverythingup}  allows us to estimate the last term on the right hand side of \eqref{eq:EVareNormalized} in the following way
	\begin{equation}
		1\le L|\braket{\delta_{x_0},\varphi_\gamma}|^2+\frac{2^{15}}{\delta^2}\e^{-2\mu_1}.
	\end{equation}
This concludes the proof.
\end{proof}
%%%%%%%%%%%%%%%%%%%%%%%%%%%%%%%%%%%%%%%%%%%%%%%%
%%%%%%%%%%%%%%%%%%%%%%%%%%%%%%%%%%%%%%%%%%%%%%%%
\begin{lemma}\label{lem:ThePointwiseDifference}
Let $\Delta>3$ such that $\mu_1\ge \ln 2$. Moreover, let $\gamma\in\mathcal V_L$. Then for all $x,x^\prime\in\mathcal V_L^N$
	\begin{equation}
		|\braket{\delta^L_{x},(\rho(\varphi^N_{L,\gamma})-\rho^N_{L,\gamma})\delta^L_{x^\prime}}|\le \frac{2^{17}}{L}
					\left\{\begin{array}{ll}
						\e^{-2\mu_1} & \textrm{if }x,x^\prime\in[\hat x_0],\\
						\e^{-\mu_1(d_L^N(x,\mathcal V_{L,1}^N)+d_L^N(x^\prime,\mathcal V_{L,1}^N))} & \textrm{else.}
					\end{array}\right.
	\end{equation}
\end{lemma}
\begin{proof}
Let again $x_0\in\widehat{\mathcal V}\cap\mathcal V_1$. We only need to discuss the case $x,x^\prime\in[ x_0]$. All other cases follow immediately from Corollary~\ref{cor:EigenfktEst}.
Let $x=T^\zeta\hat x_0$ and $x^\prime=T^\xi x_0$ for some $\xi,\zeta\in\{0,\hdots,L-1\}$. Remark \ref{rem:allarethesame} implies that
	\begin{equation}
		\braket{\delta_{x},\rho(\varphi_\gamma)\delta_{x^\prime}}
		=\e^{\frac{2\pi i}{L}(\zeta-\xi)\gamma}\braket{\delta_{ x_0},\rho(\varphi_\gamma)\delta_{ x_0}}=\e^{\frac{2\pi i}{L}(\zeta-\xi)\gamma}|\braket{\delta_{ x_0},\varphi_\gamma}|^2,
	\end{equation}
while Definition \eqref{eq:rhoinfinity} implies 
	\begin{equation}
		\braket{\delta_{x},\rho_\gamma\delta_{x^\prime}}
		=\e^{\frac{2\pi i}{L}(\zeta-\xi)\gamma}\braket{\delta_{ x_0},\rho_\gamma\delta_{ x_0}}=\frac{1}{L}\e^{\frac{2\pi i}{L}(\zeta-\xi)\gamma}.
	\end{equation}
By applying Lemma \ref{lem:bigestEntry} we obtain
	\begin{equation}
		|\braket{\delta^L_{x},(\rho(\varphi_\gamma)-\rho_{\gamma})\delta_{x^\prime}}|\le 
		\frac{2^{17}}{L}\e^{-2\mu_1}.
	\end{equation}
\end{proof}
%%%%%%%%%%%%%%%%%%%%%%%%%%%%%%%%%%%%%%%%%%%%%%%%
\begin{lemma}\label{lem:EstOfDiff}
Let $\Delta>3$ such that $\mu_1\ge \ln 2$ and let $\gamma\in\mathcal V_L$. Then, there exists an $L_0\equiv L_0(\epsilon)$ such that for all $L\geq L_0$, all $n\in\N$, $N/2<n<N$, and all $y,y^\prime\in\mathcal V^n({\Lambda_L})$, we have
	\begin{equation}\label{eq:PartTraceEst}
		|\braket{\delta^{\Lambda_L}_{y},(\rho^n_{\Lambda_L}(\varphi^N_{L,\gamma})-\rho^n_{L,\Lambda_L,\gamma})\delta^{\Lambda_L}_{y^\prime}}|
		\le \frac{2^{34}}{L}\e^{-\mu_1}\e^{-\mu_1(h^n_L(y)+h^n_L(y^\prime))} 
	\end{equation}
where $h^n_L$ was defined in Lemma \ref{lem:thebeesknees}.
\end{lemma}
\begin{proof}
Let $z_-^n:=b_{N-n,0}$ and $z_+^n:=b_{0,N-n}$. Then for all $y\in\mathcal V^n(\Lambda)$ and all $z\in\mathcal V^{N-n}(\Lambda^c)\setminus\{z_\pm^n\}$  we have that $y\cup z\notin\mathcal V_{1}$ is not a droplet configuration.
By Lemma \ref{lem:EstimateReducedDensity} we obtain for all $y,y^\prime\in\mathcal V^n(\Lambda_L)$ that
	\begin{align}
		|\braket{\delta^{\Lambda}_{y},(\rho^n_{\Lambda}(\varphi_{\gamma})-\rho^n_{\Lambda,\gamma})\delta^{\Lambda}_{y^\prime}}|&\le 
		\sum_{\eta\in\{\pm\}}|\braket{\delta_{y\cup z_\eta^n},
		(\rho(\varphi_\gamma)-\rho_{\gamma})\delta_{y^\prime\cup z_\eta^n}}|\notag\\
		&\quad+\sum_{z\in\mathcal V_\Lambda^{N-n}\setminus\{z^n_{\pm}\}} |\braket{\delta_{y\cup z},
		\rho(\varphi_\gamma)\delta_{y^\prime\cup z}}|.
	\end{align}
By Theorem~\ref{thm:EigenfktEst} and Lemma~\ref{lem:thebeesknees}, using Cauchy-Schwarz, we further estimate
	\begin{align}
		\sum_{z\in\mathcal V_\Lambda^{N-n}\setminus\{z^n_{\pm}\}} |\braket{\delta_{y\cup z},
		\rho(\varphi_\gamma)\delta_{y^\prime\cup z}}|\le \frac{2^8}{L}333\e^{-2\mu_1}\e^{-\mu_1(h^n(y)+h^n(y^\prime))}, \label{eq:Indefatigable}
	\end{align}
for all $L\geq L_0$, where $L_0\equiv L_0(\epsilon)$ was given in Lemma \ref{lem:thebeesknees}.
Moreover, according to Lemma~\ref{lem:ThePointwiseDifference} we derive the estimate
	\begin{equation}\label{eq:Hotspur}
		|\braket{\delta_{y\cup z_\eta^n},(\rho(\varphi_\gamma)-\rho_{\gamma})\delta_{y^\prime\cup z_\eta^n}}|\le \frac{2^{17}}{L}
		\left\{		\begin{array}{ll}
						\e^{-2\mu_1}																																	&	\textrm{if }y=y^\prime=y_\eta^n,\\
						\e^{-\mu_1(d(y\cup z_\eta^n,\mathcal V_{1})+d(y^\prime\cup z_\eta^n,\mathcal V_{1}))}					&	\textrm{else}
					\end{array}\right.
	\end{equation}
 for all $\eta\in\{\pm\}$. Lemma \ref{lem:thebeesknees} implies for all $\eta\in\{\pm\}$ and all $y\in\mathcal V^{n}(\Lambda)\setminus\{ y_\eta^n\}$ that
	\begin{equation}\label{eq:Renown}
		\e^{-\mu_1 d(y\cup z_\eta^n,\mathcal V_{1})}\le 333\e^{-\mu_1}\e^{-\mu_1 h^n(y)},
	\end{equation}
since in this case $y\cup z^n_\eta\notin\mathcal V_{1}^n$. On the other hand, if $y=y_\eta^n$ we have 
	\begin{equation}\label{eq:Lydia}
		\e^{-\mu_1d(y_\eta^n\cup z_\eta^n,\mathcal V_{1})}=1=\e^{-\mu_1h^n(y_\eta^n)}.
	\end{equation}
Finally, if $y,y^\prime\in\{y^n_\pm\}$ with $y^\prime\neq y$ we have
	\begin{equation}\label{eq:LtBush}
		\e^{-\mu_1(d(y\cup z_\eta^n,\mathcal V_{1})+d(y^\prime\cup z_\eta^n,\mathcal V_{1}))}	\le\e^{-\mu_1}.
	\end{equation}
By combining \eqref{eq:Indefatigable}, \eqref{eq:Hotspur}, \eqref{eq:Renown}, \eqref{eq:Lydia} and \eqref{eq:LtBush}, we conclude the proof.
\end{proof}

%%%%%%%%%%%%%%%%%%%%%%%%%%%%%%%%%%%%%%%%%%%%%%%%
%%%%%%%%%%%%%%%%%%%%%%%%%%%%%%%%%%%%%%%%%%%%%%%%
\subsection{Eigenvalue estimates}
%%%%%%%%%%%%%%%%%%%%%%%%%%%%%%%%%%%%%%%%%%%%%%%%
For this section let $n\in\N$ with $N/2<n<N$ be fixed.
The objective here is to prove the convergence of $\rho^n_{\Lambda_L}(\varphi^N_{L,\gamma}(\Delta))$ to $\rho^n_{\Lambda_L,\gamma}$ in the Schatten-quasinorm $\|\cdot\|_{1/p}$  for a any  $p\in(1,\infty)$. Firstly, we 
establish estimates for the eigenvalues of the afore mentioned operator. For an operator $A\in L(\mathbb H)$ acting on a finite dimensional Hilbert space $\mathbb H$ let $\{\lambda_j(A)\}_{j\le\textrm{dim}\mathbb H}$ denote 
the singular values of $A$ in descending order. If $A$ is a self adjoint operator, these are also the absolute values of the eigenvalues of $A$.

Let  $\Xi^n_L:\;\big\{1,\hdots,|\mathcal V^n(\Lambda_L)|\big\}\rightarrow \mathcal V^n(\Lambda_L)$ be a bijective map such that  $h^n_L\circ\Xi^n_L$ is monotonously decreasing. In a way, the function $\Xi^n_L$ orders the configurations  $y\in\mathcal V^n(\Lambda_L)$ with respect to $h^n_L(y)$.

%%%%%%%%%%%%%%%%%%%%%%%%%%%%%%%%%%%%%%%%%%%%%%%%
\begin{lemma}\label{lem:eigenvalueEst}
Let  $\Delta> 3$ such that $\mu_1\ge \ln 2$ and let $\gamma\in\mathcal V_L$. Then, there exists a $L_0>0$ such that for all $L\ge L_0$ and all $j\in\N$ with $j\le \dim \mathbb H^n_{\Lambda_L}$ holds
	\begin{equation}
		\lambda_j(\rho^n_{\Lambda_L}(\varphi^N_{L,\gamma})-\rho^n_{L,\Lambda_L,\gamma})\le \frac{2^{45}\e^{-\mu_1}}{L}\e^{-\mu_1h^n_L\circ\Xi^n_L(\lceil j/2\rceil)}.
	\end{equation}
\end{lemma}
\begin{proof}
Let $A^n:=(\rho^n_{\Lambda}(\varphi_{\gamma})-\rho^n_{\Lambda,\gamma})$ and $m:=\dim\mathbb H_{\Lambda}^n=|\mathcal V^n(\Lambda)|$. Now, we  split up $A^n$ into a sum of operators of lower rank. For all $j\in\{1,\hdots,m\}$ we define
	\begin{equation}
		A_j^n:=\braket{  \delta^{\Lambda}_{\Xi^n(j)},A^n\delta^{\Lambda}_{\Xi^n(j)}}\ket{  \delta^{\Lambda}_{\Xi^n(j)}}\!
		\bra{  \delta^{\Lambda_L}_{\Xi^n(j)}}+
		\sum_{l> j}\big(\braket{  \delta^{\Lambda}_{\Xi^n(l)},A^n  \delta^{\Lambda}_{\Xi^n(j)}}\ket{  \delta^{\Lambda}_{\Xi^n(l)}}\!
		\bra{\delta^{\Lambda}_{\Xi^n(j)}}+h.c.\big)
	\end{equation}
 These are operators of rank less than or equal to two. Moreover,
	\begin{equation}
		A^n=\sum_{j=1}^mA^n_{j}.
	\end{equation}	
First, we estimate the largest eigenvalue of $R^n_j:=\sum_{k> j}A^n_k$ for all $j\in\{0,\hdots, m\}$.  We obtain
	\begin{equation}\label{eq:Rnj}
		\lambda_1(R^n_j)\le\sup_{\psi\neq 0}\frac{\|R^n_j\psi\|_\infty}{\|\psi\|_\infty}=
		\max_{k>j}\sum_{l>j}|\braket{\delta^{\Lambda}_{\Xi^n(k)},A^n\delta^{\Lambda}_{\Xi^n(l)}}| ,
	\end{equation}
where $\|\cdot\|_\infty$ denotes the supremum norm on $\mathbb H_{\Lambda}^n\cong \C^m$. According to Lemma~\ref{lem:Sumtrace} and Lemma~\ref{lem:EstOfDiff} there exists a $L_0\equiv L_0(\alpha,\epsilon)$ such that 
	\begin{equation}\label{eq:estNorm1}
		\lambda_1(R^n_j)\le\frac{2^{34}\e^{-\mu_1}}{ L}\e^{-\mu_1h^n\circ\Xi^n(j+1)}\sum_{l>j}\e^{-\mu_1h^n\circ\Xi^n(l)}
		\le\frac{2^{45}\e^{-\mu_1}}{ L}\e^{-\mu_1h^n\circ\Xi^n(j+1)}
	\end{equation}
 for all $L\ge L_0$, where we used the monotonicity of $h^n\circ \Xi^n$ and $\mu_1\ge \ln 2$. Since $R_0^n=A^n$ this also implies 
	\begin{equation}
		\lambda_1(A^n)\le \frac{2^{45}\e^{-\mu_1}}{L}.
	\end{equation}
Let $S^n_j:=A^n-R_j^n$  for all $j\in\{0,\hdots,m\}$. Hence $\rank(S^n_j)\le\sum_{k=1}^j\rank(A^n_k)\le 2j$, and therefore also
	\begin{equation}\label{eq:rankest}
		\lambda_{2j+1}(S^n_j)=0.
	\end{equation}
The operator $A^n$ is self adjoint and $A^n=S^n_j+R^n_j$ for all $j\in\{0,\hdots,m\}$. By a well-known inequality for singular values \cite{MR1144277}, we deduce that for all $j\in\N$ with $2j+1\le m$ or $2j+2\le m$, it holds
	\begin{equation}
		\lambda_{2j+2}(A^n)\le\lambda_{2j+1}(A^n)\le \lambda_{2j+1}(S^n_j)
		+\lambda_{1}(R^n_j)\le \frac{2^{45}\e^{-\mu_1}}{L}\e^{-\mu_1h^n\circ \Xi^n(j+1)},
	\end{equation}
where we used  \eqref{eq:rankest}.  Lastly, we note that for all $j\le 2$, we have
	\begin{equation}
		\lambda_{j}(A^n)\le\lambda_1(A^n)\le \frac{2^{45}\e^{-\mu_1}}{L}\e^{-\mu_1h^n\circ\Xi^n(1)},
	\end{equation}
where we used that $h^n(\Xi^n(1))=h^n(y^n_\pm)=0$.
\end{proof}
%%%%%%%%%%%%%%%%%%%%%%%%%%%%%%%%%%%%%%%%%%%%%%%%%%%%
\begin{lemma}\label{lem:pquasiNorm}
Let $p\in(1,\infty)$ and $\Delta>3$ such that $\mu_1/p\ge \ln 2$ and let $\gamma\in\mathcal V_L$. There exists a $L_0\equiv L_0(\epsilon)>0$ such that for all $L\ge L_0$ we obtain
	\begin{equation}
		\|\rho^n_{\Lambda_L}(\varphi^N_{L,\gamma})-\rho^n_{L,\Lambda_L,\gamma}\|_{1/p}^{1/p}\le \frac{2^{56}}{L^{1/p}}\e^{-\mu_1/p}.
	\end{equation}
\end{lemma}
\begin{proof}
Let us again define $A^n:=(\rho^n_{\Lambda}(\varphi_\gamma)-\rho^n_{\Lambda,\gamma})$ and $m:=\dim \mathbb H^n_{\Lambda}$. Then
	\begin{equation}
		\|A^n\|_{1/p}^{1/p}=\sum_{j=1}^{m}\lambda_j^{1/p}(A^n).
	\end{equation}
We remark that $\{y^n_\pm\}\subseteq\{y:\;h^n(y)=0\}$. Therefore, by Lemma~\ref{lem:Sumtrace} and Lemma~\ref{lem:eigenvalueEst}, there exists a $L_0\equiv L_0(\epsilon)>0$ such that for all $L\ge L_0$ holds
	\begin{equation}
		\|A^n\|_{1/p}^{1/p}\le \frac{2^{45/p}\e^{-\mu_1/p}}{L^{1/p}}\left(2+2^{11}\e^{-\mu_1/p}\right)\le  \frac{2^{56}}{L^{1/p}}\e^{-\mu_1/p},
	\end{equation}
where we used that $\mu_1/p\ge \ln 2$.
\end{proof}

% !TEX root = main.tex

\subsection{Proof of Theorem  \ref{thm:main}} 
%%%%%%%%%%%%%%%%%%%%%%%%%%%%%%%%%%%%%%%%%%
We are now prepared to prove the logarithmically corrected area law as stated in Theorem \ref{thm:main}. As we already showed in Section~\ref{cap:Isinglimit}, the entanglement entropy of the density $\rho_{L,\gamma}^N$ in the Ising-limit ``$\Delta=\infty$" satisfies this scaling behavior.  

We use the formalism of spectral shift functions to control the difference in the entanglement entropy. For a finite dimensional vector space $\mathbb H$, the spectral shift function of a selfadjoint operator $A\in L(\mathbb H)$ and a selfadjoint perturbation $B\in L(\mathbb H)$ is given by $\xi(\cdot;A,A+B):\;\R\rightarrow\R$,
	\begin{equation}
		\xi(E;A,A+B):=\tr\{1_{\le E}(A+B)-1_{\le E}(A)\}.
	\end{equation}
According to \cite{MR1824200}, for any $p\in[1,\infty)$ the $L^p$-norm of the spectral shift functions satisfies 
	\begin{equation}\label{eq:LpSSF}
		\|\xi(\cdot;A,A+B)\|_p\le\|B\|^{1/p}_{1/p}.
	\end{equation}
%%%%%%%%%%%%%%%%%%%%%%%%%%%%%%%%%%%%%%%%%%
\begin{lemma}\label{lem:SSFonFkts}
 Let $\mathbb H$ be a finite dimensional Hilbert space, $A,B\in L(\mathbb H)$ be self adjoint operators such that $\sigma(A),\sigma(A+B)\subseteq[0,1]$. Let $p,q\in(1,\infty)$ such that $1/p+1/q=1$. Then
	\begin{equation}\label{eq:DifAAS}
		\big|\tr\{s(A+B)-s(A)\}\big|
		\le\|B\|_{1/p}^{1/p}(1+\|\ln(\cdot)1_{(0,1)}(\cdot)\|_{q}).
	\end{equation}
\end{lemma}
\begin{proof}
Kre{\u \i}n's theorem for the spectral shift function \cite{schmuedgen2012unbounded} states that
	\begin{equation}\label{eq:Krein}
		\tr\{f(A+B)-f(A)\}=\int_{\R}\;f^\prime(t)\xi(t;A,A+B) \:\dd t
	\end{equation}
for any compactly supported and smooth function $f\in C_c^\infty(\R)$. Since $s$ is not differentiable in $0$, we cannot apply this result directly. We therefore define a family of suitable auxiliary functions $(s_\eta)_{\eta\in\N}\in C_0^\infty(\R)$ such that $\lim_{\eta\rightarrow \infty}s_\eta(t)=s(t)$  for all $t\in[0,1]$. Let $\chi\in C^\infty(\R)$ be a function, such that $\chi(\R)=[0,1]$, $\chi(t)=0$ for $t\le 1/2$ and $\chi(t)=1$ for $t\ge1$. For $\eta\in\N$ and $\tau\in\R$ let
	\begin{equation}
		s_\eta(\tau):=\chi(2-\tau)\int_0^\tau s^\prime(t)\chi(\eta t)\;\dd t.
	\end{equation}
Since both $s_\eta(0)=s(0)=0$ and $\lim_{\eta\rightarrow0}\|(s^\prime_\eta-s^\prime)1_{(0,1)}\|_p=0$ for all $p\in[1,\infty)$ we conclude that $\lim_{\eta\rightarrow\infty}s_\eta(t)=s(t)$ for all $t\in[0,1]$. Both, $A$ and $A+B$ have a finite number of eigenvalues. Hence
	\begin{equation}\label{eq:Diffhheta}
		\lim_{\eta\rightarrow\infty}\big|\tr\{s_{\eta}(A+B)-s_{\eta}(A)\}-\tr\{s(A+B)-s(A)\}\big|=0.
	\end{equation}
For any $p,q\in(1,\infty)$, $1/p+1/q=1$, and $\eta\in\N$ we obtain by applying \eqref{eq:Krein} to $s_{\eta}$ that
	\begin{equation}
		\big|\tr\{s_{\eta}(A+B)-s_{\eta}(A)\}\big|\le
		\|\xi(\cdot;A,A+B)\|_p\|s^\prime_{\eta}1_{(0,1)}\|_q,
	\end{equation}
where we used that $\xi(t;A,A+B)=0$ for $t>1$.
The first term of the right hand side can be estimated by \eqref{eq:LpSSF}. We estimate the second term by
	\begin{equation}
		\|s^\prime_{\eta}1_{(0,1)}\|_q=\|s^\prime(\cdot)\chi(\eta\cdot)1_{(0,1)}(\cdot)\|_q\le\|s^\prime1_{(0,1)}\|_q\le 1+
		\|\ln(\cdot)1_{(0,1)}(\cdot)\|_q.
	\end{equation}
Together with \eqref{eq:Diffhheta} this yields \eqref{eq:DifAAS}.
\end{proof}
%%%%%%%%%%%%%%%%%%%%%%%%%%%%%%%%%%%%%%%%%%
\begin{remark}\label{rem:log}
Observe that for all $q\in(1,\infty)$ we obtain the elementary estimate
	\begin{equation}
		\|\ln(\cdot)1_{(0,1)}(\cdot)\|_q =\Gamma(q+1)^{1/q}\le 2q,
	\end{equation}
where $\Gamma$ denotes the Gamma function.
\end{remark}
%%%%%%%%%%%%%%%%%%%%%%%%%%%%%%%%%%%%%%%%%%%
\begin{proof}[Proof of Theorem \ref{thm:main}]

By \eqref{eq:sumofspectrum}, for every $E\in\sigma(H_L^N)\cap I_{1}$, there exists at least one $\gamma\in \mathcal{V}_L$ such that $E=\inf \sigma(\hat{H}_{L,\gamma}^N)$. Let $\ket{\varphi_{L,\gamma}^N}$ be the corresponding eigenvector.

Let $n\in\N$ with $N/2<n<N$ and $p,q>1$ such that $1=1/p+1/q$. Let $\Delta> 3$, such that $\mu_1(\Delta)/p\ge \ln 2$. According to Lemma~\ref{lem:pquasiNorm}, Lemma~\ref{lem:SSFonFkts} and Remark~\ref{rem:log}, there exists a $L_0^\prime\equiv L^\prime_0(\epsilon)>\e^2$ such that
	\begin{equation}\label{eq:SDiff}
		|\tr\{s(\rho^n_{\Lambda_L}(\varphi^N_{L,\gamma}))-s(\rho^n_{L,\Lambda_L,\gamma})\}|\le\frac{2^{56}}{L^{1/p}}\e^{-\mu_1/p}(1+2q)
	\end{equation}
for all $L\ge L_0^\prime$.
 We now choose $p,q>1$ to depend on $L$ as follows. Let $q\equiv q(L):=\ln(L)$ and $p\equiv p(L):=(1-1/\ln(L))^{-1}$. This implies $L^{1/p}=\e^{-1} L$. For all $L\ge\e^2$, we have $1/p\ge 1/2$ and if $\Delta>25$, then this implies $\mu_1(\Delta)/2\ge \ln 2$. For $L\ge L_0^\prime$ we bound \eqref{eq:SDiff} by
	\begin{equation}
		|\tr\{s(\rho^n_{\Lambda_L}(\varphi^N_{L,\gamma}))-s(\rho^n_{L,\Lambda_L,\gamma})\}|\le\frac{2^{57}}{L}\e^{-\mu_1/2}(1+2\ln(L)).
	\end{equation}
Corollary~\ref{cor:EigenfktEst} implies that $\mu_1(\Delta)\rightarrow\infty$ for $\Delta\rightarrow\infty$. Therefore, there exists a $\Delta_0\ge 25$ such that 
	\begin{equation}\label{eq:Deltatoinfty}
		2^{57}\e^{-\mu_1/2}\le1/2
	\end{equation}
for all $\Delta\ge\Delta_0$. Hence, by applying \eqref{eq:Srho0} as well as \eqref{eq:Deltatoinfty}, we obtain
	\begin{equation}
		\tr\{s(\rho^n_{\Lambda_L}(\varphi^N_{L,\gamma}))\}\ge\tr\{s(\rho^n_{L,\Lambda_L,\gamma})\}-
		|\tr\{s(\rho^n_{\Lambda_L}(\varphi^N_{L,\gamma}))-s(\rho^n_{L,\Lambda_L,\gamma})\}|\ge \frac{\ln L-1}{L}.
	\end{equation}
This implies for the entanglement entropy that
	\begin{equation}
		S(\varphi^N_{L,\gamma},\Lambda_L)\ge\sum_{\genfrac..{0pt}{}{n\in\N,}{N/2<n<N}}
		\tr\{s(\rho^n_{\Lambda_L}(\varphi^N_{L,\gamma}))\}\ge (N/2-1)\frac{\ln L-1}{L}.
	\end{equation}
We notice, that $\lim_{L\rightarrow\infty}\frac{N/2-1}{L}=\epsilon/2$. 
Hence, for all $\Delta\ge\Delta_0$ we have
	\begin{equation}
		\liminf_{L\rightarrow\infty}\frac{S(\varphi^N_{L,\gamma},\Lambda_L)}{\ln L}\ge\frac{\epsilon}{2}.
	\end{equation}
\end{proof}
% !TEX root = main.tex

\appendix
\section{}

%%%%%%%%%%%%%%%%%%%%%%%%%%%%%%%%%%%%%%%%%%%%%%%%%%%%%%%%
\subsection{Uniqueness of fiber operator ground states}
%%%%%%%%%%%%%%%%%%%%%%%%%%%%%%%%%%%%%%%%%%%%%%%%%%%%%%%%
%%%%%%%%%%%%%%%%%%%%%%%%%%%%%%%%%%%%%%%%%%%%%%%%%%%%%%%%%%%
\begin{lemma}\label{lem:Fasereinduetigkeit} 
Let $\Delta>2$, let $L,N\in\N$ with $1<N<L-1$. Then for any $\gamma\in\mathcal V_L$, the operator $\hat{H}_{L,\gamma}^N$ has exactly one eigenvalue in $[1-\frac{1}{\Delta},1]$ and no eigenvalue in $(1,2-\frac{2}{\Delta})$.	
\end{lemma}
\begin{proof} \label{lem:unqiueEV}
By Lemma \ref{lem:WAW}, we get
	\begin{equation}
		\hat{H}_{L,\gamma}^N=-\frac{1}{2\Delta}\hat{A}^N_{L,\gamma}+\hat{W}_{L,\gamma}^N\geq\left(1-\frac{1}{\Delta}\right)\hat{W}_{L,\gamma}^N\:.
	\end{equation}
There exists exactly one element $\hat{x}_0\in\widehat{\mathcal{V}}_L^N\cap \mathcal V_{L,1}^N$. It satisfies $W(\hat{x}_0)=1$. For any other $\hat{x}\in	\widehat{\mathcal{V}}_L^N\setminus\{\hat{x}_0\}$ we have $W(\hat{x})\geq 2$. 
Let $\phi_{L,\gamma,0}^N\in\mathbb S_{L,\gamma}^N$ be defined by $\phi_{L,\gamma,0}^N(\sigma,\hat x):=\delta_{\gamma,\sigma}\delta_{\hat x_0,\hat x}$.

Hence, the operator
	\begin{equation}
		\hat{H}_{L,\gamma}^N+\left(1-\frac{1}{\Delta}\right) \ket{\phi_{L,\gamma,0}^N}\!\bra{\phi_{L,\gamma,0}^N}\geq \left(1-\frac{1}{\Delta}\right)
		\left( \hat{W}_{L,\gamma}^N+ \ket{\phi_{L,\gamma,0}^N}\!\bra{\phi_{L,\gamma,0}^N}\right)\geq \left(2-\frac{2}{\Delta}\right) 
	\end{equation}
is a rank-one perturbation of $\hat{H}_{L,\gamma}^N$. Therefore the unperturbed operator $\hat H_{L,\gamma}^N$ has at most one eigenvalue below $(2-\frac{2}{\Delta})$.
% MinMax Prinzip Kato?
On the other hand, since
$\braket{\phi_{L,\gamma,0}^N,\hat{H}_{L,\gamma}^N\phi_{L,\gamma,0}^N}=1$, there exists at least one eigenvalue which is less than or equal to $1$. Since for $\Delta>2$, we get $1<2-\frac{2}{\Delta}$, this concludes the proof.
\end{proof}

%%%%%%%%%%%%%%%%%%%%%%%%%%%%%%%%%%%%%%%%%%%%%%%%%%%%%%%%
For $\gamma=0$, it follows from the explicit structure of the fiber operator $\hat{H}_{L,0}^N$ that it has a unique ground state $\hat{\varphi}_{L,0}^N$ which can be chosen to be strictly positive. The same is true for the original operator $H_L^N$. This will allow us to conclude that $\varphi_{L,0}^N:=(\mathfrak{F}_L^N)^*\hat{\varphi}_{L,0}^N$ is the ground state of $H_L^N$. The main tool for our result will be an idea presented in \cite{YangYang}, where the existence of a strictly positive ground state for the XXZ model on the ring was established, however let us also point out that this piece of the proof also follows from the Allegretto-Piepenbrink theorem shown in \cite{HaesKell}. 

%%%%%%%%%%%%%%%%%%%%%%%%%%%%%%%%%%%%%%%%%%%%%%%%%%%%%%%%
\begin{lemma}\label{lem:absolutergroundstate}
Let $N,L\in \N$, $0<N<L$. Moreover, let $E_0\equiv E_0(L,N,\Delta)=\inf\sigma(\hat{H}_{L,0})$. Then, $E_0$ is non-degenerate and the corresponding eigenvector $\hat{\varphi}_{L,0}^N\in\mathbb{S}_{L,0}^N$ can be chosen such that $\|\hat{\varphi}_{L,0}^N\|=1$ and $\hat{\varphi}_{L,0}^N(0,\hat{x})>0$ for all $\hat{x}\in\widehat{\mathcal{V}}_L^N$.  In addition, $\varphi_{L,0}^N:=(\mathfrak{F}_L^N)^*\hat{\varphi}_{L,0}^N$ is the unique ground state of $H_L^N$. 
\end{lemma}
\begin{proof} Firstly, note that if we choose the constant $C>2N\geq \|W_L^N\|=\|\hat{W}^N_{L,0}\|$, we then get that the matrix representations of both operators $A_1:=(C\idty_{\mathbb H_L^N}-H_L^N)$ and $A_2:=(C\idty_{\mathbb S_{L,0}^N}-\hat{H}_{L,0}^N)$ have only non--negative entries. Moreover, note that since $A_1$ and $A_2$ are irreducible, we can choose $D\geq \dim(\mathbb H_L^N)$ large enough, such that the matrix entries of $A_1^D$ and $A_2^D$ will all be strictly positive. Hence, by the Perron-Frobenius Theorem, the largest eigenvalue of each of these operators $A_1^D$ and $A_2^D$ is positive, non-degenerate and the corresponding eigenfunctions can be chosen to be strictly positive. Let $\varphi_{L,0}^N$ and $\hat{\varphi}_{L,0}^N$ denote the eigenfunctions for $A_1^D$ and $A_2^D$ respectively, that satisfy these properties. Clearly, $\varphi_{L,0}^N$ and $\hat{\varphi}_{L,0}^N$ will then be the eigenfunctions of $H_L^N$ and $\hat{H}_{L,0}^N$ corresponding to the respective minima $E_0$ and $\hat{E}_0$ of the spectra. Now, since $H_L^N$ and $\hat{H}_L^N$ are unitarily equivalent via the Fourier transform $\mathfrak{F}_L^N$, the function $(\mathfrak{F}_L^N)^*\hat{\varphi}_{L,0}^N$ is also an eigenfunction of $H_L^N$ and thus $\hat{E}_0\in\sigma(H_L^N)$. However, from the explicit form of $(\mathfrak{F}_L^N)^*$ as given in \eqref{eq:FourierAdj}, one sees that since $\hat{\varphi}_{L,0}^N\in\mathbb S_{L,0}^N$ we get that $(\mathfrak{F}_L^N)^*\hat{\varphi}_{L,0}^N$ is a strictly positive eigenfunction of $H_L^N$. Thus, we conclude that $(\mathfrak{F}_L^N)^*\hat{\varphi}_{L,0}^N=\varphi_{L,0}^N$ and consequently $E_0=\hat{E}_0$.
\end{proof}

\subsection{An auxiliary result}

The following lemma is an adaptation of a similar result in \cite[Thm.\ 6.1]{ARFS19}. 

%%%%%%%%%%%%%%%%%%%%%%%%%%%%%%%%%%%%%%%%%%%%%%%%%%%%%
\begin{lemma}\label{lem:geomsum} Let $N\in\N$ and
	\begin{equation}
	\mathcal X^N:=\{\chi=(\chi_1,\hdots,\chi_N)\subseteq\N_0^N:\;\chi_1\le \hdots\le \chi_N\}.
	\end{equation}
	Then, for all $\mu\ge\ln 2$ we have
	\begin{equation}
	\sum_{\chi\in\mathcal X^N\setminus\{0\}}\e^{-\mu |\chi|_1}\le 30\e^{-\mu},
	\end{equation}
	where $|\cdot|_1$-denotes the $\ell^1$-norm of $\Z^N$.
\end{lemma}
\begin{proof}
	Let  $\Psi:\;\N^{N}_0\rightarrow\mathcal X^N$ with
	\begin{equation}
	x\mapsto\Psi(x):=(\Psi_1(x),\hdots,\Psi_N(x))\:,
	\end{equation}
	where for each $j\in\{1,2,\dots,N\}$, we have defined $\psi_j(x):=\sum_{i=1}^jx_i$. Note that $\Psi$ is a bijection.
	
	For any $x\in \N^{N}_0$, we therefore get
	\begin{equation}
	\sum_{j=1}^N\Psi_j(x)=\sum_{k=1}^N(N-k+1)x_k
	\end{equation}
	and $\Psi$ is a bijection, we have
	\begin{equation}\label{eq:geomsum}
	\sum_{\chi\in \mathcal X^N}\e^{-\mu |\chi |_1}=
	\sum_{x\in\N^N_0}\e^{-\mu|\Psi(x)|_1}=\sum_{x\in\N^N_0}\prod_{k=1}^N\e^{-\mu (N-k+1)x_k}\:,
	\end{equation}
	which yields 
	\begin{equation}
	\sum_{\chi\in \mathcal X^N}\e^{-\mu |\chi|_1}=\prod_{k=1}^N\sum_{y=0}^\infty \e^{-\mu y(N-k+1) }=\prod_{k=1}^N\frac{1}{1-\e^{-\mu (N-k+1)}}.
	\end{equation}
	This gives us the following estimate, which is uniform in $N$:
	\begin{equation} \label{eq:infprod}
	\sum_{\chi\in \mathcal X^N}\e^{-\mu |\chi|_1}\le \prod_{m=1}^\infty\frac{1}{1-\e^{-\mu m}}\le\exp\Big(\frac{2\e^{-\mu}}{1-\e^{-\mu}}\Big)\:,
	\end{equation}
	where we have used that $\ln(1-\lambda)^{-1}\le 2\lambda$, whenever $\lambda\in(0,1/2)$.
	Hence,
	\begin{equation}
	\sum_{\chi\in \mathcal X^N\setminus\{0\}}\e^{-\mu |\chi|_1}\le\exp\Big(\frac{2\e^{-\mu}}{1-\e^{-\mu}}\Big)-1\le 4\e^2\e^{-\mu},
	\end{equation}
	since $\e^x-1\le x\e^x$ for all $x\ge 0$ and $\e^{-\mu}\le 2^{-1}$ for $\mu\ge\ln 2$.
\end{proof}

%\bibliography{bibliographyXXZ}
%\bibliographystyle{mybib}

\end{document}